\documentclass{ws-ijmpd}
\usepackage[super,compress]{cite}
\usepackage[dvips,breaklinks]{hyperref}
\hypersetup{colorlinks,urlcolor=black,citecolor=black,linkcolor=black,filecolor=black}
\usepackage{breakurl}
\usepackage[utf8]{inputenc}
\usepackage{rotating}     
\usepackage{graphicx}    
\usepackage{float}
\usepackage{siunitx}
\usepackage{subfigure}
\usepackage[dvipsnames]{xcolor} 
\usepackage{hyperref}
\usepackage[capitalise]{cleveref} 
\usepackage{mathtools} 

\begin{document}
\markboth{Zhen-xiang Hao}
{Spot size estimation of flat-top beams in space-based gravitational wave detectors}

\title{Spot size estimation of flat-top beams in space-based gravitational wave detectors}
\author{$^{1,3}$Zhen-Xiang Hao, $^2$Tim Haase, $^{4,5}$Hong-Bo Jin, $^3$Ya-Zheng Tao, $^{2}$Gudrun Wanner\footnote{Corresponding author: gudrun.wanner@aei.mpg.de}, $^{1,3,5}$Ruo-Xi Wu, $^{1,3,5}$Yue-Liang Wu\footnote{Corresponding author: ylwu@itp.ac.cn}}
\address{$^1$Institute of Theoretical Physics, Chinese Academy of Sciences, No.55 Zhongguancun East Road, Haidian District, Beijing, P.R.China 100190\\
            $^2$Max Planck Institute for Gravitational Physics (Albert Einstein Institute) and 
Institute for Gravitational Physics of the Leibniz Universit\"at Hannover, Callinstr. 38, 30167 Hannover, Germany\\
             $^3$University of Chinese Academy of Sciences, No.19(A) Yuquan Road, Shijingshan District, Beijing, P.R.China 100049\\
             $^4$National Astronomical Observatories, Chinese Academy of Sciences, 20A Datun Road, Chaoyang District, Beijing, P.R.China 100101\\
             $^5$Hangzhou Institute for Advanced Study, University of Chinese Academy of Sciences, Hangzhou, Zhejiang, P.R.China 310024}
\maketitle

\begin{abstract}
Motivated by the necessity of a high-quality stray light control in the detection of the gravitational waves in space, the spot size of a flat top beam generated by the clipping of the Gaussian beam (GB) is studied. By adopting the mode expansion method (MEM) approach to simulating the beam, a slight variant of the definition of the mean square deviation (MSD) spot size for the MEM beam is proposed. This enables us to quickly estimate the spot size for arbitrary propagation distances. Given that the degree of clipping is dependent on the power ratio within the surface of an optical element, the power ratio within the MSD spot range is used as a measure of spot size. The definition is then validated in the cases of simple astigmatic Gaussian beam and nearly-Gaussian beam profiles. As a representative example, the MSD spot size for a top-hat beam in a science interferometer in the detection of gravitational waves in space is then simulated. As in traditional MSD spot size analysis, the spot size is divergent when diffraction is taken into account. A careful error analysis is carried out on the divergence and in the present context, it is argued that this error will have little effect on our estimation. Using the results of our study allows an optimal design of optical systems with top-hat or other types of non-Gaussian beams. Furthermore, it allows testing the interferometry of space-based gravitational wave detectors for beam clipping in optical simulations. The present work will serve as a useful guide in the future system design of the optical bench and the sizes of the optical components.

\end{abstract}
\ccode{PACS Nos.: 42.60.-v, 07.87.+v, 04.80.Nn}

\section{Introduction}
In the detection of gravitational waves in spacetime, the heterodyne interferometry is required to track the relative displacement between two spacecraft that are on the order of $10^6\ \SI{}{kms}$ apart at the picometer level \cite{2017SPIE10565E..2XD,2020ResPh..1602918L,Ruan:2020smc}. At this level of precision, many noises need to be suppressed to enhance the signal to noise ratio of gravitational wave sources. Among these noises is the stray light, which is generated by a number of channels, one of which is the beam clipping when a beam of laser light impinges on optical elements of the optical bench.

In the science phase of the mission, a Gaussian beam originating from a spacecraft propagates to a distant spacecraft several million kilometers away \cite{2017SPIE10565E..2XD,2020ResPh..1602918L,Ruan:2020smc}. Upon arriving at the distant spacecraft, a small part of the Gaussian beam wavefront is clipped and generates a top-hat beam that is propagated into the spacecraft for interferometric measurement. Within the optical bench, another clipping effect occurs in the sense that, on impinging on an optical element, the finite size of the element entails that at the edge of an element diffraction will occur and power loss of the beam is inevitable. This in turn will generate a potential source of stray light and at the picometer level ghost beam interference might occur and disturb the gravitational wave measurement. 

The aim of the present work is to numerically track the time evolution of the spot size of the beam in a LISA or TAIJI type optical bench. With the stray light problem in mind, we shall first propose to use the power ratio within the MSD spot range as a measure of spot size. Here, the MSD spot size for the MEM beam is a slight variant of the definition given by M. A. Porras et al\cite{1992ApOpt..31.6389P}. After certain validations of the proposed definition, the spot size of a flat top beam propagated within an optical bench of a LISA or TAIJI type mission is then studied and our analysis is expected to be useful for the future system design of the laser metrology subsystem. Common to all MSD spot size studies of a diffracted beam, divergence inevitably occurs \cite{1990SPIE.1224....2S,1995OptL...20..124M,1996OptCo.123..679P,2000ApOpt..39.3914A}. A careful error analysis is carried out to understand the impact of the divergence on our result and it is argued that the divergence will have little effect on our conclusion  and hopefully this renders our results more trustworthy. 

The paper is structured as follows. In \cref{sec:method}, in terms of the power ratio within a MSD spot range as the quality measurement of the spot size, a slight variant of the definition of the MSD spot size of a MEM beam \cite{1992ApOpt..31.6389P} is suggested. In \cref{sec:applications}, we validate the formula for the MSD spot size of a MEM beam and the numerical code in the cases of a simple astigmatic Gaussian beam and Nearly-Gaussian beams which can appear in laboratory experiments. In \cref{top_general}, we investigate the  MSD spot size in a top-hat beam case and compare this definition with the spot size calculated from the analytic divergence angle for the top-hat beam \cite{Drege2000}. After that, in section \cref{top_si}, we calculate the estimated value of the MSD spot size in the case of a science interferometer in the detection of gravitational waves in space. Finally, we perform a careful error analysis for the MSD spot size of the top-hat beam in a science interferometer in \cref{err}.

\section{MSD spot size for MEM beams}\label{sec:method}

There are a number of spot size definitions, catering to different problems and situations \cite{siegman1998maybe}. In this section, we shall put forward a definition of spot size which is a slight variant of that given by Porras et al. \cite{1992ApOpt..31.6389P}, again mainly motivated by the stray light problem. Further, we define the power ratio within the surface of an optical element as a measure of the MSD spot size.  

Consider a general beam that propagates along the $z$-axis, the local electric field can be written as
\begin{equation}
 	E(x,y,z,t)=u(x,y,z)\exp{\left(i\omega t\right)}, 
\end{equation} 
where $x$ and $y$ are the Cartesian coordinates in the beam cross-section, $\omega$ is the angular frequency, $t$ is the time and $u=u(x,y,z)$ is the complex amplitude.

Further, the MSD spot size is defined as \cite{1992ApOpt..31.6389P}
\begin{equation}
 \begin{split}
w_j(z)&=2\sqrt{\frac{\iint_{-\infty}^{+\infty} (j-s_j(z))^2I(x,y,z)dxdy}{P_{in}}}\\
&=\sqrt{\frac{\iint_{-\infty}^{+\infty} j^2I(x,y,z)dxdy}{P_{in}}-s_j(z)^2},\label{MSD_spot_general}
\end{split}
\end{equation}
which is two times of the square root of the variance of the intensity profile normalized by the beam power. 
 $j$ represents $x$ or $y$ for $x$ or $y$ direction, $P_{in}=\iint_{-\infty}^{+\infty}I(x,y,z)dxdy$ is the total power of the beam, $I(x,y,z)=E(x,y,z)E^*(x,y,z)$ is the intensity of the beam for point $(x,y,z)$ and $s_j(z)$ is the mean value of the transversal position of the beam for one dimension, which is also called ``energy center'' \cite{Vlasov1971} 
\begin{equation}
s_j(z)=\frac{\iint_{-\infty}^{+\infty}jI(x,y,z)dxdy}{P_{in}}.\label{cen_general}
\end{equation}

Based on the fact that Hermite-Gaussian (HG) modes are a complete orthonormal set of basis functions, any electric field may be expanded in terms of an infinite superposition of HG modes \cite{2016LRR....19....1B}. This is called the mode expansion method (MEM). As for real applications, the infinite series needs to be truncated at a finite number of modes as an approximation. The complex amplitude $u=u(x,y,z)$ may be approximated as
\begin{equation}
 u(x,y,z)\approx \sum_{m=0}^{N}\sum_{n=0}^{N-m}a_{mn}u_{mn}(x,y,z).\label{superhgdecom_finite}
\end{equation}
For notation convenience, we rewrite $u_{mn}(x,y,z)$ and $a_{mn}$ as
\begin{equation}
	u_{mn}(x,y,z)=u_{mn}=|u_{mn}|\exp{\left(i\phi_ {mn}\right)},\label{umn_simple}
\end{equation}
\begin{equation}
	a_{mn}=|a_{mn}|\exp{\left(i\beta_{mn}\right)},
\end{equation}
where the max mode order $N$ is the max value of the sum of mode order $m$ and $n$, $a_{mn}=\iint_{-\infty}^{+\infty}u(x,y,z)u_{mn}^*(x,y,z)dxdy$ is a constant called the complex coefficient for $u_{mn}(x,y,z)$ and  $u_{mn}(x,y,z)$ is the complex amplitude for $(m,n)$ order of symmetric HG modes
 \begin{equation}
 \begin{split}
 	u_{mn}(x,y,z)=&\frac{c_{mn}}{w(z)}H_m\left(\frac{\sqrt{2}x}{w(z)}\right)H_n\left(\frac{\sqrt{2}y}{w(z)}\right)\exp{\left(-\frac{x^2+y^2}{w^2(z)}\right)}\\
	&\cdot\exp{\left(-ikz-ik\frac{x^2+y^2}{2R(z)}+i(m+n+1)\zeta(z)\right)}.
	\label{hmn_complex_amp} 
\end{split}	
 \end{equation}

For simplicity, here we use the symmetric HG modes as decomposition bases. The symmetry here means that the beam parameters $w(x)$, $R(z)$ and $\xi(z)$ for $x$ and $y$ directions are equal. The parameter $w(z)$ is the spot size of the fundamental Gaussian mode $u_{00}(x,y,z)$. It is defined as  $w(z)=w_0\sqrt{1+\left(\frac{z}{z_r}\right)^2}$, where $w_0$ is the waist, $z_r=\frac{\pi{w_0}^2}{\lambda}$ is the Rayleigh range, $\lambda$ is the wavelength and $k=\frac{2\pi}{\lambda}$ is the wave number. The radius of curvature is $R(z)=z\left(1+\left(\frac{z_r}{z}\right)^2\right)$ and the normalization constant is $c_{mn}=\left(\pi m!n!2^{m+n-1}\right)^{-\frac{1}{2}}$. The Gouy phase $\zeta(z)$ of the fundamental Hermite-Gaussian mode is
\begin{equation}
	\zeta(z)=\arctan\frac{z}{z_r}.
\end{equation}
The beam generated by this method is called MEM beam. In this paper, MEM processing is performed with IfoCAD \cite{2012OptCo.285.4831W}. The total power $P_{MEM}$ of the MEM beam can be written as
\begin{equation}
	P_{MEM}=\iint_{-\infty}^{+\infty}Idxdy= \sum_{m=0}^{N}\sum_{n=0}^{N-m}|a_{mn}|^2\label{coe_norm}. 
\end{equation}	
M. A. Porras et al. calculated the MSD spot size and energy center for the superposition of higher-order HG modes \cite{1992ApOpt..31.6389P}, but they leave out the phase term $\phi_{mn}$ in \cref{umn_simple} for the HG mode bases. This leads to the formulas of MSD spot size and energy center being only valid in the waist plane of the basic HG modes of the MEM beam. Here we add the phase term in the representation of HG modes and substitute \cref{superhgdecom_finite} into \cref{cen_general}. Then, the energy center $\left(s_x(z),s_y(z)\right)$  can be rewritten as
\begin{subequations}
\begin{align}
s_x(z)=\frac{w(z)}{P_{MEM}} \sum_{m=0}^{N}\sum_{n=0}^{N-m}|a_{mn}||a_{(m+1)n}|\cos{(\beta_{mn}-\beta_{(m+1)n}-\zeta(z))}\sqrt{m+1}\label{xbahg},\\
s_y(z)=\frac{w(z)}{P_{MEM}} \sum_{m=0}^{N}\sum_{n=0}^{N-m}|a_{mn}||a_{m(n+1)}|\cos{(\beta_{mn}-\beta_{m(n+1)}-\zeta(z))}\sqrt{n+1}\label{ybahg}.
\end{align}\label{aver_pos}
\end{subequations}

The MSD spot size of the MEM beam in x and y directions can be rewritten as
\begin{subequations}
\begin{gather}
\begin{multlined}
	w_x(z)^2=\frac{w(z)^2}{P_{MEM}}\left( \sum_{m=0}^{N}\sum_{n=0}^{N-m}|a_{mn}|^2(2m+1)+2 \sum_{m=0}^{N}\sum_{n=0}^{N-m}|a_{mn}||a_{(m+2)n}|\right.\\
	 \left.\cdot\cos{(\beta_{mn}-\beta_{(m+2)n}-2\zeta(z))}\sqrt{(m+2)(m+1)}\right)-4s_x(z)^2
	\label{wxhg},
\end{multlined}
\\
\begin{multlined}
	w_y(z)^2=\frac{w(z)^2}{P_{MEM}}\left( \sum_{m=0}^{N}\sum_{n=0}^{N-m}|a_{mn}|^2(2n+1)+2 \sum_{m=0}^{N}\sum_{n=0}^{N-m}|a_{mn}||a_{m(n+2)}|\right.\\
	 \left.\cdot\cos{(\beta_{mn}-\beta_{m(n+2)}-2\zeta(z))}\sqrt{(n+2)(n+1)}\right)-4s_y(z)^2
	\label{wyhg}.
\end{multlined}
\end{gather}\label{msd_spot}
\end{subequations}

The above definition of spot size will be the basis for subsequent discussions in our work. 
MSD spot size may be used to calculate the spot size for general beams except for hard-edge diffraction beams \cite{1990SPIE.1224....2S,1992ApOpt..31.6389P}. It has many advantages: 1. It may be used to define a good beam quality factor which characterizes the global spatial behaviour of a laser beam \cite{1990SPIE.1224....2S}; 2. This definition automatically agrees with the definition of the spot size for fundamental GB, HG beam and Laguerre-Gaussian (LG) beam \cite{1980ApOpt..19.1027C,1983ApOpt..22..643P}; 3. The MSD spot size satisfies the ABCD law and the corresponding beam quality factor is invariable when the beam propagates through the purely real ABCD system (the paraxial optical system that can be described by an ABCD ray transfer matrix) \cite{Simon1988}; 4. The MSD spot size always is a hyperbolic function of propagation distance $z$ \cite{1994OptCo.109....5P}.  These advantages in general are not shared by other spot size definitions for general light beams \cite{1994OptCo.109....5P}.

A key question for applications in material processing and fibre transmission is the amount of power inside a spot region \cite{weber1992some}. One typical criterion of the spot size definition is the power ratio within the spot range \cite{1980ApOpt..19.1027C,1983ApOpt..22..643P}. For detection of gravitational waves in space, we need to do the clipping test for the optical system to probe the degree of the stray light coming from clipping. The clipping degree is dependent on the power ratio within the surface of optical elements. As a preliminary investigation, we calculate the MSD spot size and the corresponding power ratio concentrated inside the MSD spot range for the beam type which we are interested. We use the following definition to calculate the power ratio concentrated inside the MSD spot range
	\begin{equation}
		\epsilon_{P_{spot}}=\frac{\int_{s_y(z)-w_y(z)}^{s_y(z)+w_y(z)}\int_{s_x(z)-w_x(z)}^{s_x(z)+w_x(z)}Idxdy}{P_{MEM}}.\label{eq_pr}
	\end{equation}

For the Gaussian beam, HG beam and LG beam, the fractional beam energy concentrated inside the MSD spot range is independent of $z$. As for general beams, the power ratio within the MSD spot range is dependent on $z$ in the near field. As for the very far-field, the power ratio within the MSD spot range is also independent of $z$. These will be discussed in \ref{appen: pr}.

  The original formulas \cref{MSD_spot_general,cen_general} to estimate the spot size and beam center for the beam passing through the optical system for the detection of gravitational waves in space are computationally expensive. By comparison, \cref{xbahg,ybahg,wxhg,wyhg} are computationally simple. The second advantage is that the finite decomposition for MEM can eliminate the divergence problem of MSD spot size for diffracted beams even though the MSD spot size is dependent on the decomposition parameters. The paper by S. N. Vlasov et al. \cite{Vlasov1971} shows the MSD spot size for the non-diffracted beam is a half-hyperbola function of the propagation distance. The MSD spot size for MEM beams here can be regarded as changing the intensity profile rather than changing the integral limitation of \cref{MSD_spot_general}. Based on this fact, we can use the same method used by S. N. Vlasov et al. \cite{Vlasov1971} and M. A. Porras et al. \cite{1992ApOpt..31.6389P} to prove the MSD spot size and energy center for the MEM beam here still is the half-hyperbola function of the propagation distance and obeys the ABCD law in the ABCD system in diffraction cases. This is the third advantage.
	
\section{Validation of MSD spot size definition and numerical code}\label{sec:applications}
 In this section, we will subject the MSD spot size definition to tests in cases of a simple astigmatic Gaussian beam as well as Nearly-Gaussian beams which can appear in laboratory experiments. At the same time, the tests serve to calibrate our numerical code. The numerical calculation of analytical expression in this paper is performed by C++ code with double type precision.
 
\subsection{Simple astigmatic Gaussian beam}\label{sec:non_diff}

The MSD spot size definition agrees with the definition of the spot size for Gaussian, HG and LG beams \cite{1980ApOpt..19.1027C, 1983ApOpt..22..643P}. If a Gaussian beam impacts on a curved surface at an angle, the beam becomes a simple astigmatic Gaussian beam, which has an elliptical spot \cite{1969ApOpt...8..975M,alda2003}. Considering the simple astigmatic Gaussian beam as an expansion of the basic Gaussian and HG beams, the MSD spot size should agree with the analytic spot size here.

 The electric field of the input simple astigmatic Gaussian beam is \cite{kochkina2013stigmatic}
\begin{equation}
\begin{split}
E_{in}=u_{in}\exp{\left(i\omega t\right)}=&\sqrt{\frac{P}{\pi}}\frac{1}{\sqrt{w_xw_y}}\exp{(-\frac{x^2}{w_x^2}-\frac{y^2}{w_y^2})}\\
&\times\exp{ \left(i\left(-k\left(\frac{x^2}{2R_x}+\frac{y^2}{2R_y}\right)+\frac{1}{2}\left(\zeta_x+\zeta_y\right)-kz+\omega t\right)\right)}.
\end{split}
\end{equation}
Here, the beam propagates along $+z$ direction and the parameter $w_{x,y}$ is the spot size for x or y direction of the simple astigmatic Gaussian beam. It is defined as  $w_{x,y}=w_{0_{x,y}}\sqrt{1+\left(\frac{z}{z_{r_{x,y}}}\right)^2}$, using the Rayleigh range $z_{r_{x,y}}=\frac{\pi w_{0_{x,y}}^2}{\lambda}$ and the waist $w_{0_{x,y}}$, where $\lambda$ is the wavelength and $k=\frac{2\pi}{\lambda}$ is the wave number. The radius of curvature is $R_{x,y}=z\left(1+\left(\frac{z_{r_{x,y}}}{z}\right)^2\right)$ and the Gouy phase is $\zeta_{x,y}=\arctan\frac{z}{z_{r_{x,y}}}$. The total power of this beam is $P$. Here we set $w_{0_x}=\SI{1}{mm}$, $w_{0_y}=\SI{3}{mm}$, $P=1$ and $\lambda=\SI{1064}{nm}$. This beam can be decomposed into the superposition of a finite number of HG modes by using \cref{superhgdecom_finite}. Here we use two different waists $\SI{0.8}{mm}$ and $\SI{0.6}{mm}$ for the basic HG modes. The direction and center of these HG modes are the same as the input beam. The decomposition happens in the waist plane. 
\begin{figure}[H]
	\centering
	\subfigure[]{
	\begin{minipage}[t]{0.45\linewidth}
		\centering
		\includegraphics[width=0.9\textwidth]{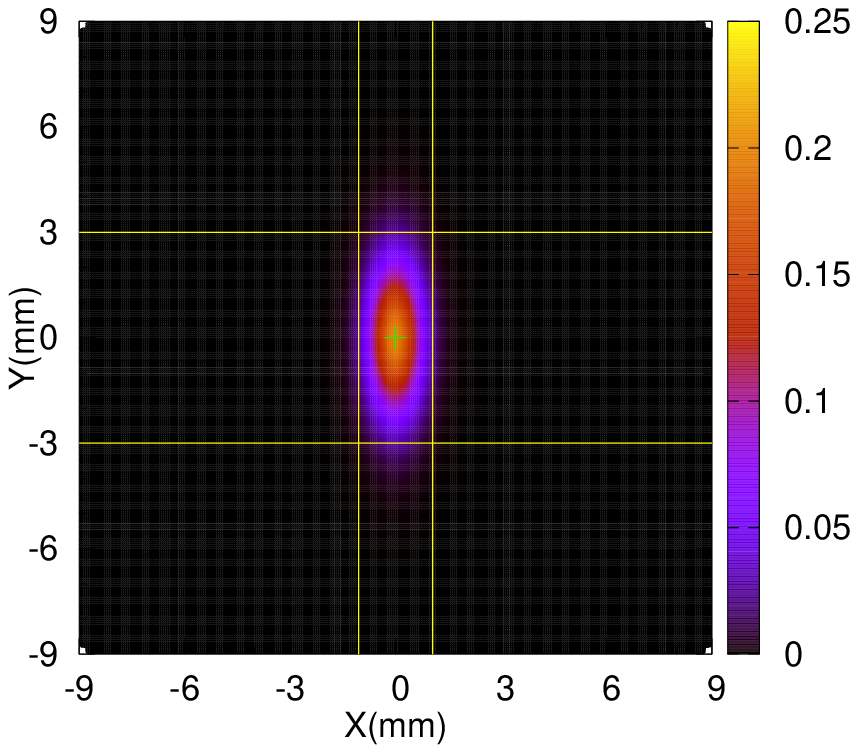}\label{intensity_sagb_1000}
		
	\end{minipage}
	}
	\subfigure[]{
	\begin{minipage}[t]{0.4\linewidth}
		\centering
		\includegraphics[width=0.9\textwidth]{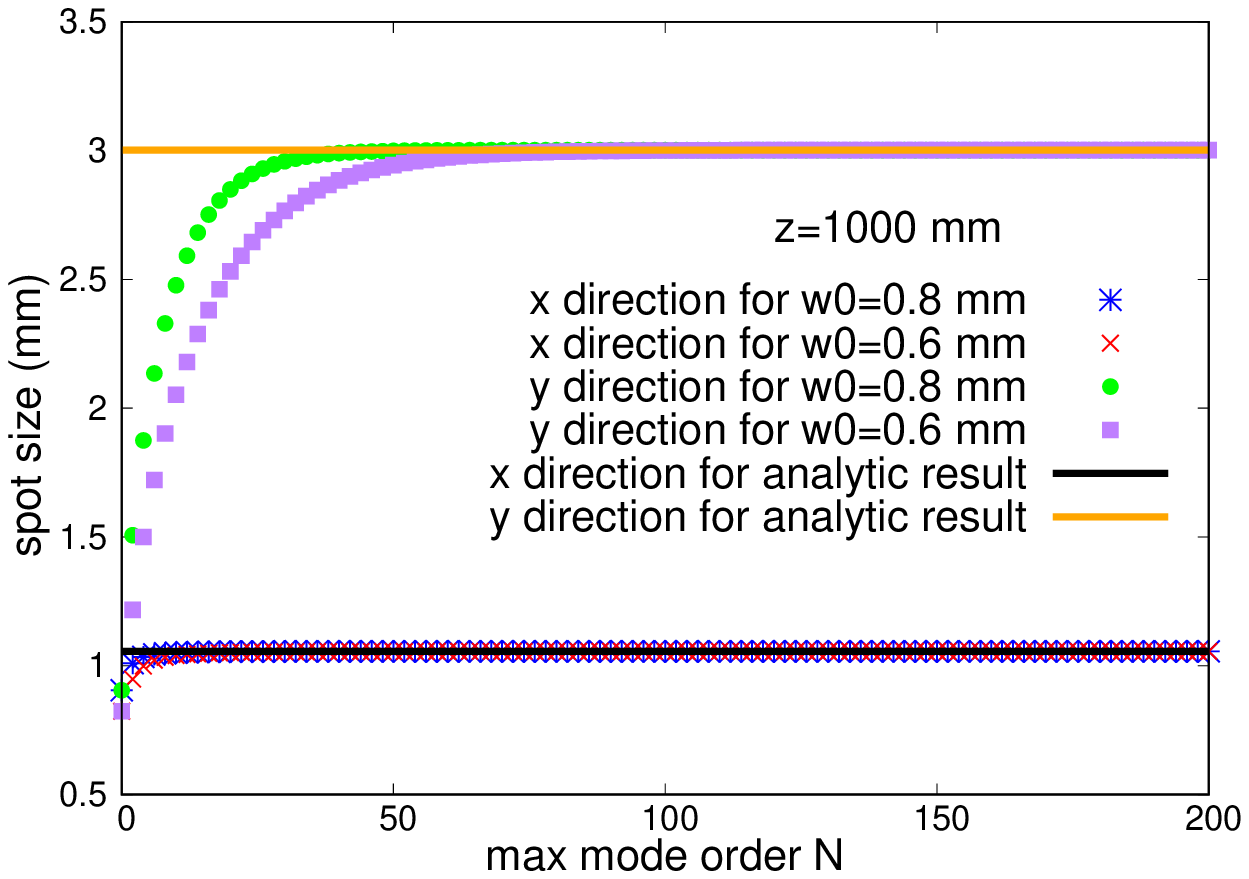}\label{spot_sagb_1000_scan_ord}
		
	\end{minipage}
	}
	\subfigure[]{
	\begin{minipage}[t]{0.4\linewidth}
		\centering
		\includegraphics[width=0.9\textwidth]{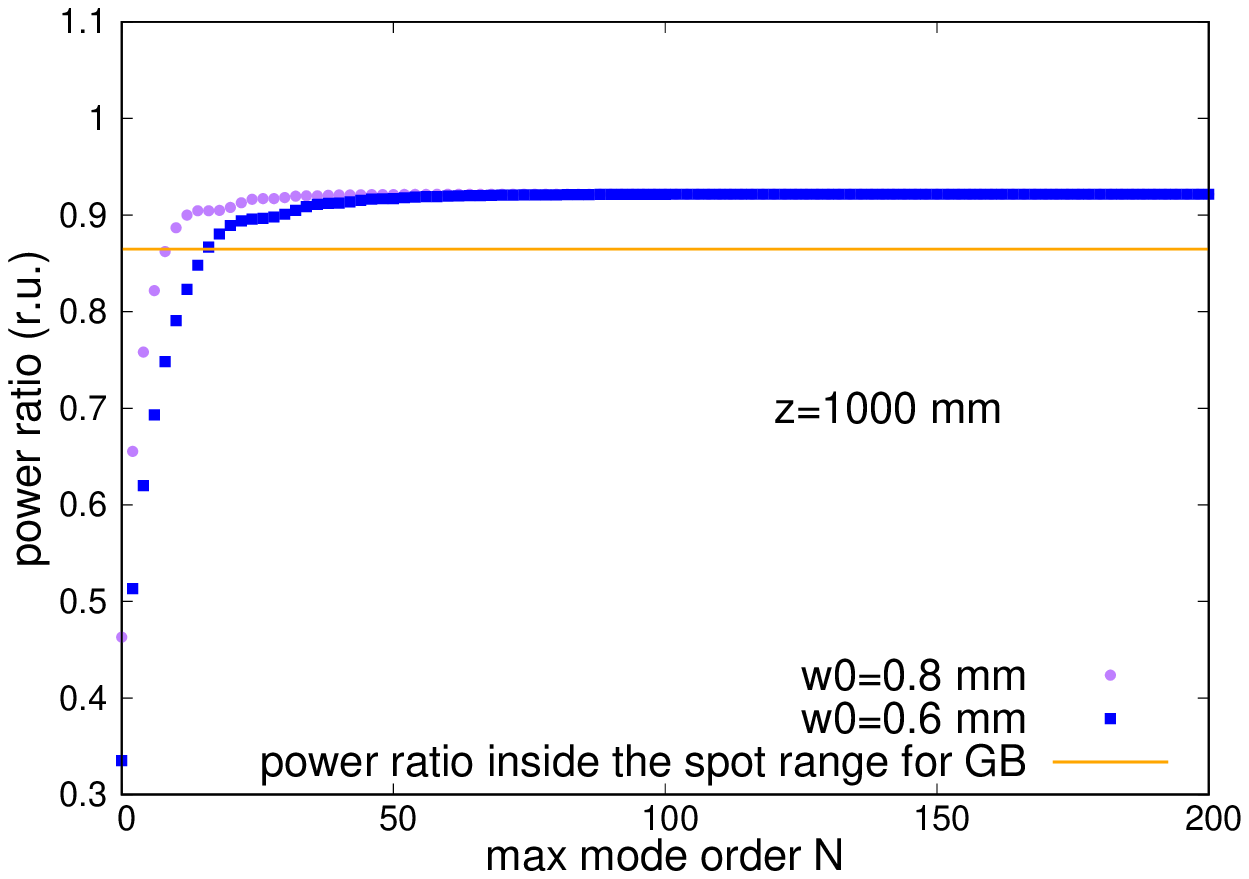}\label{pr_sagb_1000_scan_ord}
		
	\end{minipage}
	}
	\subfigure[]{
	\begin{minipage}[t]{0.4\linewidth}
		\centering
		\includegraphics[width=0.9\textwidth]{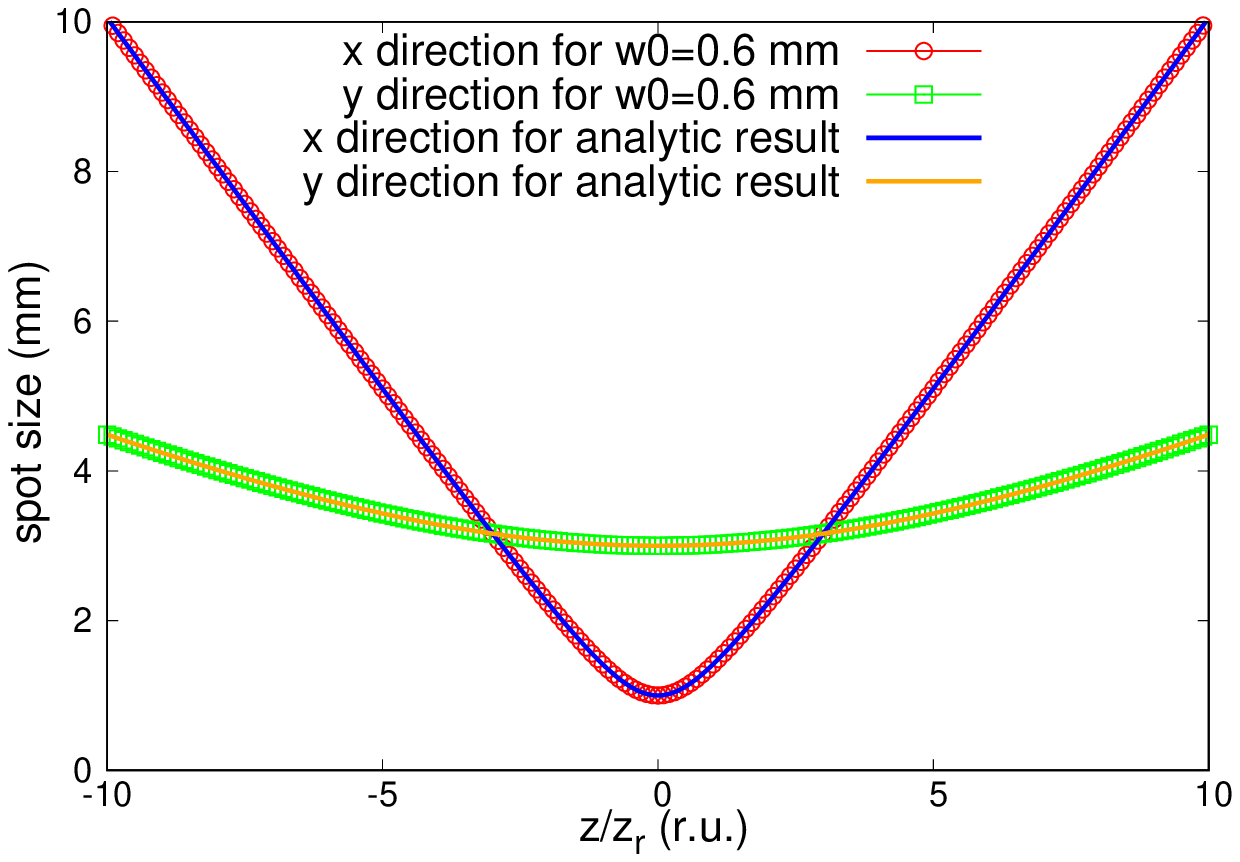}\label{spot_sagb_1000_scan_z}
		
	\end{minipage}
	}
	\centering
	\caption{(a). The intensity profile for $z= \SI{1000}{mm}$. (b). The estimated values of MSD spot size for different max mode order $N$ compared with the analytic spot size for $z= \SI{1000}{mm}$. (c). The power ratio within the MSD spot size for different max mode order $N$ for $z= \SI{1000}{mm}$. (d). The estimated values of spot size of MEM beam compared with the analytic spot size for different propagation distances.}\label{elliptical}
\end{figure}

\Cref{intensity_sagb_1000} shows the intensity profile for $z= \SI{1000}{mm}$ of the MEM beam which represents the simple astigmatic Gaussian beam. The yellow line in \cref{intensity_sagb_1000} is the boundary of the MSD spot for the MEM beam and the green point is the estimated beam center $(0,0)$. The waist of the basic HG modes for the MEM beam is $w_0=\SI{0.6}{mm}$ and the max mode order is $N=200$. \Cref{spot_sagb_1000_scan_ord,pr_sagb_1000_scan_ord} show the variations for the MSD spot size and the power ratio within the MSD spot range with different max mode order $N$.  The waist of the basic HG modes for the MEM beam is $w_0=\SI{0.6}{mm}$ or $w_0=\SI{0.8}{mm}$ and $z= \SI{1000}{mm}$. The blue points in \cref{spot_sagb_1000_scan_ord} shows the MSD spot size for different max mode order $N$ for $w_0=\SI{0.8}{mm}$ in x direction, the red points show the MSD spot size for $w_0=\SI{0.6}{mm}$ in x direction, the green points show the MSD spot size for $w_0=\SI{0.8}{mm}$ in y direction and the purple points show the MSD spot size for $w_0=\SI{0.6}{mm}$ in y direction. The black line and orange line in \cref{spot_sagb_1000_scan_ord} show the analytic spot size for this simple astigmatic Gaussian beam in x direction and y direction respectively. The purple points in \cref{pr_sagb_1000_scan_ord} show the power ratio within the MSD spot range for different max mode orders N for $w_0=\SI{0.8}{mm}$, the blue points show the power ratio within the MSD spot range for $w_0=\SI{0.6}{mm}$ and the orange line shows the power ratio inside the circular spot range for the Gaussian beam. When the max mode order $N$ is large enough, the MSD spot size is coincident with the analytical spot size of the input simple astigmatic Gaussian beam, no matter what the waist of the basic HG modes is. As shown in \cref{spot_sagb_1000_scan_z}, the MSD spot size is coincident with the analytic spot size for the simple astigmatic Gaussian beam for different $z$.  The max mode order $N$ is 200 and the waist of the basic HG modes $w_0$ is $\SI{0.6}{mm}$. The red line in \cref{spot_sagb_1000_scan_z} shows the MSD spot size for different $z$ in x direction, the green line shows the MSD spot size in y direction, the blue line shows the analytical spot size in x direction and the orange line shows the analytical spot size in y direction. The MSD spot size is a hyperbolic function of $z$. The fractional beam energy concentrated inside the MSD spot range of this situation is $92.15\%$. If the max mode order $N$ is large enough, the waist $w_0$ of the basic HG modes will have little influence on the estimated value of MSD spot size and the power ratio within the MSD spot range. 

Another non-diffracted beam concerning a shifted HG beam will be discussed in \ref{sec:shift_HG}.
\subsection{Nearly-Gaussian beams}\label{sec: near_gb}
 In this subsection, we will test the MSD spot size definition for some non-perfect beams as they can appear in laboratory experiments. We use the MEM to decompose these non-perfect beams and then calculate the MSD spot size and the fractional beam energy concentrated inside the MSD spot range for these beams.
\subsubsection{Halo Beam}
\begin{figure}[H]
	\centering
	\subfigure[]{
	\begin{minipage}[t]{0.45\linewidth}
		\centering
		\includegraphics[width=0.95\textwidth]{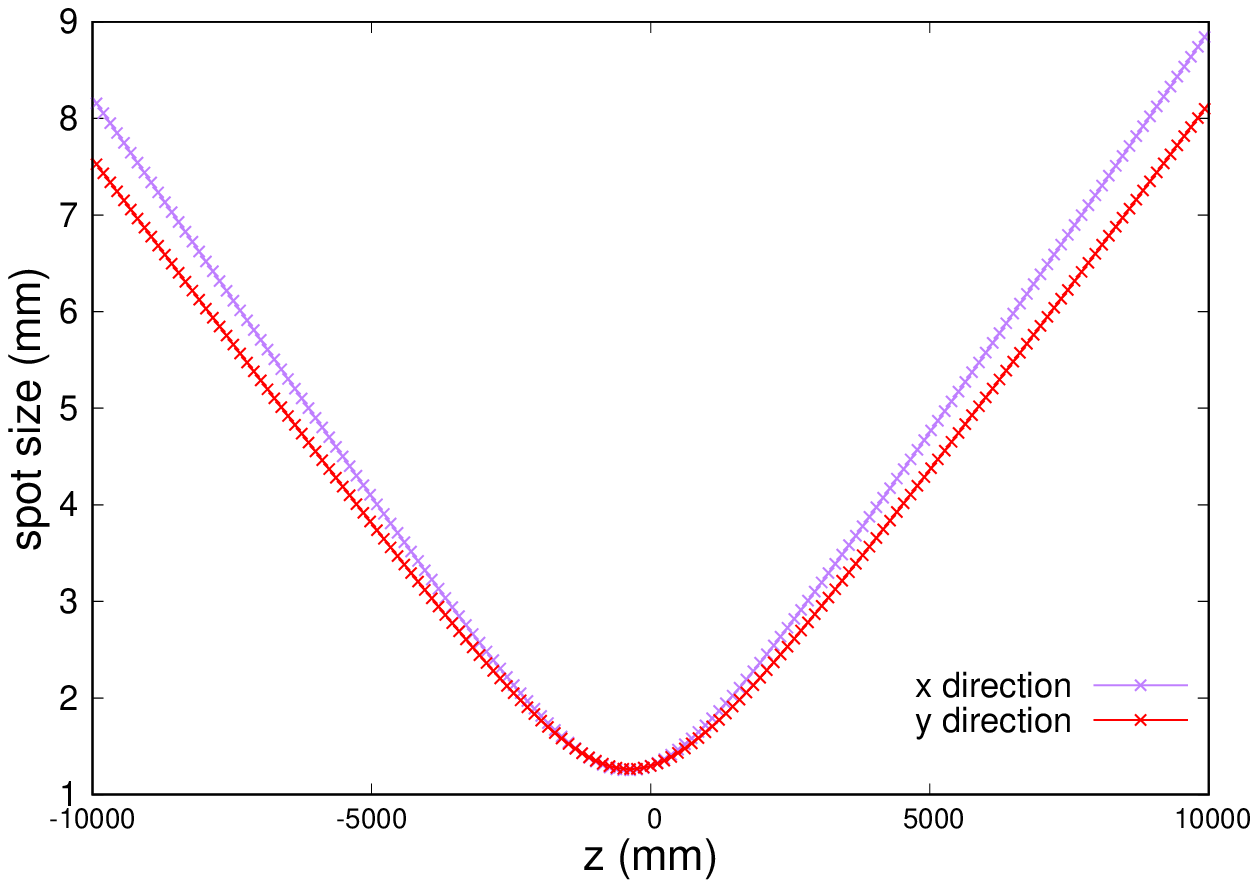}\label{spot_halo}
		
	\end{minipage}
	}
	\subfigure[]{
	\begin{minipage}[t]{0.45\linewidth}
		\centering
		\includegraphics[width=0.95\textwidth]{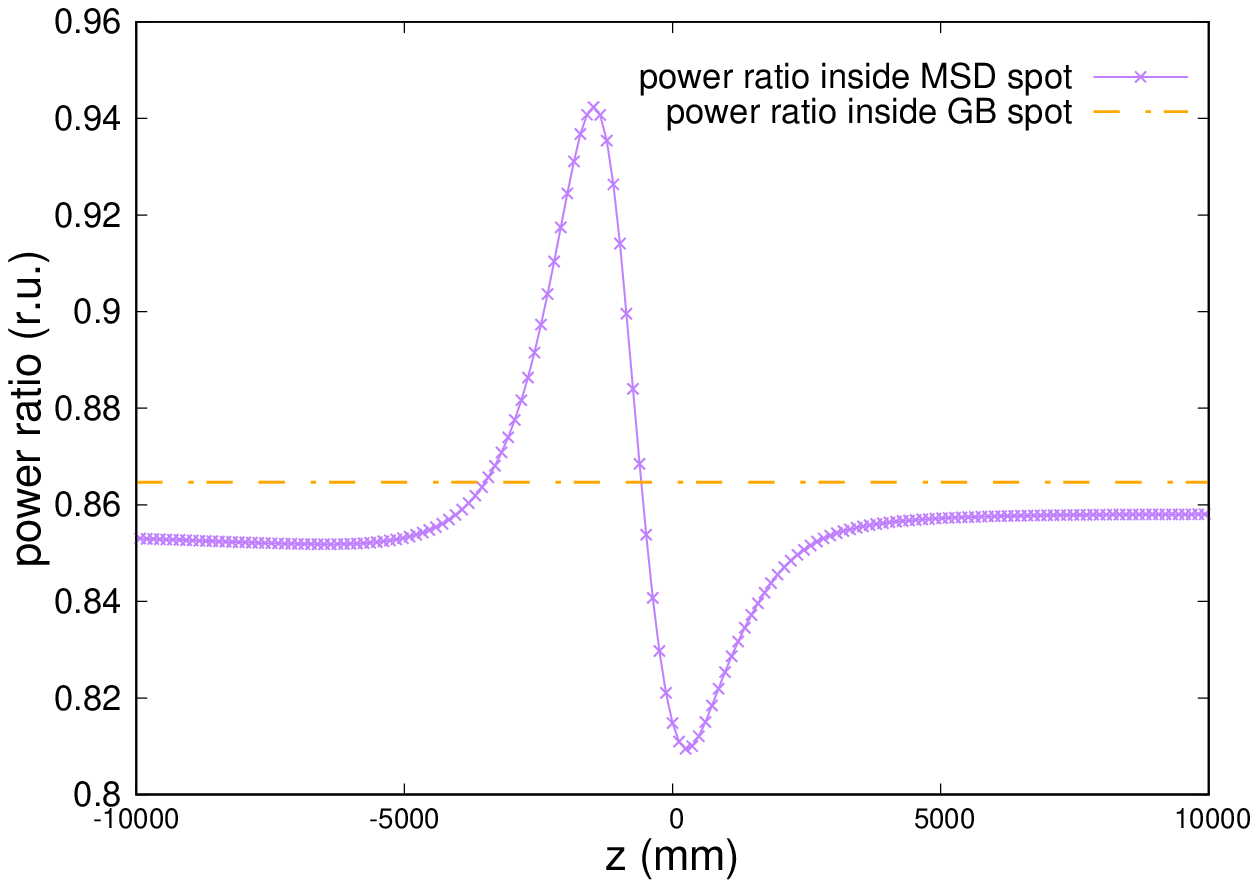}\label{pr_halo}
		
	\end{minipage}
	}
	\centering
	\caption{(a) The MSD spot size of halo beam for different $z$. (b)The fractional beam energy concentrated inside the MSD spot for different $z$.}\label{halo_z}
\end{figure}

A halo beam might be produced by touching a fiber end to a monolithic fiber collimator with optical contact gel in the laboratory. If there is a small gap between the collimator and the fiber, the intensity profiles will deform like a halo as illustrated in \cref{halo}:
\begin{figure}[H]
	\centering
	\subfigure[]{
	\begin{minipage}[t]{0.45\linewidth}
		\centering
		\includegraphics[width=0.95\textwidth]{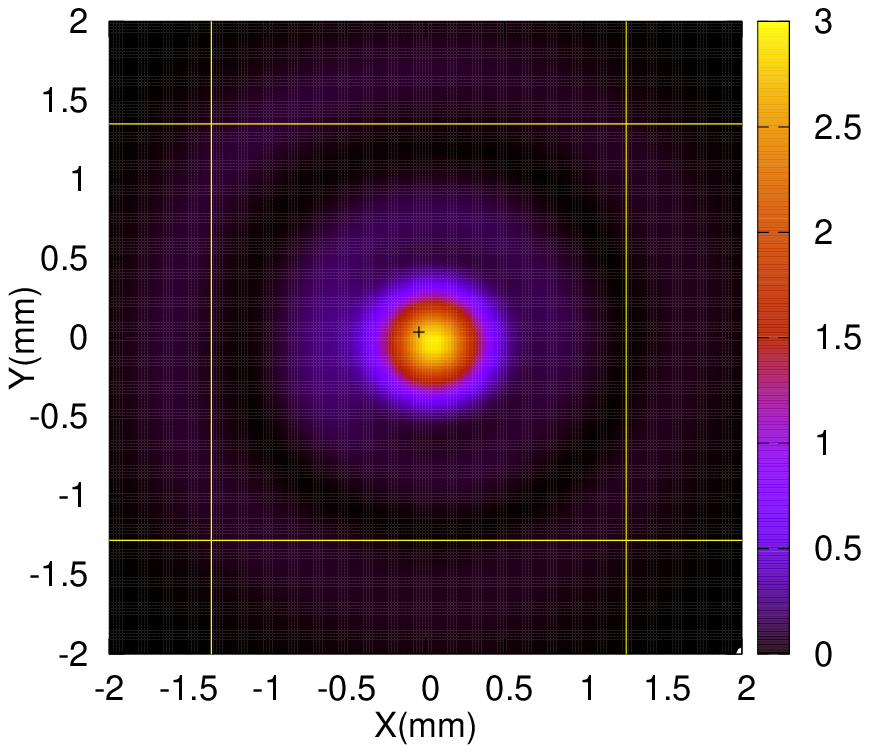}\label{halo_939}
		
	\end{minipage}
	}
	\subfigure[]{
	\begin{minipage}[t]{0.45\linewidth}
		\centering
		\includegraphics[width=0.95\textwidth]{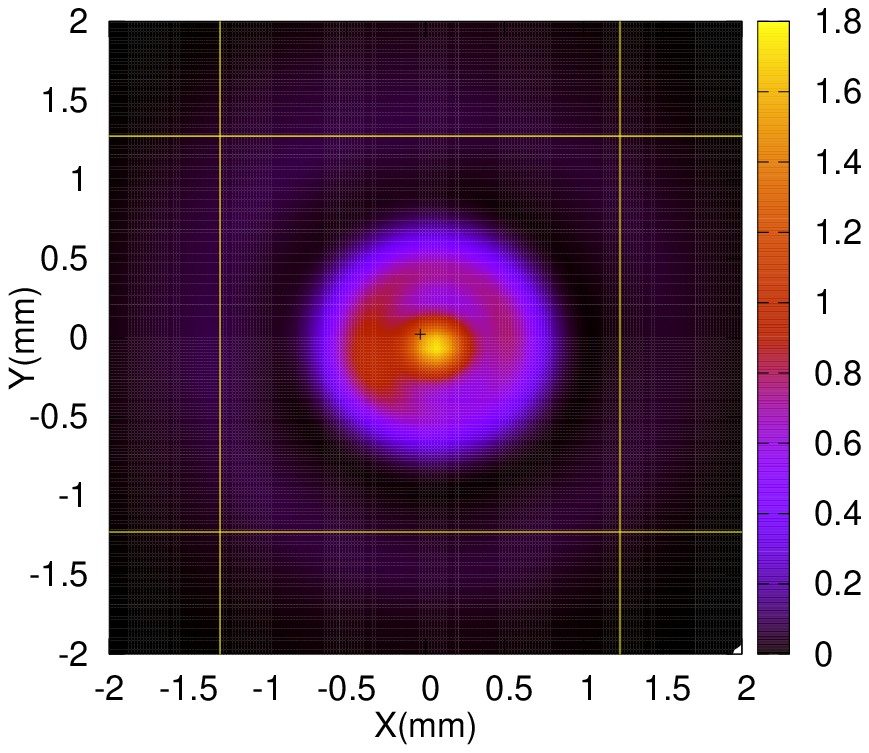}\label{halo_501}
		
	\end{minipage}
	}
	\centering
	\caption{(a) The intensity profile for $z= \SI{75.865}{mm}$ and the yellow line is the boundary of the MSD spot.   (b) The intensity profile for $z= \SI{-362.135}{mm}$ and the yellow line is the boundary of the MSD spot.}\label{halo}
\end{figure}
The "plus" points shown in \cref{halo_939,halo_501} mark the energy centers. The energy center in \cref{halo_939} is $(-0.04281,0.03558)$. It is $(-0.03492,0.02280)$ in \cref{halo_501}. The fractional beam energy concentrated inside the spot size shown in \cref{halo_939} is $81.2142\%$ and $84.0180\%$ in \cref{halo_501}. \Cref{spot_halo} shows the MSD spot size for different $z$ and \cref{pr_halo} shows the fractional beam energy concentrated inside the MSD spot for different $z$. The MSD spot size for the halo beam satisfies the hyperbolic law and the fractional beam energy concentrated inside the MSD spot size is always bigger than $80.94\%$. The waist of these HG modes is $w_0=\SI{644.314}{\mu m}$ and the waist plane of these HG modes are all located in the cross-section of $z=\SI{0}{mm}$. The decomposition results of the MEM beam for this halo beam are shown in \ref{sec:table_halo}.
\subsubsection{Nearly-Gaussian beam \uppercase\expandafter{\romannumeral2}}
\begin{figure}[H]
	\centering
	\subfigure[]{
	\begin{minipage}[t]{0.45\linewidth}
		\centering
		\includegraphics[width=0.95\textwidth]{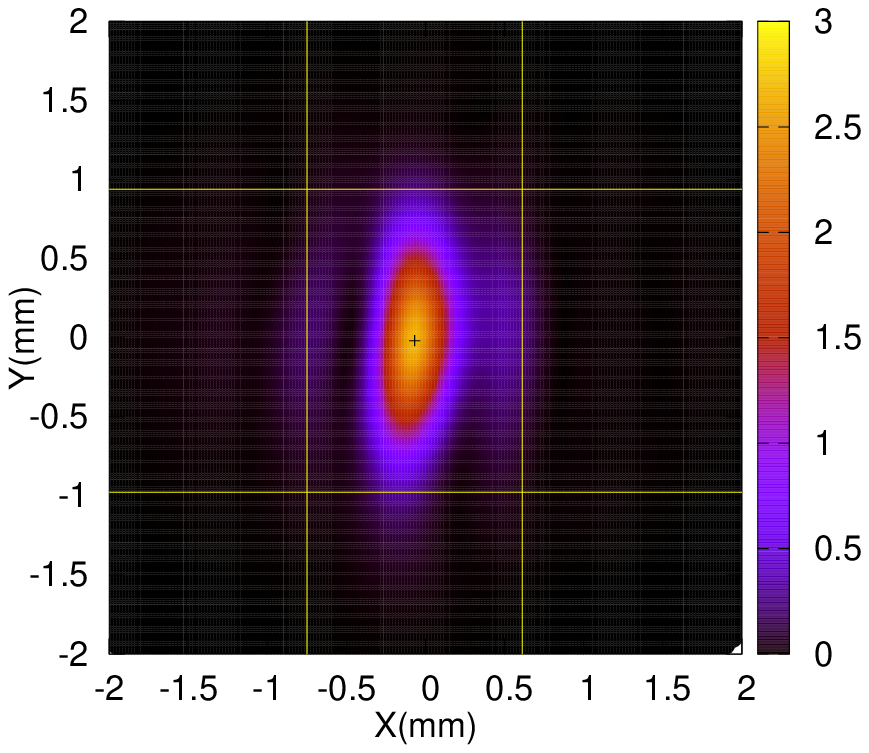}\label{gerald_240}
		
	\end{minipage}
	}
	\subfigure[]{
	\begin{minipage}[t]{0.45\linewidth}
		\centering
		\includegraphics[width=0.95\textwidth]{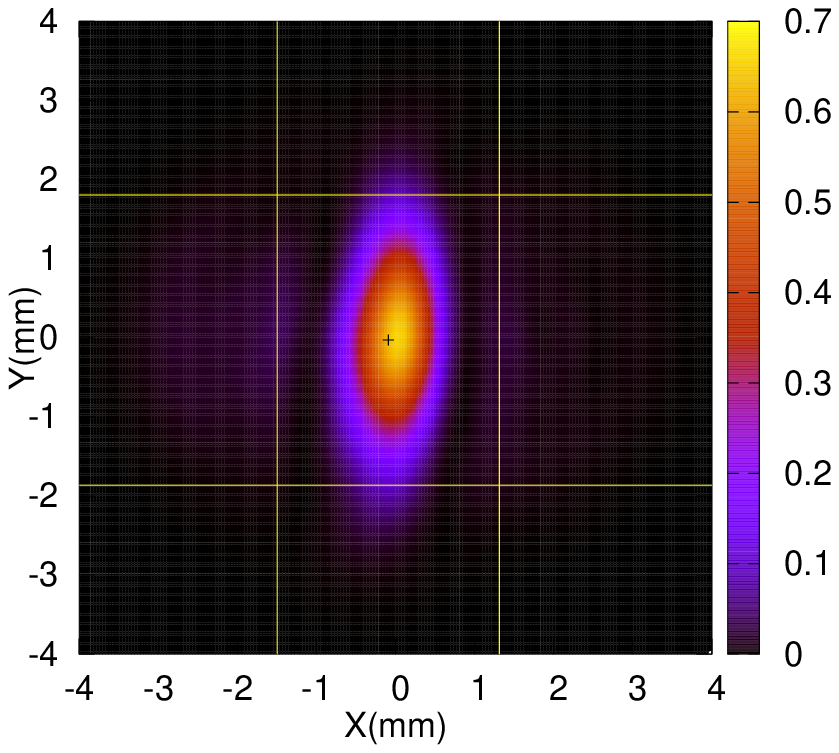}\label{gerald_840}
		
	\end{minipage}
	}
	\centering
	\caption{(a) The intensity profile for $z= \SI{481.101}{mm}$ and the yellow line is the boundary of the MSD spot.   (b) The intensity profile for $z= \SI{1081.101}{mm}$ and the yellow line is the boundary of the MSD spot.}\label{near_gb2}
\end{figure}

\begin{figure}[H]
	\centering
	\subfigure[]{
	\begin{minipage}[t]{0.45\linewidth}
		\centering
		\includegraphics[width=0.95\textwidth]{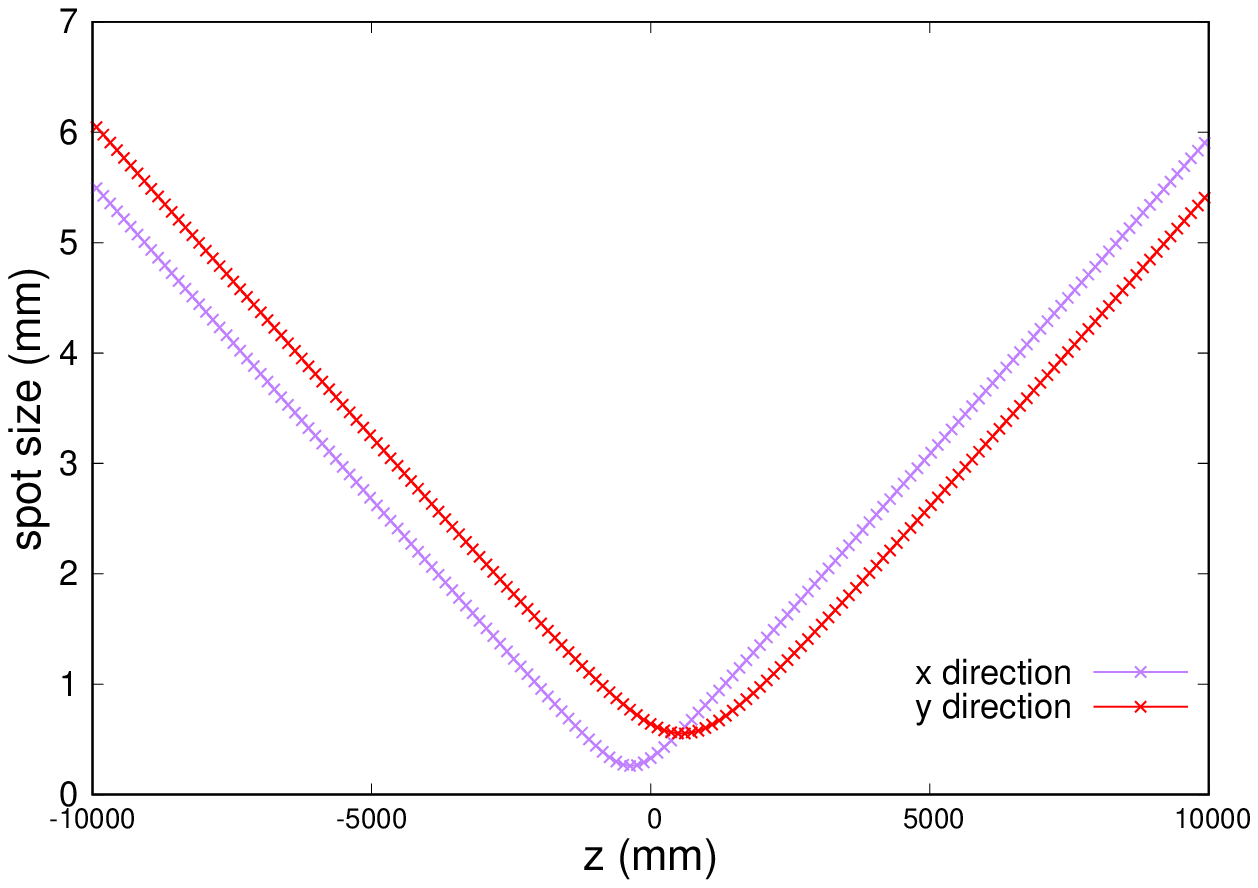}\label{spot_gerald}
		
	\end{minipage}
	}
	\subfigure[]{
	\begin{minipage}[t]{0.45\linewidth}
		\centering
		\includegraphics[width=0.95\textwidth]{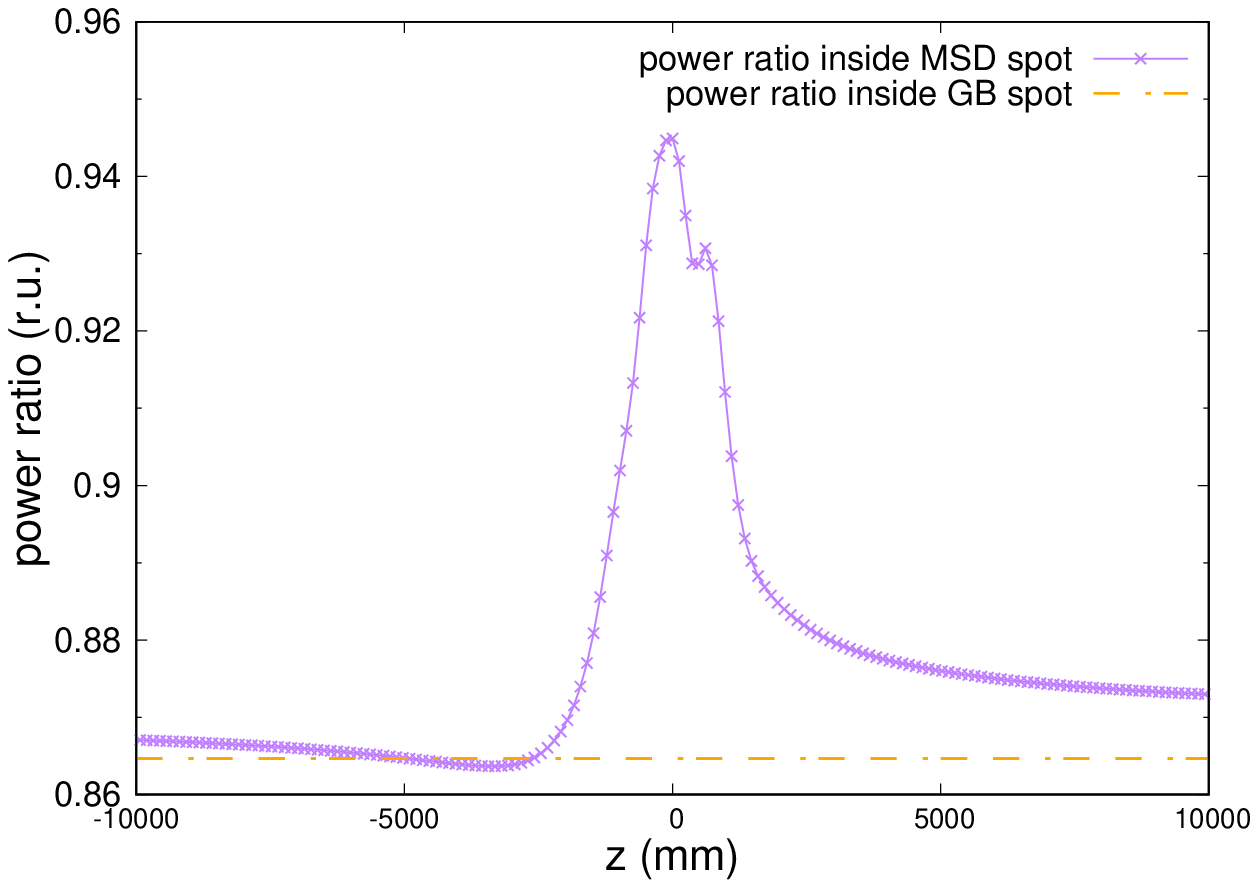}\label{pr_gerald}
		
	\end{minipage}
	}
	\centering
	\caption{(a) The MSD spot size of nearly-Gaussian beam \uppercase\expandafter{\romannumeral2} for different $z$. (b)The fractional beam energy concentrated inside the MSD spot for different $z$.}\label{gerald_z}
\end{figure}

Another non-perfect beam is shown in \cref{near_gb2}.
The "plus" points shown in \cref{gerald_240,gerald_840} mark the energy centers. The energy center in \cref{gerald_240} is $(-0.0691484,-0.0196754)$. It is $(-0.0927757,-0.0306717)$ in \cref{gerald_840}. The fractional beam energy concentrated inside the spot size shown in \cref{gerald_240} is $89.6420\%$ and $88.0395\%$ in \cref{gerald_840}. \Cref{spot_gerald} shows the MSD spot size for different $z$ and \cref{pr_gerald} shows the fractional beam energy concentrated inside the MSD spot for different $z$. The waist of these HG modes is $w_0=\SI{399.465}{\mu m}$ and the waist planes of these HG modes are all located in the cross-section of $z=\SI{0}{mm}$. The decomposition results of the MEM beam for this beam are shown in \ref{sec:table_near_gb2}.

\section{MSD spot size and power ratio within the MSD spot range for top-hat beams in detection of gravitational waves in space.}\label{sec:si}

In this section, we shall enter the core of our work and trace the evolution of the spot size of a top-hat beam propagating 
within an optical bench for a LISA or TAIJI type mission. The spot size together with the corresponding power ratio within this spot range will also be estimated. To this end, we shall first address in the following subsection the issue of hard edge diffraction of a flat top beam. The MSD spot size for a top-hat beam in a science interferometer in a LISA or TAIJI type optical bench will then be simulated. Some error analysis will then be presented to render our calculations more trustworthy. 

\subsection{Spot size for a top-hat beam}\label{top_general}

In this section, we investigate the performance of the MSD spot size in the top-hat beam case. Unlike the previously discussed cases, diffraction occurs at the edge of the beam and this generates divergence for the MSD spot size of a top-hat beam. The analysis will then be applied in the next subsection when we look at the propagation of a flat top beam within a distance scale dictated by the size of an optical bench in detection of gravitational waves in space. 

Consider a  top-hat beam which may be regarded as a plane wave truncated by a circular aperture. The complex amplitude of the top-hat beam maybe written as
\begin{equation}
u_{top}(x,y,0)=\left\{ 
\begin{aligned}
&\sqrt{\frac{P}{\pi}}\frac{1}{r},\ x^2+y^2\leq r^2\\
&0,\ x^2+y^2>r^2,
\end{aligned}
\right.
\end{equation}
where $r$ is the initial radius of the top-hat beam and $P$ is the total power. Here we assume the beam propagates along $+z$ direction and the wavelength of this top-hat beam is $\lambda=\SI{1064}{nm}$.

As discussed in \cref{sec:method}, the top-hat beam can be decomposed by MEM and the propagation of this MEM beam can be used to represent the propagation of the top-hat beam. The beam center and the propagation direction of these basic HG modes are the same as the top-hat beam in this section. The waist planes of these basic HG modes are located at the initial plane of the top-hat beam. The total power of the input top-hat beam is $P=1$ a.u.. The initial radius $r$ of the top-hat beam is $\SI{2.5}{mm}$ here which is coincident with the top-hat beam in the paper by \mbox{L. D’Arcio} et al.\cite{2017SPIE10565E..2XD}

One obvious shortcoming of the MSD spot size for the hard-edge diffraction beam is, that it is divergent \cite{1990SPIE.1224....2S}. The word divergent here means that the MSD spot size for the hard-edge diffraction beam is infinite (\cref{MSD_spot_general}$\rightarrow \infty$). Rounding off the spatial intensity distribution in some reasonable way such as with a super-Gaussian edge is one choice to solve this divergence \cite{1990SPIE.1224....2S}. Alternatively, another choice is to truncate the limitation of the integration for MSD spot size to eliminate the large angle and evanescent waves \cite{1990SPIE.1224....2S}. Many criteria are proposed for the truncated limitation such as general truncated second-order moment method \cite{1995OptL...20..124M}, asymptotic approximation method \cite{1996OptCo.123..679P} and self-convergent beam width method \cite{2000ApOpt..39.3914A} and many more. However, all of them introduce another free parameter which is changed by hand and can influence the final value of the MSD spot size. There is no general guidance to find a suitable value for the free parameter.  These truncated methods introduce another disadvantage that they can not guarantee that the hyperbolic law and ABCD law for propagation still holds. 

The MSD spot size is sensitive to the diffracted wings even if there are only few fractional parts of beam energy inside these wings \cite{1996OptCo.123..679P}. Limiting the interval of integration of the MSD spot size definition to eliminate the divergence \cite{1996OptCo.123..679P} can be regarded as only reserving a few dominating wings. The diffracted beam may also be decomposed by MEM as the previous section shows. The higher the HG mode order is, the more outside wings can be represented. In a sense, using a finite number of HG modes to represent the diffracted beam is similar to limiting the interval of the integration. Compared with the MSD spot size of the diffracted beam obtained by limiting the upper and lower limits of the integral, the advantages of the MSD spot of the MEM beam are that it still satisfies the ABCD law during passing through the ABCD system and still exhibits the hyperbolic function of propagation distance.

\begin{figure}[H]
	\centering
\subfigure[]{
	\begin{minipage}[t]{0.35\linewidth}
		\centering
		\includegraphics[width=1.0\textwidth]{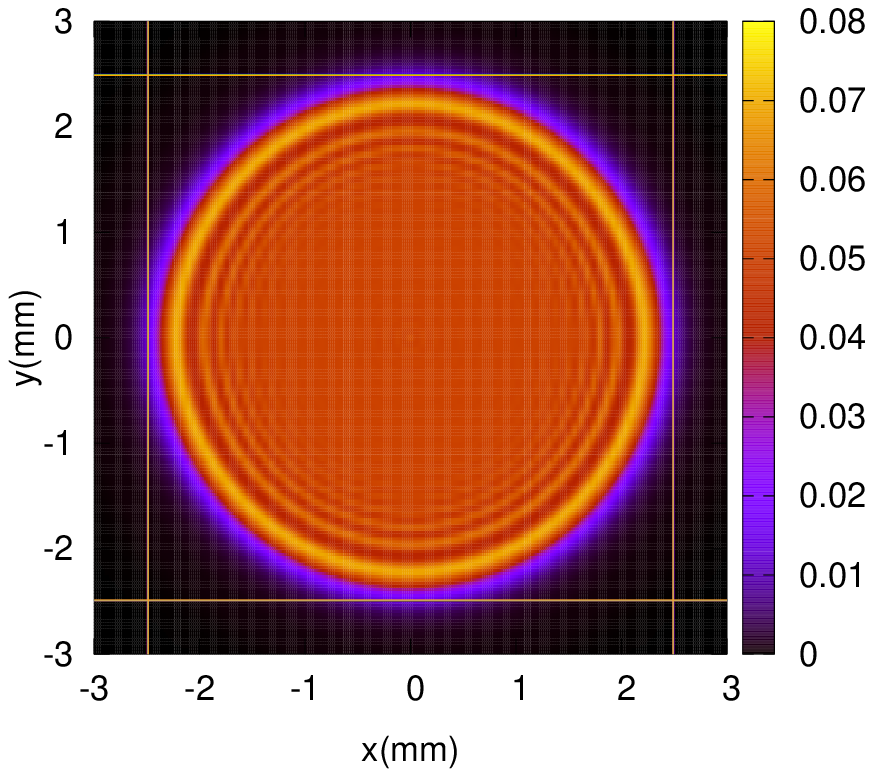}\label{top_hat_25_inten_100_3d}
		
	\end{minipage}
	}
	\subfigure[]{
	\begin{minipage}[t]{0.35\linewidth}
		\centering
		\includegraphics[width=1.0\textwidth]{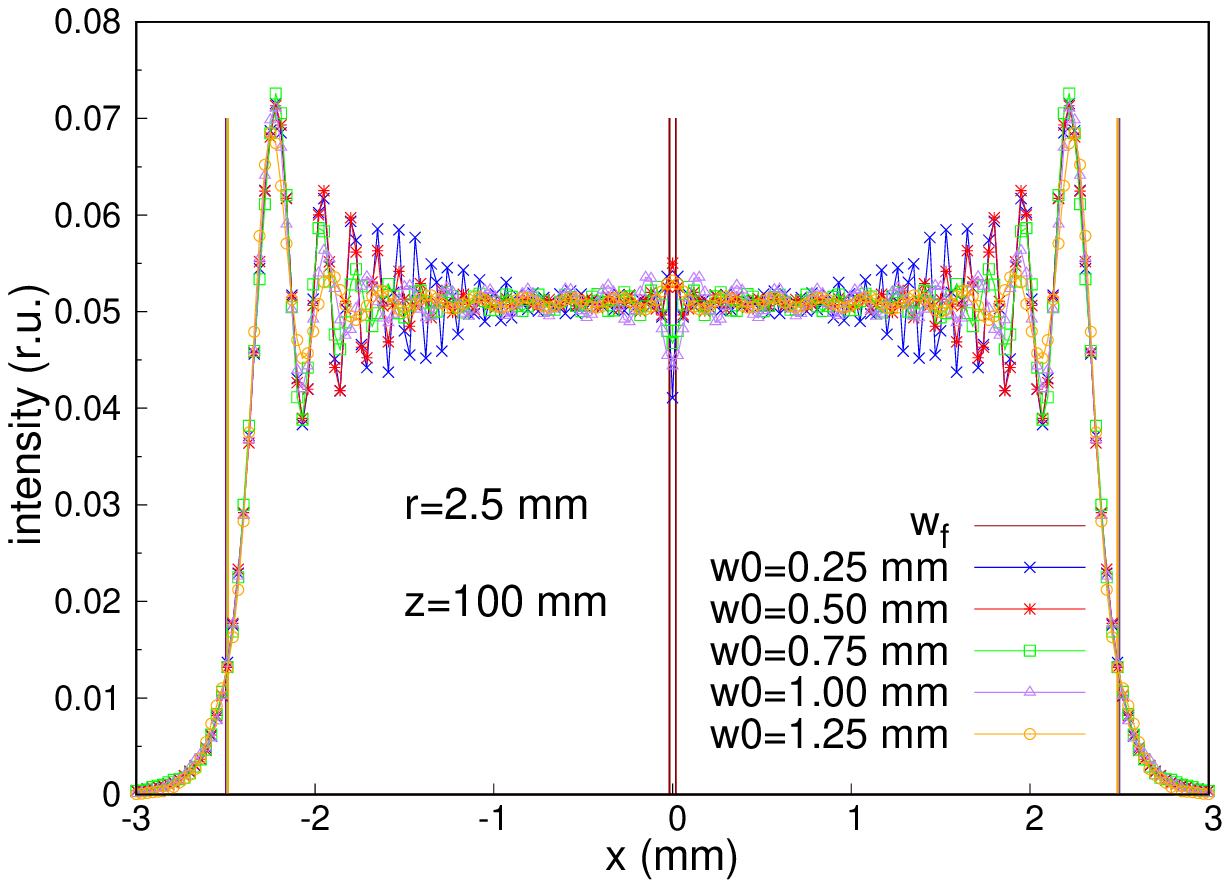}\label{top_hat_25_inten_100_2d}
		
	\end{minipage}
	}
	\subfigure[]{
	\begin{minipage}[t]{0.35\linewidth}
		\centering
		\includegraphics[width=1.0\textwidth]{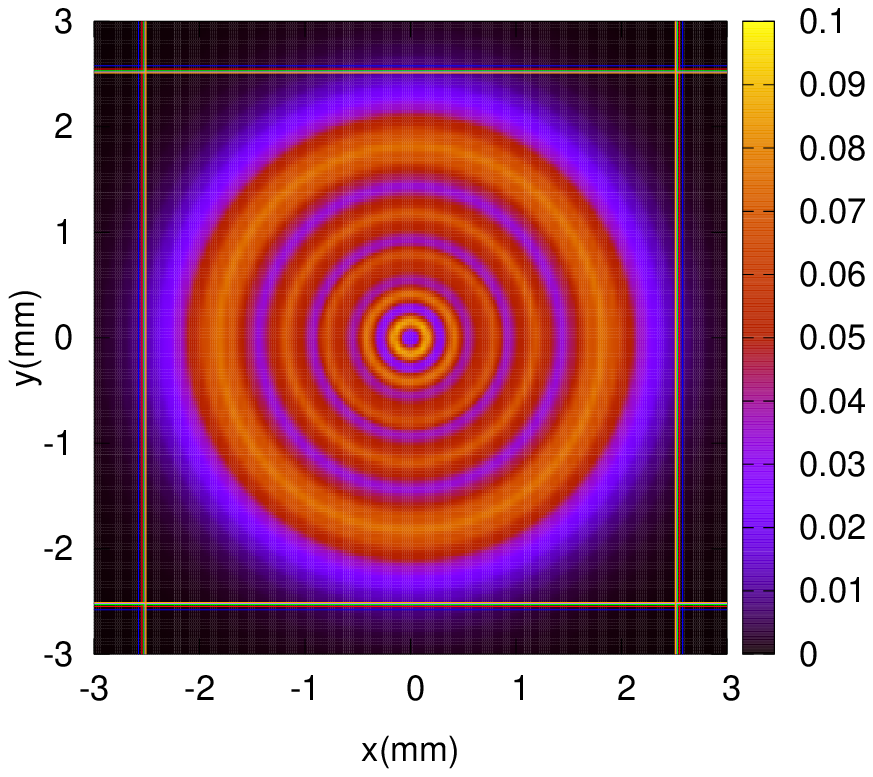}\label{top_hat_25_inten_600_3d}
		
	\end{minipage}
	}
	\subfigure[]{
	\begin{minipage}[t]{0.35\linewidth}
		\centering
		\includegraphics[width=1.0\textwidth]{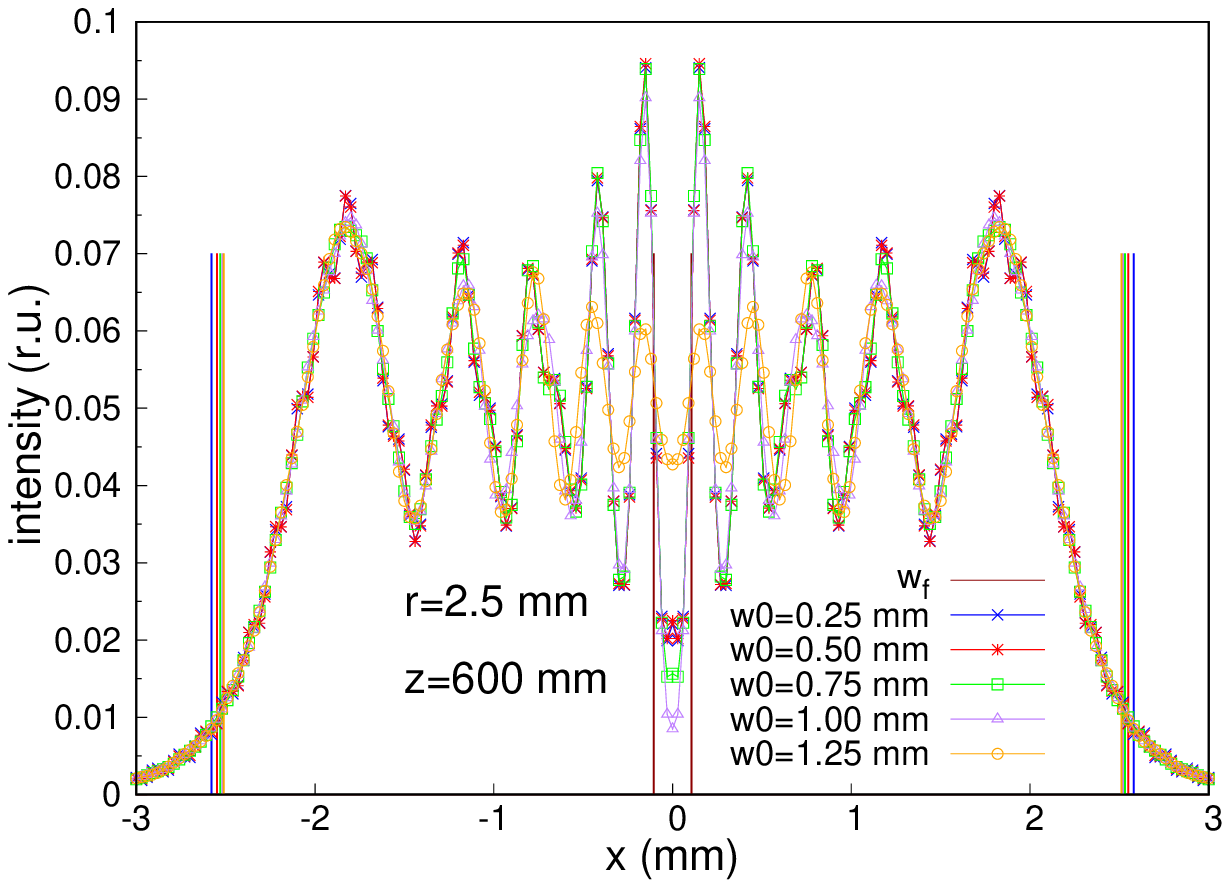}\label{top_hat_25_inten_600_2d}
		
	\end{minipage}
	}
	\subfigure[]{
	\begin{minipage}[t]{0.35\linewidth}
		\centering
		\includegraphics[width=1.0\textwidth]{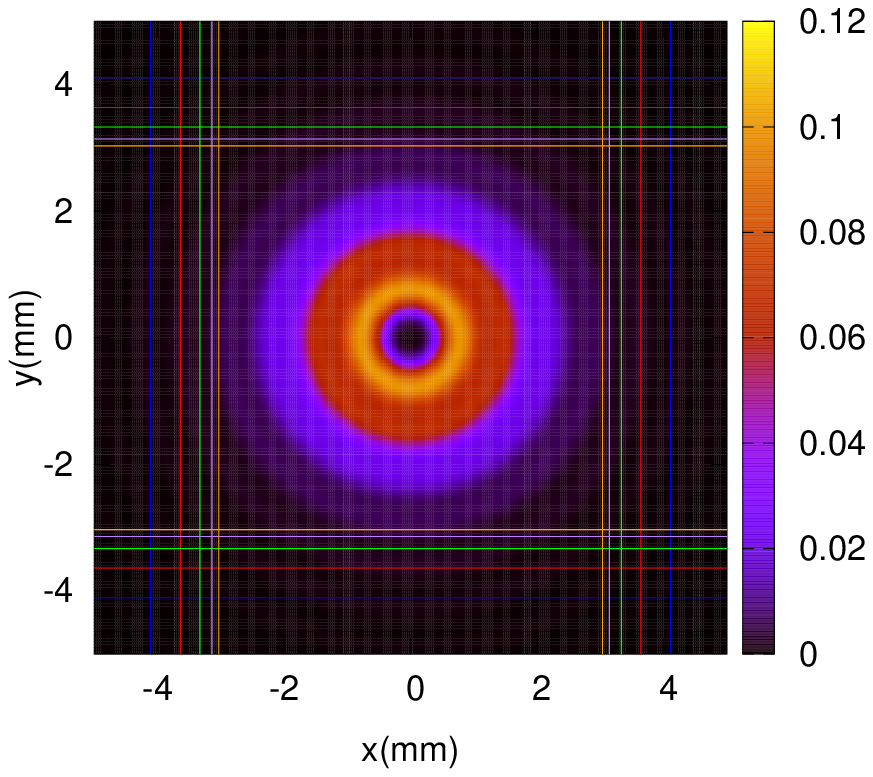}\label{top_hat_25_inten_3000_3d}
		
	\end{minipage}
	}
	\subfigure[]{
	\begin{minipage}[t]{0.35\linewidth}
		\centering
		\includegraphics[width=1.0\textwidth]{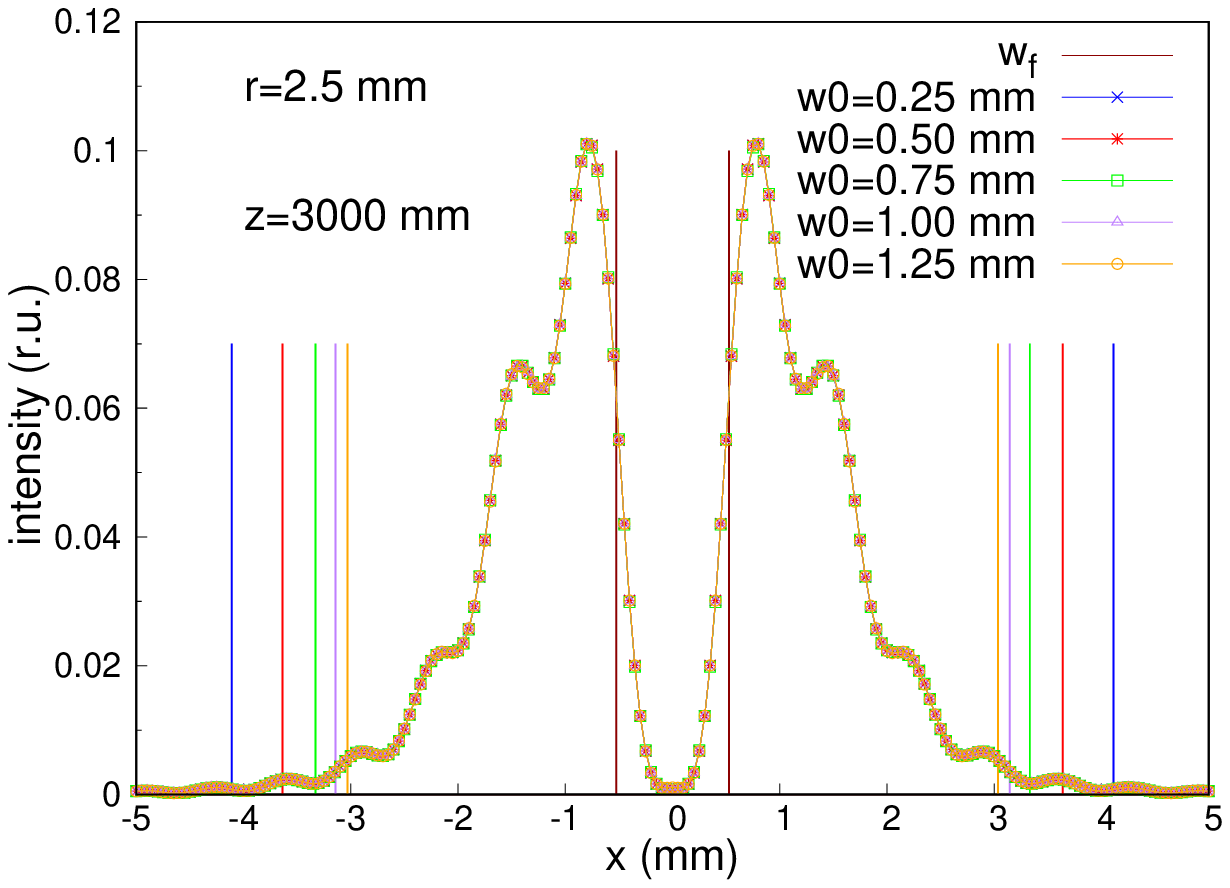}\label{top_hat_25_inten_3000_2d}
		
	\end{minipage}
	}

	\centering
	\caption{The intensity profiles of the simulated top-hat beam for the propagation distance $z=\SI{100}{mm}$ (a), $z=\SI{600}{mm}$ (c) or $z=\SI{3000}{mm}$ (e). The intensity distributions of the simulated top-hat beam along x-axis for the propagation distance $z=\SI{100}{mm}$ (b), $z=\SI{600}{mm}$ (d) or $z=\SI{3000}{mm}$ (f). }\label{top-hat_inten}
\end{figure}

 \Cref{top_hat_25_inten_100_3d,top_hat_25_inten_600_3d,top_hat_25_inten_3000_3d} show the intensity profiles of the MEM beams which represent the top-hat beam for the propagation distances $z=\SI{100}{mm}$ , $z=\SI{600}{mm}$ and $z=\SI{3000}{mm}$. The waist of the basic HG modes of the MEM beam is $w_0=\SI{0.5}{mm}$ and the max mode order $N$ of the MEM beams is $N=200$. \Cref{top_hat_25_inten_100_2d,top_hat_25_inten_600_2d,top_hat_25_inten_3000_2d} show the intensity distributions of the MEM beams along the x-axis of  \cref{top_hat_25_inten_100_3d,top_hat_25_inten_600_3d,top_hat_25_inten_3000_3d}.  The vertical lines in these subfigures are the edges of the MSD spot range for different MEM beams. The blue vertical lines in these subfigures represent the MSD spot size for the MEM beam with $w_0=\SI{0.25}{mm}$, the red lines represent $w_0=\SI{0.5}{mm}$, the green lines represent $w_0=\SI{0.75}{mm}$, the purple lines represent $w_0=\SI{1.0}{mm}$ and the orange lines represent $w_0=\SI{1.25}{mm}$. The dark-red vertical line is special and represents the spot size $w_f$ calculated from the paper by E. M. Drège et al.\cite{Drege2000} A typical divergence angle $\theta_f$ is defined as the distance between the central brightest point and the point where the intensity is $\frac{1}{e^2}$ of the maximum intensity value over the large propagation distance $z$. Furthermore, the spot size $w_f$ is defined as the divergence angle $\theta_f$ times the propagation distance $z$. This means the spot size $w_f$ represents the distance between the central brightest point and the point where the intensity is $\frac{1}{e^2}$ of the maximum intensity value in the far-field. In the far-field, this spot size $w_f$ and the far-field divergence $\theta_f$ angle is a function of wavelength $\lambda$ and the radius of the circular aperture $R$ \cite{Drege2000}
\begin{subequations}
\begin{align}
\theta_{f}=\frac{2.5838\lambda}{2\pi R},\\
w_{f}=\theta_f\cdot z.
\end{align}
\end{subequations}

\begin{figure}[H]
	\centering
\subfigure[]{
	\begin{minipage}[t]{0.35\linewidth}
		\centering
		\includegraphics[width=1.0\textwidth]{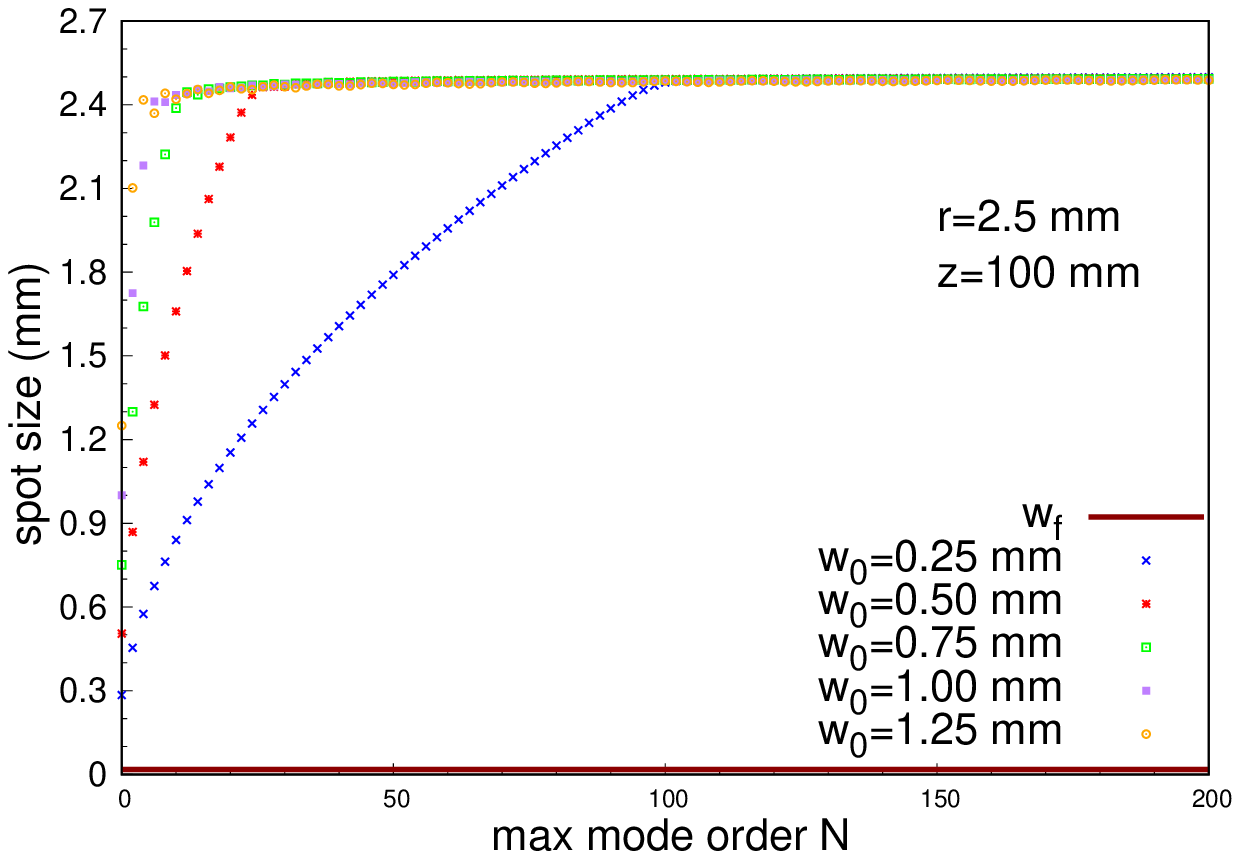}\label{top_hat_25_spot_100}
		
	\end{minipage}
	}
	\subfigure[]{
	\begin{minipage}[t]{0.35\linewidth}
		\centering
		\includegraphics[width=1.0\textwidth]{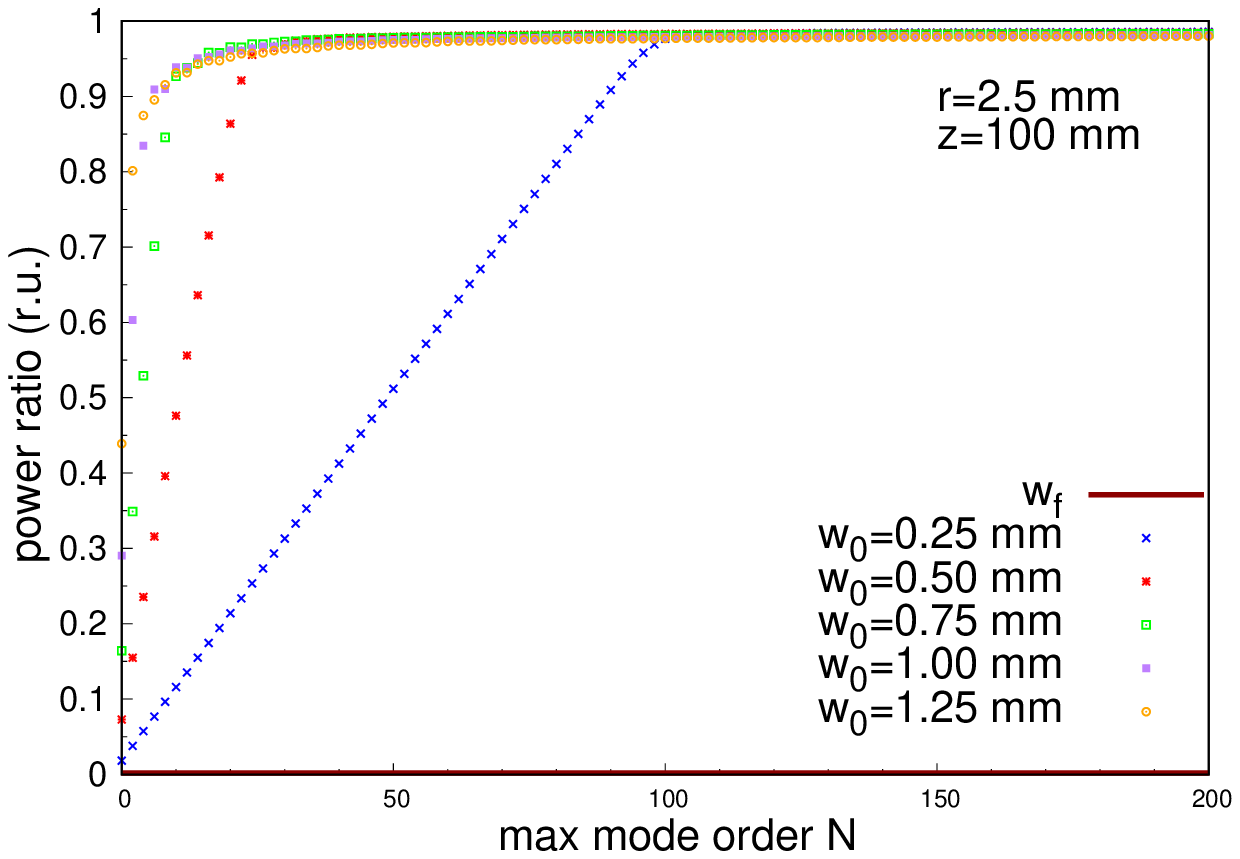}\label{top_hat_25_tpr_100}
		
	\end{minipage}
	}
\subfigure[]{
	\begin{minipage}[t]{0.35\linewidth}
		\centering
		\includegraphics[width=1.0\textwidth]{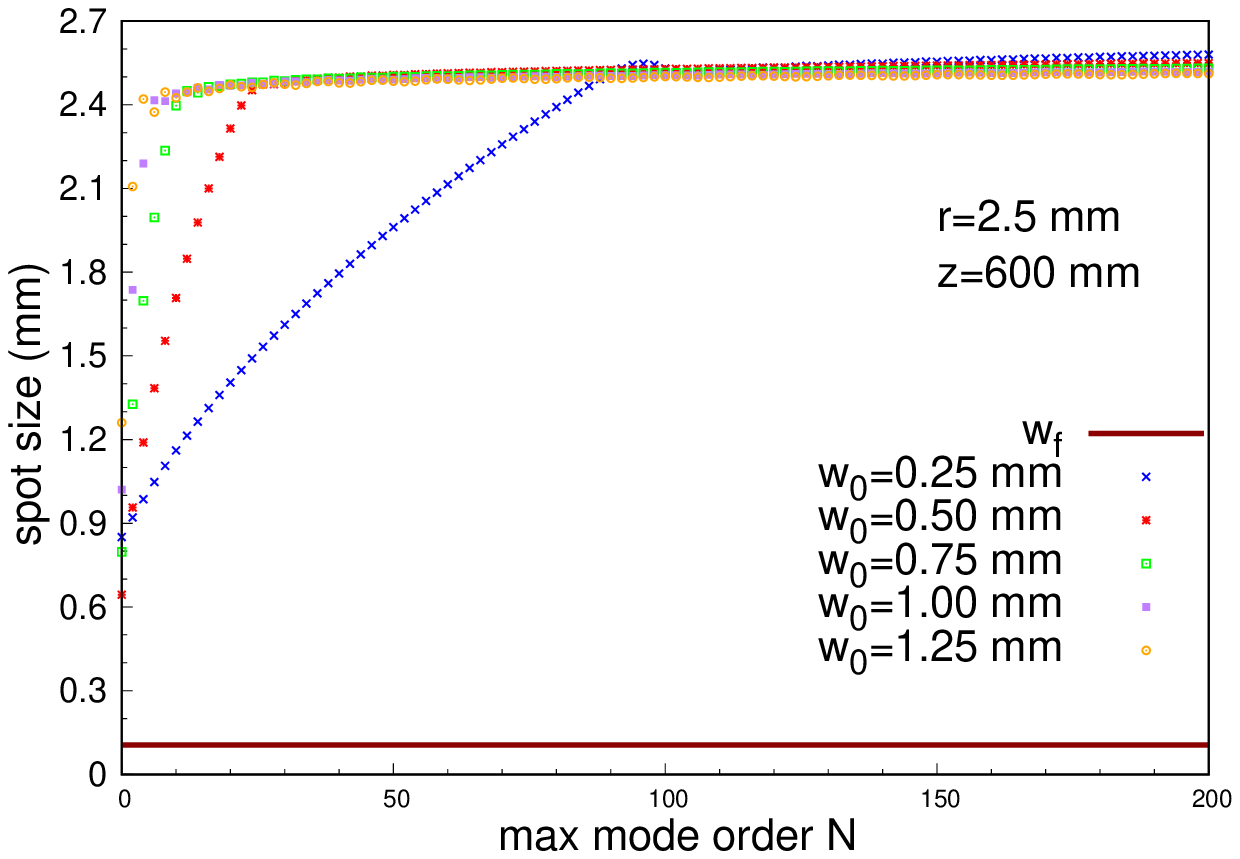}\label{top_hat_25_spot_600}
		
	\end{minipage}
	}
	\subfigure[]{
	\begin{minipage}[t]{0.35\linewidth}
		\centering
		\includegraphics[width=1.0\textwidth]{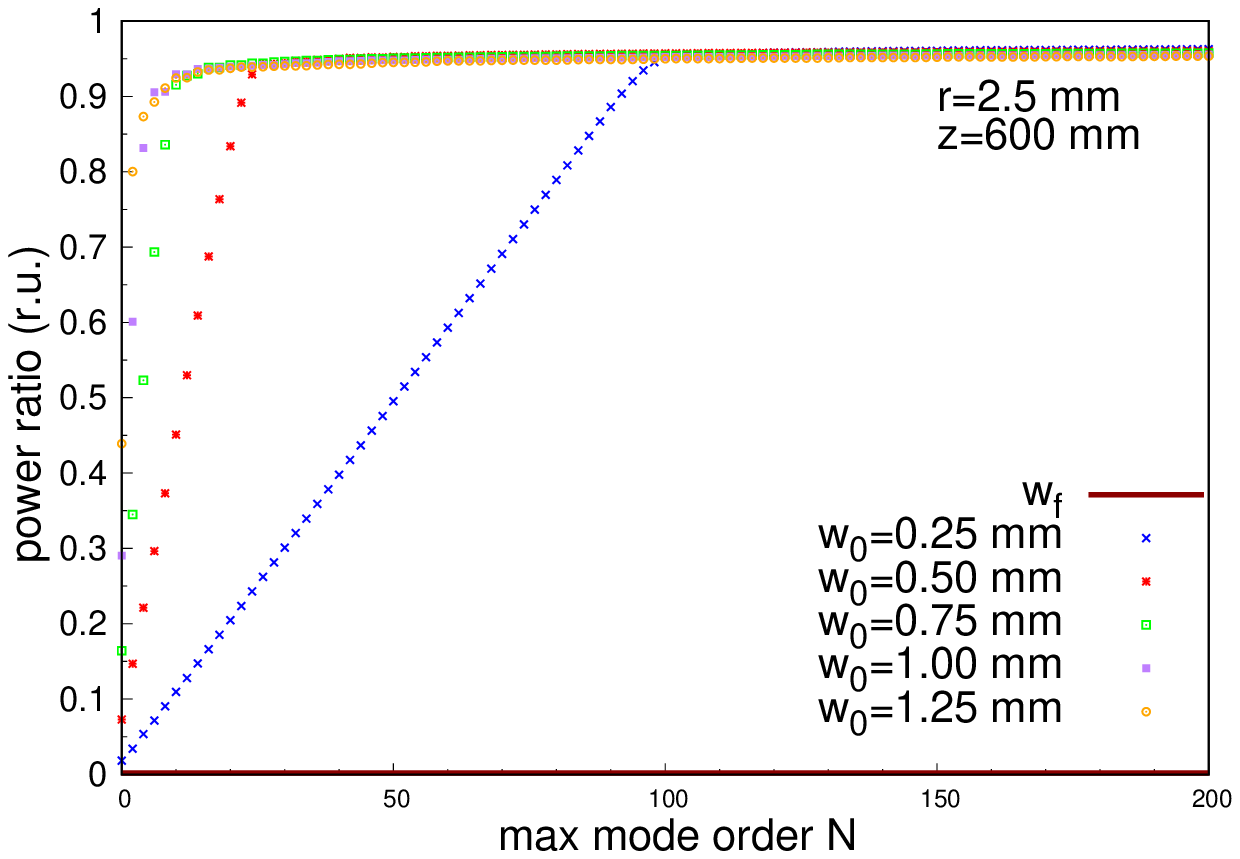}\label{top_hat_25_tpr_600}
		
	\end{minipage}
	}
	\subfigure[]{
	\begin{minipage}[t]{0.35\linewidth}
		\centering
		\includegraphics[width=1.0\textwidth]{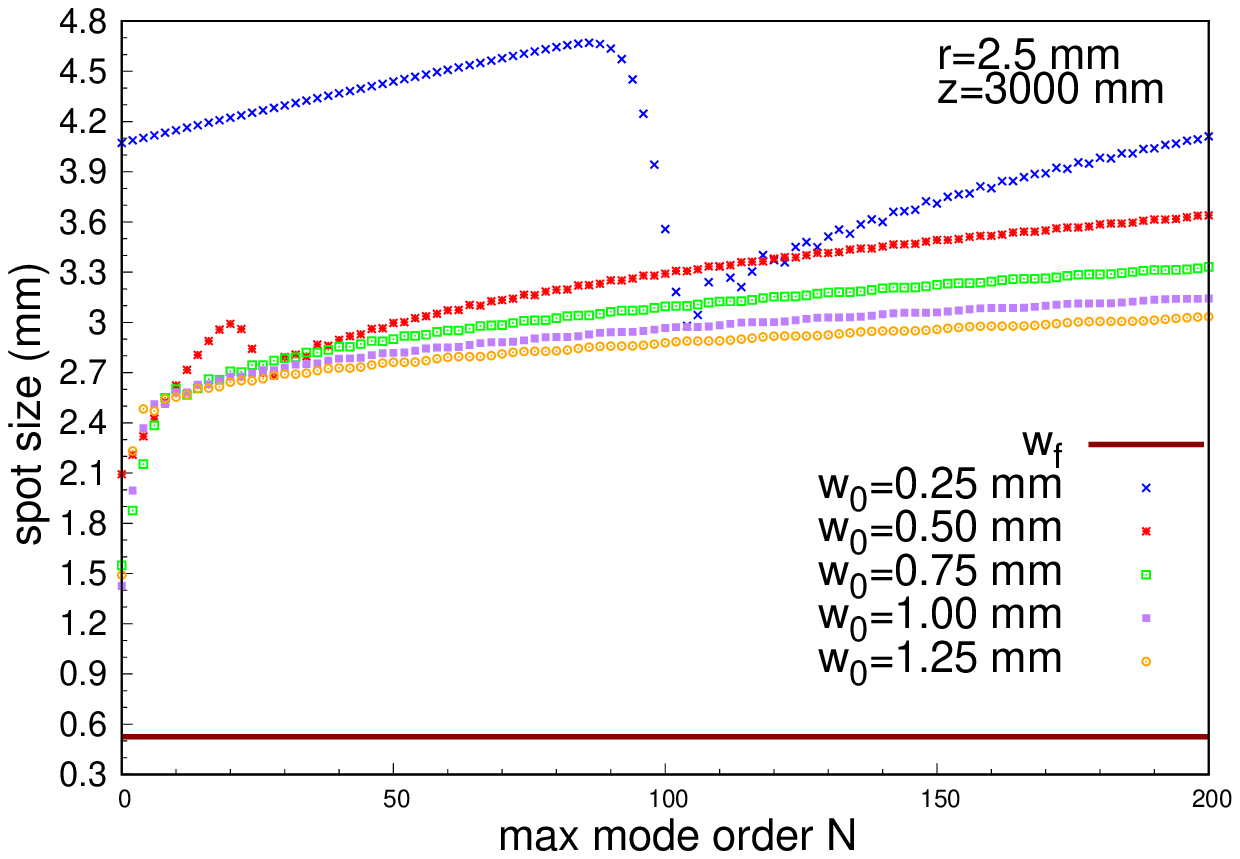}\label{top_hat_25_spot_3000}
		
	\end{minipage}
	}
	\subfigure[]{
	\begin{minipage}[t]{0.35\linewidth}
		\centering
		\includegraphics[width=1.0\textwidth]{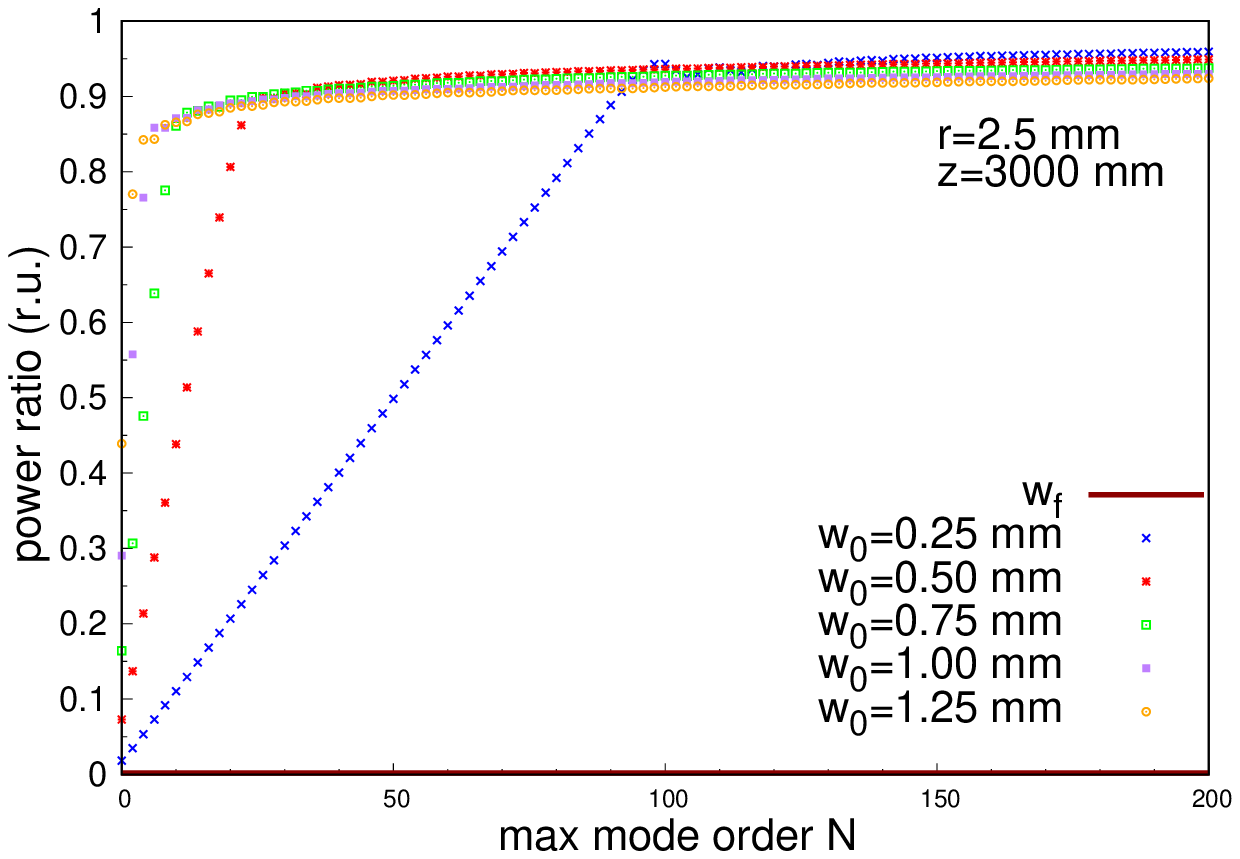}\label{top_hat_25_tpr_3000}
		
	\end{minipage}
	}
	\centering
	\caption{The MSD spot size of the simulated top-hat beam for different max mode order $N$ with the propagation distance $z=\SI{100}{mm}$ (a), $z=\SI{600}{mm}$ (c) or $z=\SI{3000}{mm}$ (e). The power ratio within the MSD spot range of the simulated top-hat beam for different max mode order $N$ with the propagation distance $z=\SI{100}{mm}$ (b), $z=\SI{600}{mm}$ (d) or $z=\SI{3000}{mm}$ (f).}\label{top-hat_spot_tpr}
\end{figure}

The max mode orders $N$ in the above cases are fixed to be 200. As a matter of fact, the max mode order $N$ may also influence the MSD spot size of a top-hat beam. \Cref{top_hat_25_spot_100,top_hat_25_spot_600,top_hat_25_spot_3000} show the variation of the spot size with the max mode order $N$ for the propagation distances $z=\SI{100}{mm}$, $z=\SI{600}{mm}$ and $z=\SI{3000}{mm}$. The variation of the corresponding power ratio within the MSD spot range are shown in \cref{top_hat_25_tpr_100,top_hat_25_tpr_600,top_hat_25_tpr_3000} for the propagation distances $z=\SI{100}{mm}$ , $z=\SI{600}{mm}$ and $z=\SI{3000}{mm}$. 

As mentioned before, the MSD spot size is infinite in hard-edge diffraction situations such as the top-hat beam here. Using the MEM, we can get a finite MSD spot size for the diffracted beam. However, this finite MSD spot size is dependent on the max mode order $N$ and the waist $w_0$ of the basic HG modes. For any fixed propagation distance $z$, the MSD spot size diverges with the increasing mode order $N$. As \cref{top-hat_inten} shows, all these spot ranges are large enough to contain the main part of the intensity patterns except for $w_f$. The spot size calculated from the analytical divergence angle of the top-hat beam $w_f$ is so small that we can not use it here.

The Fresnel number for this top-hat beam is 
\begin{equation}
F(z)=\frac{r^2}{z\lambda}.
\end{equation}
The Fresnel numbers for propagation distances $z=\SI{100}{mm}$, $z=\SI{600}{mm}$ and $z=\SI{3000}{mm}$ are $F(100)=58.7406$, $F(600)=9.7901$, $F(3000)=1.9580$. All of these three cross-sections are located in the near field of this top-hat beam because the Fresnel numbers are bigger than $1$. When the max mode order $N$ is large enough, the estimated spot ranges for these MEM beams are almost the same for very near fields such as $z=\SI{100}{mm}$ and $z=\SI{600}{mm}$. As for the not very near field such as $z=\SI{3000}{mm}$, the estimated spot ranges for different MEM beams are obviously different. The differences of the power ratio within the corresponding MSD spot range are small for $z=\SI{100}{mm}$, $z=\SI{600}{mm}$ and $z=\SI{3000}{mm}$. The larger the waist for the basic HG modes is, the smaller the MSD spot size and power ratio within the MSD spot range are. Diffracted aberration Gaussian beam as another diffracted beam case will be discussed in \ref{diffraction_aberr}.
\subsection{Spot size estimation for a top-hat beam in Science Interferometer}\label{top_si}

In this subsection, we are in a position to numerically track the evolution of the spot size of a flat top beam for the science interferometer in an optical bench for a LISA or TAIJI type mission. Further we will also argue that, despite of the divergence behaviour of the MSD spot size for a flat top beam, our analysis still draws a reliable conclusion. 

\begin{figure}[H]
	\centering
	\includegraphics[width=0.6\textwidth]{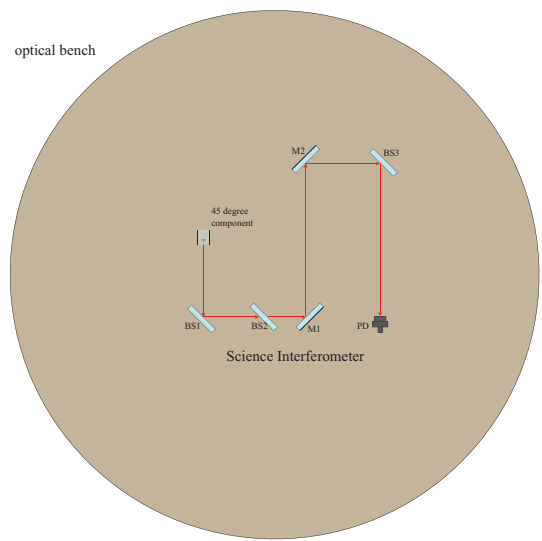}
	\centering
	\caption{Science interferometer qualitatively derived from paper\cite{2017SPIE10565E..2XD}, shown only starting from the $45$ degree component delivering the received beam from the telescope onto the optical bench, neglecting imaging systems and redundancy.}\label{science_interferometer}
\end{figure}

Consider one of the optical designs for the LISA optical bench \cite{2017SPIE10565E..2XD} and we take a top-hat beam of a science interferometer as a representative example of optical beam propagation within the optical bench for our investigation. In \cref{science_interferometer}, the science interferometer subsystem of this optical bench is shown. The initial radius of the input top-hat beam in the exit pupil plane is $\SI{2.5}{mm}$ \cite{2017SPIE10565E..2XD}. Given that the optical bench design of neither of both missions is available yet, we can only estimate that the propagation distances in the science interferometer on the optical bench will be in the order of tens of centimeters. Consequently, we test here the spot size evolution for propagation distances up to $\SI{60}{cm}$. We use \cref{msd_spot,eq_pr} to calculate the MSD spot size and the corresponding power ratio within the spot range here.

\begin{figure}[H]
	\centering
	\subfigure[]{
	\begin{minipage}[t]{0.33\linewidth}
		\centering
		\includegraphics[width=1.0\textwidth]{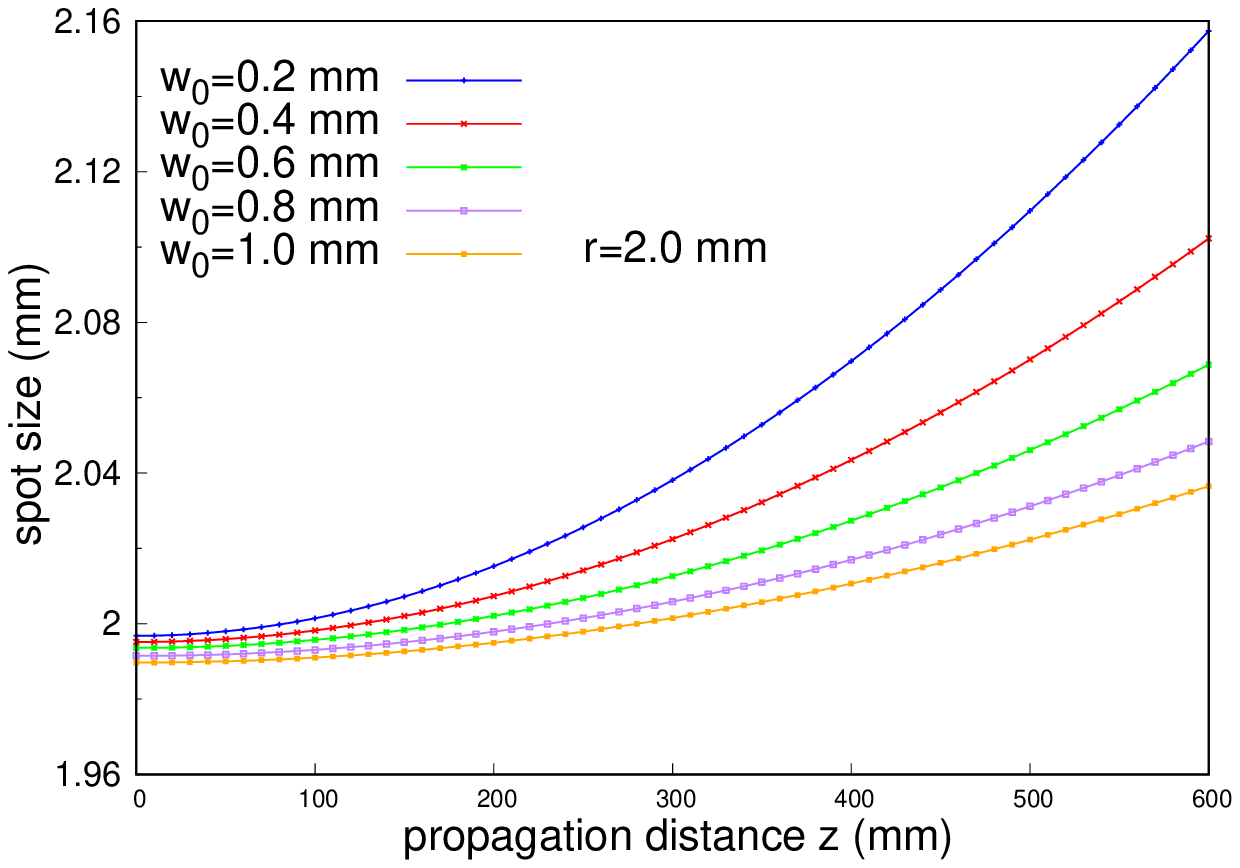}\label{top_hat_sp_20}
		
	\end{minipage}
	}
		\subfigure[]{
	\begin{minipage}[t]{0.33\linewidth}
		\centering
		\includegraphics[width=1.0\textwidth]{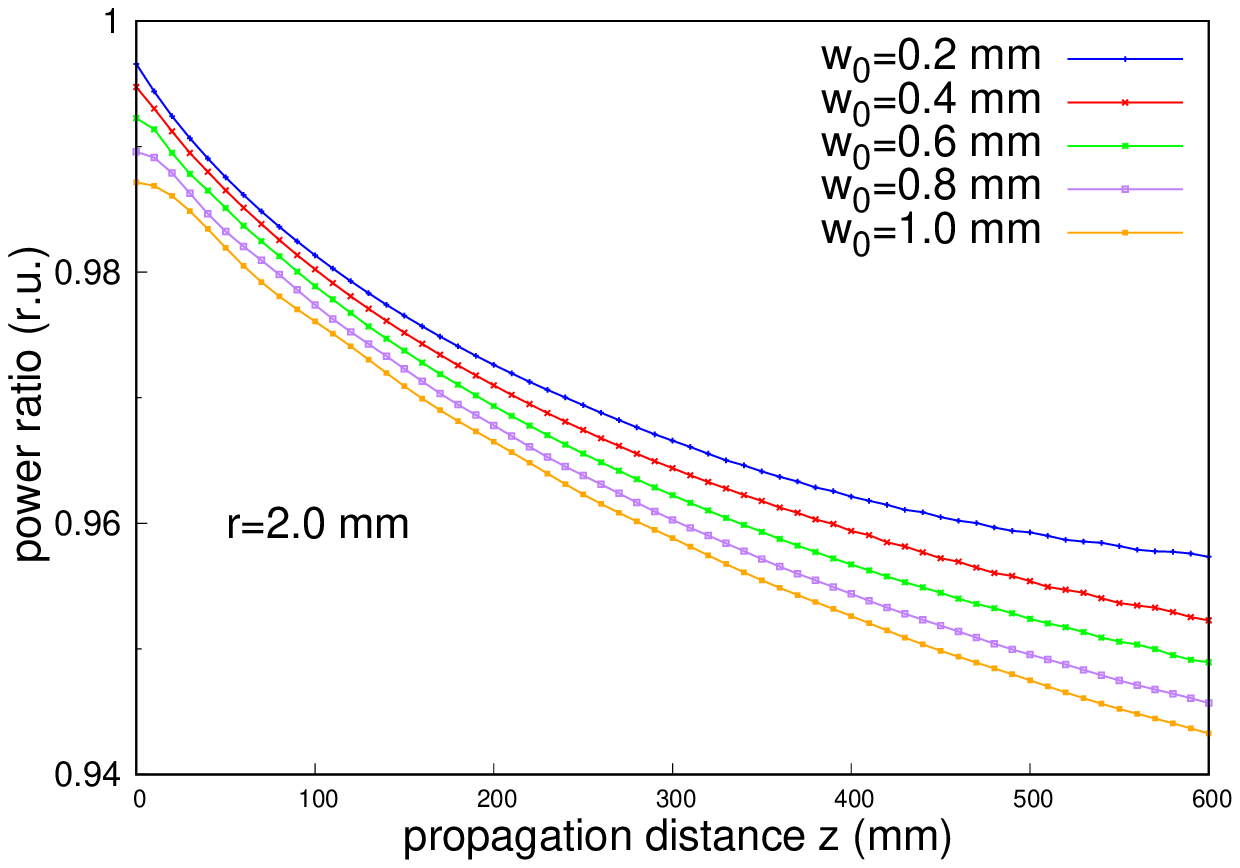}\label{top_hat_tpr_20}
		
	\end{minipage}
	}
		\subfigure[]{
	\begin{minipage}[t]{0.33\linewidth}
		\centering
		\includegraphics[width=1.0\textwidth]{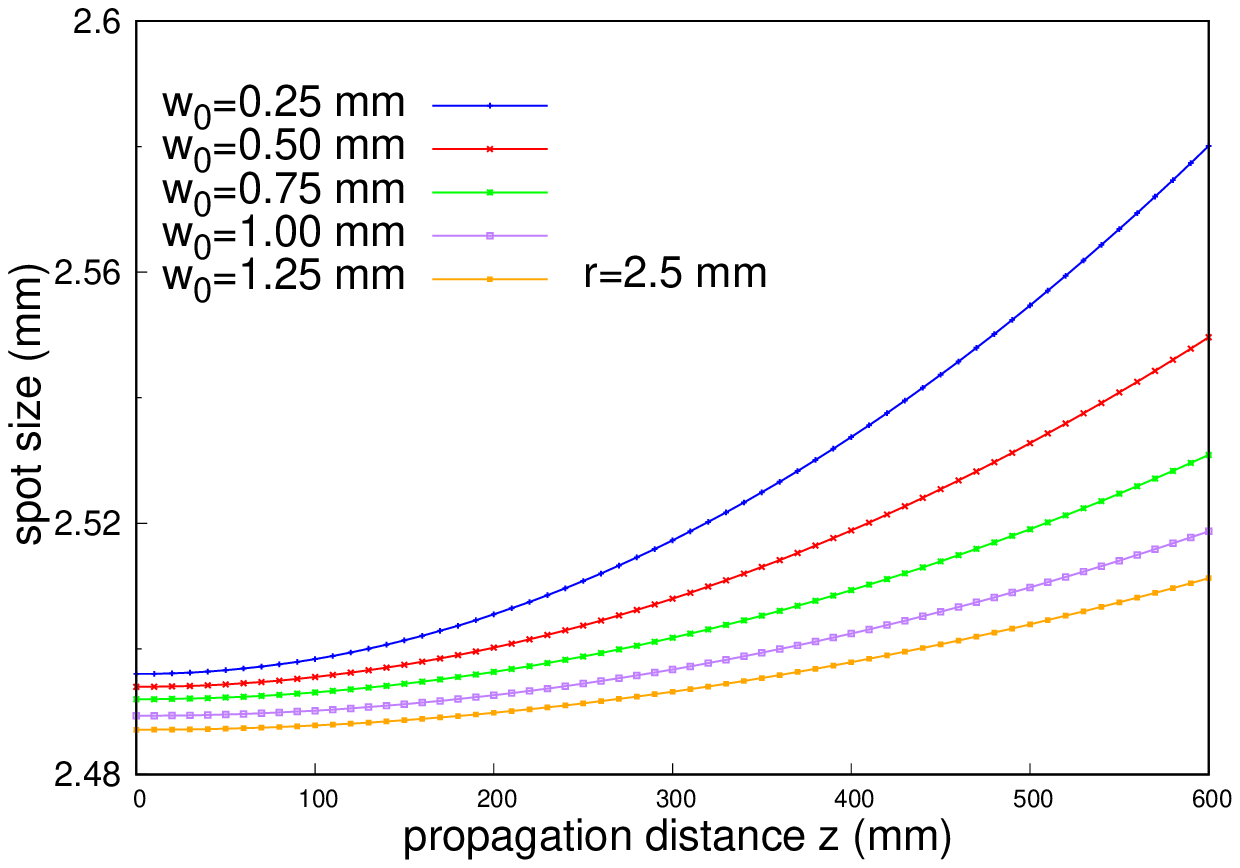}\label{top_hat_sp_25}
		
	\end{minipage}
	}
		\subfigure[]{
	\begin{minipage}[t]{0.33\linewidth}
		\centering
		\includegraphics[width=1.0\textwidth]{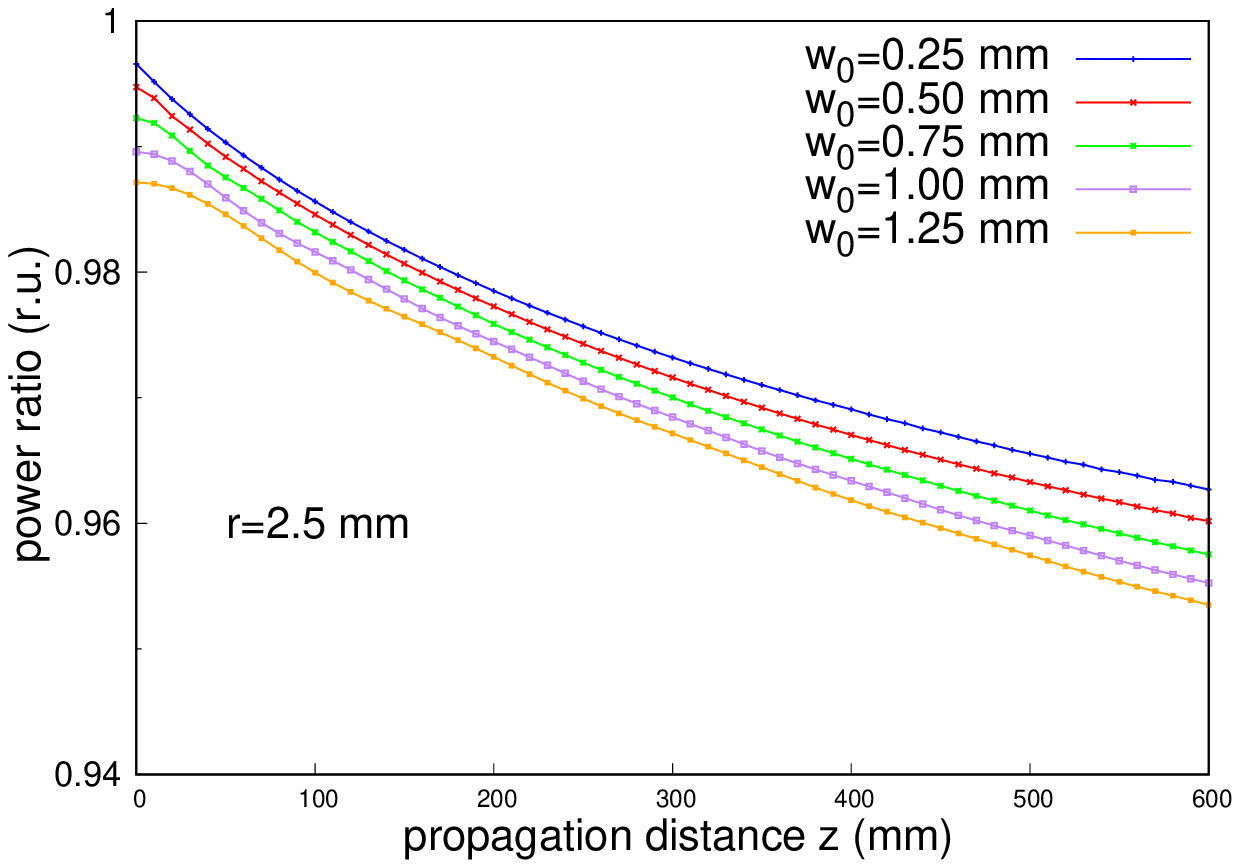}\label{top_hat_tpr_25}
		
	\end{minipage}
	}
		\subfigure[]{
	\begin{minipage}[t]{0.33\linewidth}
		\centering
		\includegraphics[width=1.0\textwidth]{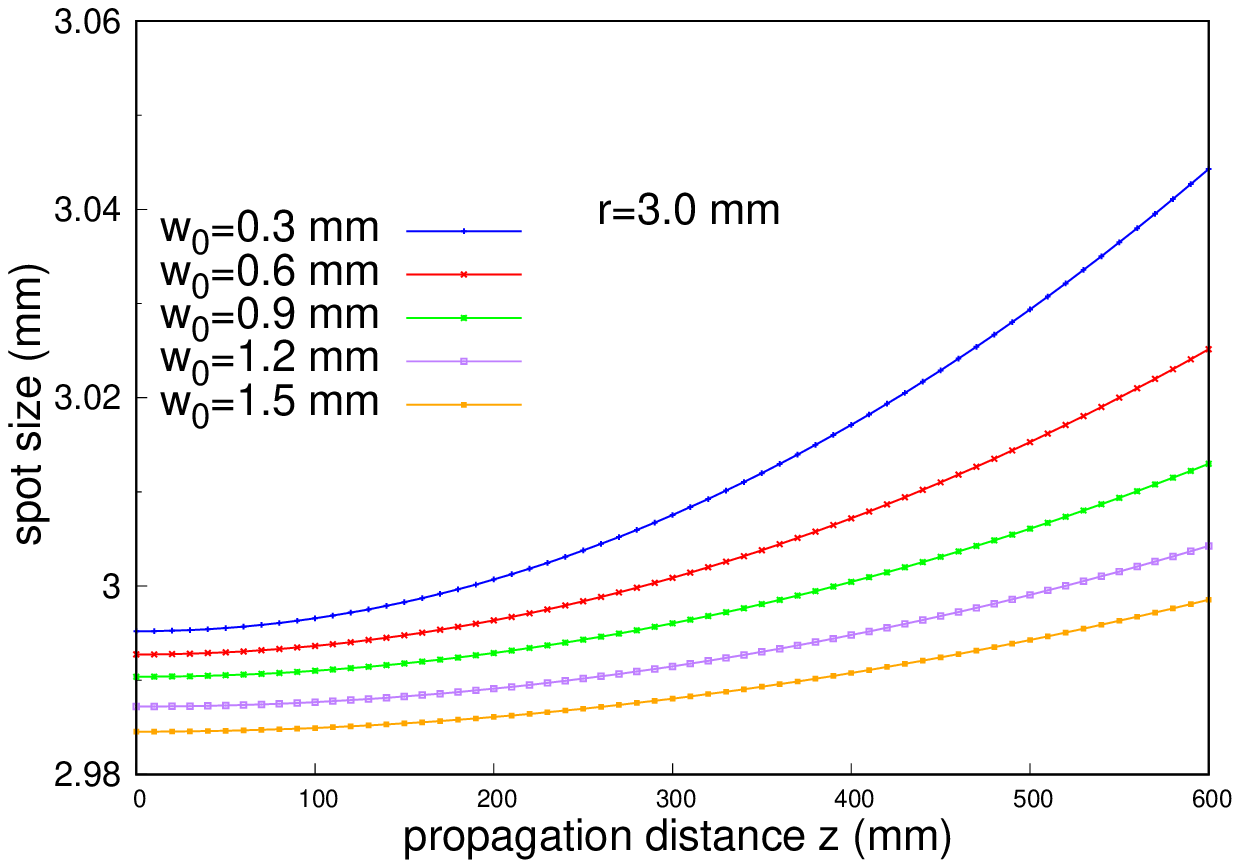}\label{top_hat_sp_30}
		
	\end{minipage}
	}
		\subfigure[]{
	\begin{minipage}[t]{0.33\linewidth}
		\centering
		\includegraphics[width=0.99\textwidth]{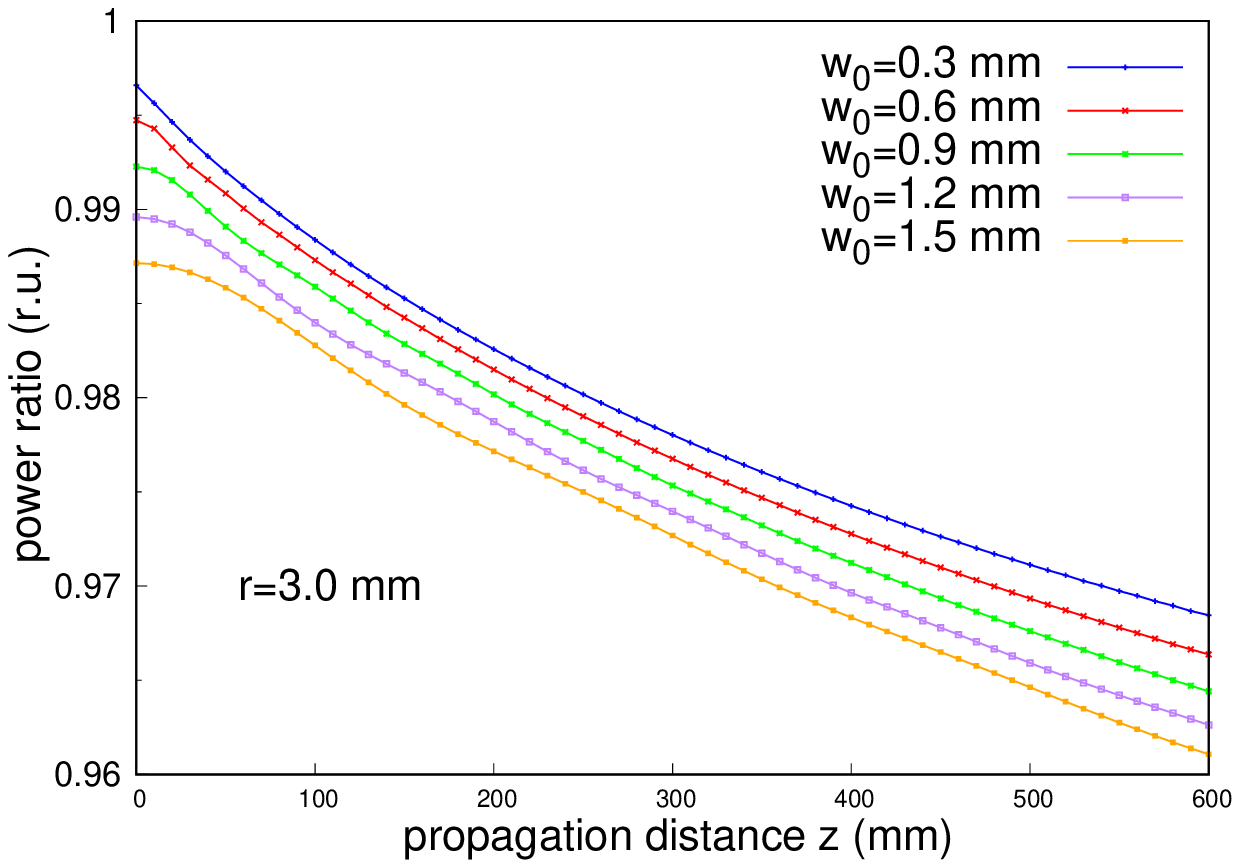}\label{top_hat_tpr_30}
		
	\end{minipage}
	}
	\centering
	\caption{The MSD spot sizes for different propagation distance $z$ for the top-hat beams with initial radii $r=\SI{2.0}{mm}$ (a),$r=\SI{2.5}{mm}$ (c) or $r=\SI{3.0}{mm}$ (e). The power ratio within the MSD spot range for different propagation distance $z$ of the top-hat beams with different initial radii $r=\SI{2.0}{mm}$ (b),$r=\SI{2.5}{mm}$ (d) or $r=\SI{3.0}{mm}$ (f).  }\label{top_hat_si}
\end{figure}

\Cref{top_hat_sp_20,top_hat_sp_25,top_hat_sp_30} show the MSD spot sizes for different propagation distances $z$ of different MEM beams which represent the top-hat beam with initial radii $r=\SI{2}{mm}$, $\SI{2.5}{mm}$ or $\SI{3}{mm}$ respectively. The settings of these MEM beams are the same except for the waist size $w_0$ of basic HG modes in the same subfigure. The max mode order $N$ of these MEM beams is equal to $200$. The origin and direction of the basic HG modes of MEM beams are the same as the origin and direction of the input top-hat beam.

\Cref{top_hat_tpr_20,top_hat_tpr_25,top_hat_tpr_30} show the power ratio within the MSD spot range for different propagation distances $z$ of different MEM beams which represent the top-hat beam with initial radii $r=\SI{2}{mm}$, $\SI{2.5}{mm}$ or $\SI{3}{mm}$ respectively. The settings of these MEM beams are the same as the settings in \cref{top_hat_sp_20,top_hat_sp_25,top_hat_sp_30}. 

For certain MEM beams, the MSD spot size increases with the propagation distance and it's the hyperbolic function of propagation distance $z$ \cite{1994OptCo.109....5P}. For certain propagation distances $z$, the MSD spot size increases as the waist size $w_0$ of the basic HG modes of the MEM beam decreases. As we discussed before, even though the MSD spot size of the MEM beam representing diffracted beams is finite, it is not unique for the corresponding diffracted beam. The MSD spot size growth rate increases with propagation distance. Furthermore, this growth rate increases as the waist size $w_0$ of basic HG beams decreases or the initial radius $r$ of the top-hat beam decreases.

For certain MEM beams, the power ratio within the MSD spot range decreases as the propagation distance $z$ increases here. For certain propagation distances $z$, the power ratio within the MSD spot range decreases as the waist size $w_0$ of basic HG modes increases. The growth rate of the power ratio within the MSD spot range increases with decreasing initial radii $r$ of the top-hat beam.

\subsection{Error analysis}\label{err}
In the previous section, we qualitatively discussed the performance of MSD spots for a science interferometer. In this part, we use error analysis to quantitatively discuss the performance of MSD spot size. The estimated errors of the MSD spot size and the estimated errors of the power ratio within the MSD spot range are small enough to be negligible for the top-hat beam initial radius from $2\SI{-3}{mm}$ in the propagation distance range $z\in[0,600]$ $\SI{}{mm}$.  The power ratios within the MSD spot ranges are large enough for the top-hat beam initial radius from $2\SI{-3}{mm}$ in the propagation distance range $z\in[0,600]$ $\SI{}{mm}$. This means that the estimated values of MSD spot size work well in a science interferometer in the detection of gravitational waves in space. 

Unphysical decomposition parameters such as the waist $w_0$ of basic HG modes can influence the value of the MSD spot size of an MEM beam. It is necessary to estimate the effect of the unphysical decomposition parameter $w_0$ on the MSD spot size of a top-hat beam in a science interferometer. Here we define the estimated error of the MSD spot size of the top-hat beam as twice the ratio of the difference between the maximum value of the MSD spot size for a certain propagation distance and the minimum value of the MSD spot size for the same propagation distance to the sum of this maximum value and minimum value 
\begin{equation} 
err_{spot}(z)=2\frac{w_{max}(z)-w_{min}(z)}{w_{max}(z)+w_{min}(z)},
\end{equation}
where $w_{max}(z)$ is the maximum value of the MSD spot size for propagation distance $z$ and $w_{min}(z)$ is the minimum value of the MSD spot size for propagation distance $z$.

Similar to the estimated error definition of the MSD spot, we define the estimated error of the power ratio within the MSD spot range as
 \begin{equation} 
err_{p_r}(z)=2\frac{p_{r_{max}}(z)-p_{r_{min}}(z)}{p_{r_{max}}(z)+p_{r_{min}}(z)},
\end{equation}
where $p_{r_{max}}(z)$ is the maximum value of the power ratio within the MSD spot range for propagation distance $z$ and $p_{r_{min}}(z)$ is the minimum value of the power ratio within the MSD spot range for propagation distance $z$.

\begin{figure}[H]
	\centering
	\subfigure[]{
	\begin{minipage}[t]{0.4\linewidth}
		\centering
		\includegraphics[width=0.95\textwidth]{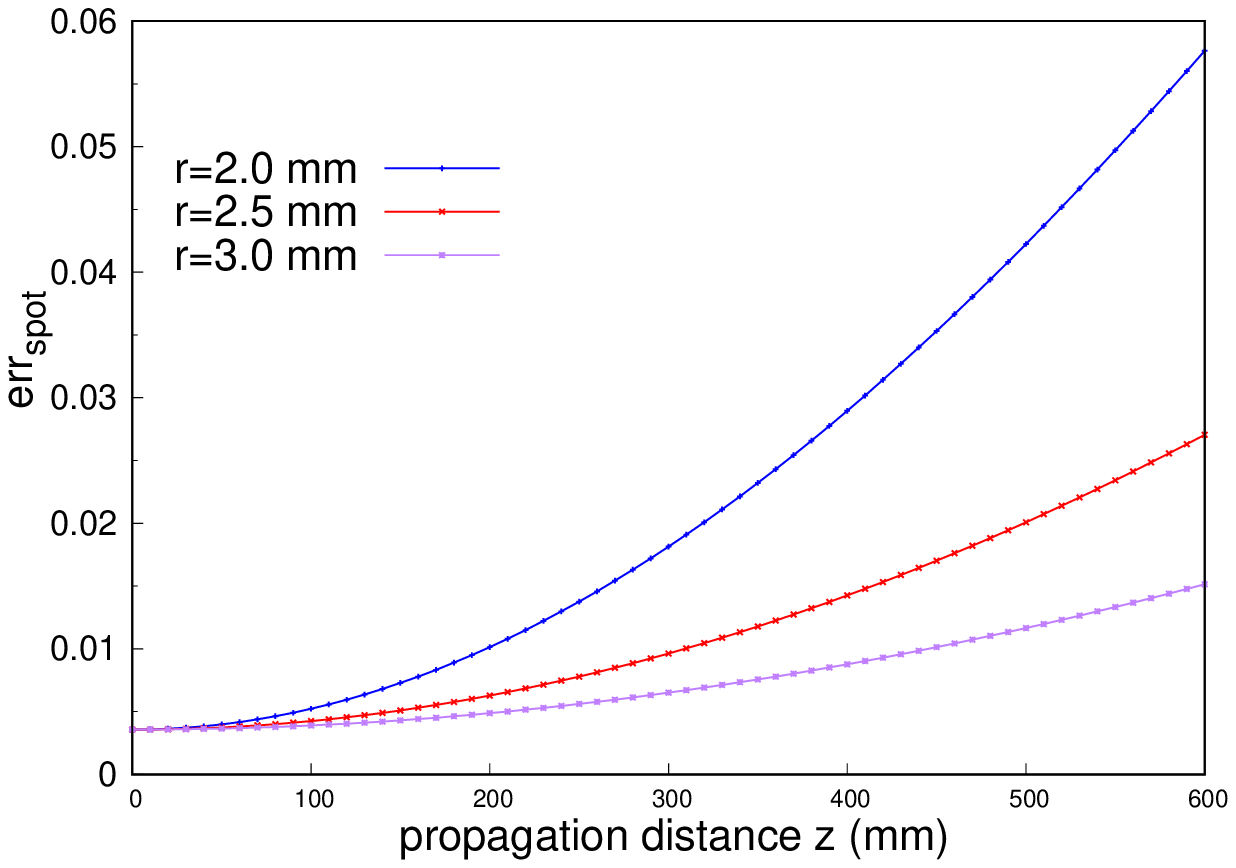}\label{err_sp}
		
	\end{minipage}
	}
		\subfigure[]{
	\begin{minipage}[t]{0.4\linewidth}
		\centering
		\includegraphics[width=0.95\textwidth]{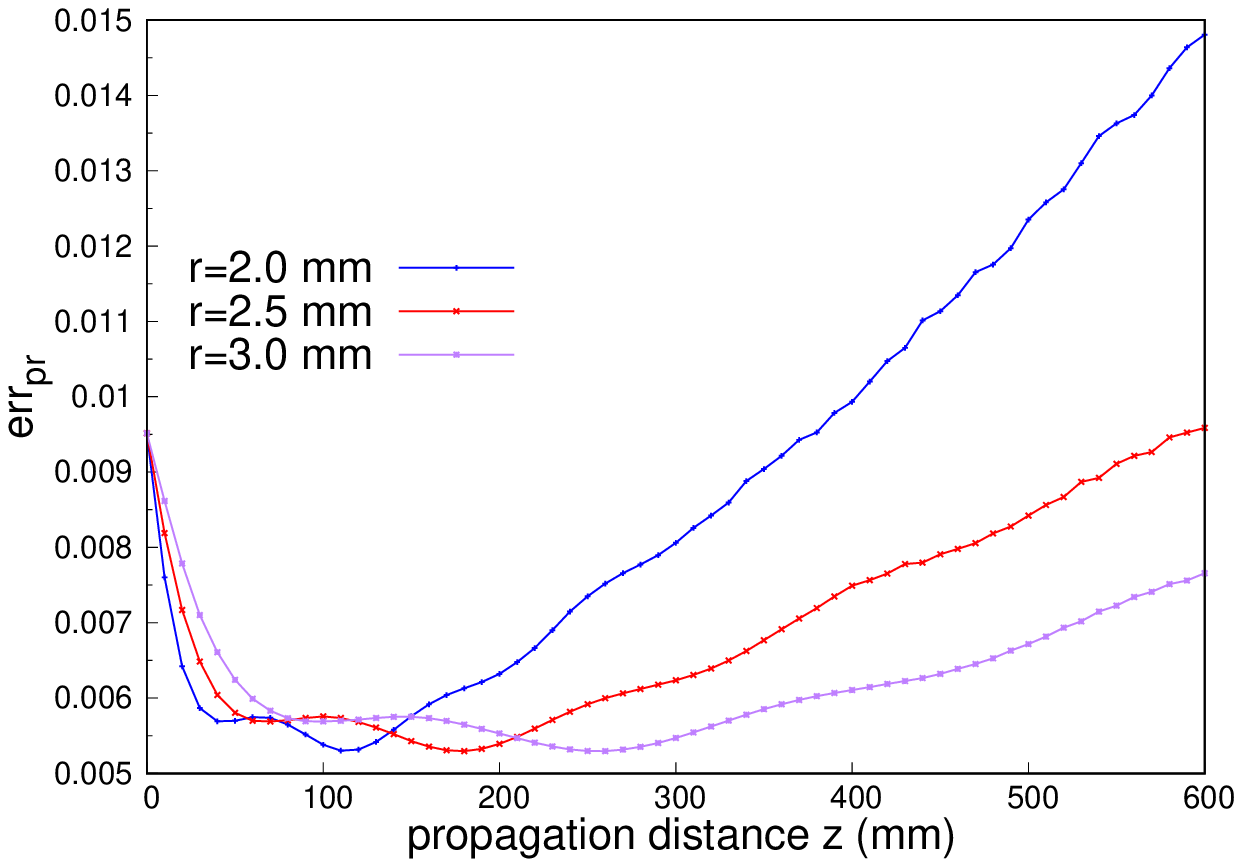}\label{err_pr}
		
	\end{minipage}
	}
	\centering
	\caption{(a) The estimated error of MSD spot sizes for different propagation distance $z$. (b) The estimated error of the power ratio within the MSD spot range for different propagation distance $z$.  }\label{err_sp_pr}
\end{figure}
\Cref{err_sp} shows the estimated error of the MSD spot size for different propagation distances $z$ of different MEM beams which represent the top-hat beam with the initial radii $r=\SI{2}{mm}$, $\SI{2.5}{mm}$ and $\SI{3}{mm}$. It is obvious that the estimated error of MSD spot size for the top-hat beam increases with the propagation distance $z$. The maximum estimated error of MSD spot size in these simulations of the top-hat beam with $\SI{2}{mm}$ initial radius is $5.764\%$ for $z=\SI{60}{cm}$ and the max value of MSD spot size in \cref{top_hat_sp_20} is $\SI{2.157}{mm}$ for $z=\SI{60}{cm}$. For the top-hat beam with $\SI{2.5}{mm}$ initial radius, the maximum estimated error of MSD spot size is $2.705\%$ for $z=\SI{60}{cm}$ and the max value of MSD spot size in \cref{top_hat_sp_25} is $\SI{2.58}{mm}$ for $z=\SI{60}{cm}$. As for the top-hat beam with $\SI{3}{mm}$ initial radius, the maximum estimated error of MSD spot size is $1.514\%$ for $z=\SI{60}{cm}$ and the max value of MSD spot size in \cref{top_hat_sp_30} is $\SI{3.044}{mm}$ for $z=\SI{60}{cm}$. The max estimated error of the MSD spot size decreases when the initial radius of the top-hat beam increases. For the science interferometer that we investigated here, the initial radius $r$ of the input top-hat beam is $\SI{2.5}{mm}$, and the corresponding maximum estimated error of MSD spot size is $2.705\%$ for $z=\SI{60}{cm}$. The estimated error of the MSD spot size for this situation is so small that we can use this MSD spot size to represent the spot size of the top-hat beam here. If the initial radius r is changed to $\SI{2}{mm}$ or $\SI{3}{mm}$, the estimated error of the MSD spot size would still be small enough to use. 

\Cref{err_pr} shows the estimated errors of the power ratio within the MSD spot range for different propagation distances $z$ of different MEM beams which represent the top-hat beams with the initial radii $r=\SI{2}{mm}$, $\SI{2.5}{mm}$ and $\SI{3}{mm}$. The maximum estimated error of the power ratio within the MSD spot range of the top-hat beam with $\SI{2}{mm}$ initial radius is $1.481\%$ for $z=\SI{60}{cm}$. For the top-hat beam with $\SI{2.5}{mm}$ initial radius, the maximum estimated error of the power ratio within the MSD spot range of the top-hat beam is $0.959\%$ for $z=\SI{60}{cm}$. As for the top-hat beam with $\SI{3}{mm}$ initial radius, the maximum estimated error of MSD spot size is $0.951\%$ for $z=\SI{0}{cm}$. The fractional part of the beam energy concentrated inside the MSD spot range is always larger than $94.326\%$ for the top-hat beam with $\SI{2}{mm}$ initial radius  in \cref{top_hat_tpr_20}, $95.351\%$ for the top-hat beam with $\SI{2.5}{mm}$ initial radius in \cref{top_hat_tpr_25} and $95.66\%$ for the top-hat beam with $\SI{3}{mm}$ initial radius in \cref{top_hat_tpr_30}. The maximum estimated error of the power ratio within the MSD spot range for the top-hat beam with $\SI{2.5}{mm}$ initial radius is the smallest in these three cases. If the initial radius $r$ is changed from $\SI{2.5}{mm}$ to $\SI{2}{mm}$ or $\SI{3}{mm}$, the estimated error of the MSD spot size would still be small enough to use. 

There are two reasons why the power inside the MSD spot range does not equal the power of the input top-hat beam. First of all, the power loss appears in the MEM process because of the finite decomposition mode order. This power loss is independent of the propagation distance $z$ and it increases with the waist size $w_0$ of basic HG modes. Secondly, a top-hat beam is a diffracted beam, parts of the diffracted wings always appear outside a finite range. This power loss is dependent on the propagation distance $z$. At the same time, we notice that the MSD spot size increases with the propagation distance $z$. As mentioned before, the power ratio within the MSD spot range decreases as the propagation distance $z$ increases. This means that the rate of increase of the MSD spot size is slightly slower than the diffusion rate of the diffracted wings.

In the above discussion, we find that the estimated errors of the MSD spot size, as well as the estimated errors of the power ratio within the MSD spot range, are small and the power ratios within the MSD spot range are large enough for the top-hat beam initial radius from $2\SI{-3}{mm}$ in the propagation distance range $z\in[0,600]$ $\SI{}{mm}$. 

The optical elements on the LISA Pathfinder optical bench had a size of $15\times20\times7\SI{}{mm}$ \cite{2004SPIE.5500..164B}. The largest spot size for the top-hat beam in a science interferometer here is $\SI{2.580}{mm}$. The usual practice of the optical element's half width is designed to be approximately equal or slightly larger than three times the spot sizes of the input beams to avoid the clipping problem \cite{1990SPIE.1224....2S}. Using this rule, the size of the optical elements can be $\SI{15.48}{mm}$. 
Based on this size and the size of the optical elements for LISA Pathfinder, we suggest that also space-based gravitational wave detectors like LISA and TAIJI will use comparable component sizes. 

Depending on the optical designs, i.e. particularly the aperture sizes and potential use of imaging optics, the given equations in \cref{sec:method} can be used to derive the optimal size of all mirrors and beam splitters. This optimal size means that the components are small and lightweight but at the same time, large enough to prevent beam clipping. The present work will serve as a useful guide in the future system design of the optical bench and the sizes of the optical components. 

\section{ Conclusion}

In this paper, a slightly different variant of the definition of spot size is proposed, with which we calculate the MSD spot size of a light beam for arbitrary propagation distances. Even with diffraction taken into account, the definition still upholds the hyperbolic law and ABCD law of the MSD spot size. Though the MSD spot size for a diffracted MEM beam is dependent on the decomposition parameters such as the max mode order and the waist of the basic HG modes, it is shown that, for different reasonable choices of parameters, the estimated error for the MSD spot size and the power ratio within the MSD spot range will have little effect on our drawn conclusion. With the help of this method, optical systems such as the interferometers in space based gravitational wave detectors can be efficiently tested for beam clipping.

 Our study results provide a method to allow for an optimal design of optical systems with top-hat or other types of non-Gaussian beams. Furthermore, it allows testing the interferometry of space-based gravitational wave detectors for beam clipping in optical simulations. The present work will serve as a useful guide in the future system design of the optical bench and the sizes of the optical components.

\section{Acknowledgements}
 We thank Prof. Yun-Kau Lau and Apl. Prof. Gerhard Heinzel for the useful discussions. This work has been supported in part by the National Key Research and Development Program of China under Grant No.2020YFC2201501, the National Science Foundation of China (NSFC) under Grants No. 12147103 (special fund to the center for quanta-to-cosmos theoretical physics), No. 11821505, the Strategic Priority Research Program of the Chinese Academy of Sciences under Grant No. XDB23030100, and the Chinese Academy of Sciences (CAS) and the Max Planck Society (MPG) in the framework of the LEGACY cooperation on low-frequency gravitational wave astronomy (M.IF.A.QOP18098). Likewise, we gratefully acknowledge the German Space Agency, DLR and support by the Federal Ministry for Economic Affairs and Energy based on a resolution of the German Bundestag (FKZ50OQ1801) as well as the Deutsche Forschungsgemeinschaft (DFG) funding the Cluster QuantumFrontiers (EXC2123, Project ID 390837967) for funding the work contributions by Gudrun Wanner. We gratefully acknowledge DFG for funding the Collaborative Research Centres CRC 1464: TerraQ – Relativistic and Quantum-based Geodesy and the Clusters of Excellence PhoenixD (EXC 2122, Project ID 390833453). In this paper, the part of the numerical computation is finished by TAIJI Cluster.
\appendix
\section{power ratio within the MSD spot range}\label{appen: pr}
Here, we deduce the relationship between the power ratio within the MSD spot range and $z$. The intensity of the MEM beam can be represented as
\begin{equation}
\begin{split}
	I=  &\sum\limits_{m_1=0}^{N}\sum\limits_{n_1=0}^{N-m_1}|a_{m_1n_1}|^2|u_{m_1n_1}(x,y,z)|^2+2 \sum\limits_{m_1=0}^{N}\sum\limits_{n_1=0}^{N-m_1} \sum\limits_{m_2=0}^{N}\sum\limits_{n_2=0}^{N-m_2}|a_{m_1n_1}||a_{m_2n_2}|\\ 
	&\cdot  |u_{m_1n_1}(x,y,z)||u_{m_2n_2}(x,y,z)|\\
	&\cdot\cos{(\beta_{m_1n_1}+\phi_{m_1n_1}-\beta_{m_2n_2}-\phi_{m_2n_2})}
	\quad | \text{for} \ (m_2\neq m_1 \ \text{or}\  n_2\neq n_1),
\end{split}
\end{equation}
where $(m_1,n_1)$, $(m_2,n_2)$ are two of the basic HG modes of the MEM beam and $N$ is the max mode order of the MEM beam. The integrand of \cref{eq_pr} is 
\begin{equation}
\begin{split}
\frac{I}{P_{MEM}}=&\frac{1}{\sum\limits_{m_1=0}^{N}\sum\limits_{n_1=0}^{N-m_1}|a_{m_1n_1}|^2}\frac{1}{w(z)^2}\exp{\left(-2\frac{x^2+y^2}{w(z)^2}\right)}\\
&\cdot\left(\sum\limits_{m_1=0}^{N}\sum\limits_{n_1=0}^{N-m_1}|a_{m_1n_1}|^2 c_{m_1n_1}^2H_{m_1}^2\left(\frac{\sqrt(2)x}{w(z)}\right)H_{n_1}^2\left(\frac{\sqrt(2)y}{w(z)}\right)\right.\\
& +2\sum\limits_{m_1=0}^{N}\sum\limits_{n_1=0}^{N-m_1} \sum\limits_{m_2=0}^{N}\sum\limits_{n_2=0}^{N-m_2}|a_{m_1n_1}||a_{m_2n_2}|c_{m_1n_1}c_{m_2n_2}\\
&\cdot H_{m_1}\left(\frac{\sqrt(2)x}{w(z)}\right)H_{n_1}\left(\frac{\sqrt(2)y}{w(z)}\right)H_{m_2}\left(\frac{\sqrt(2)x}{w(z)}\right)H_{n_2}\left(\frac{\sqrt(2)y}{w(z)}\right)\\
&\left. \cdot \cos{\left(\beta_{m_1n_1}-\beta_{m_2n_2}+\left(m_1+n_1-m_2-n_2\right)\zeta(z)\right)} \right) .
\end{split}
\end{equation}
Rewrite \cref{xbahg,ybahg,wxhg,wyhg} as
\begin{subequations}
\begin{align}
s_x(z)=f_1(z)w(z),\\
s_y(z)=f_2(z)w(z),\\
w_x(z)=f_3(z)w(z),\\
w_y(z)=f_4(z)w(z).
\end{align}
\end{subequations}
$f_1(z)$, $f_2(z)$, $f_3(z)$ and $f_4(z)$ are dependent on $z$ because of the Gouy phase part. Set $u=\frac{\sqrt{2}x}{w(z)}$ and $v=\frac{\sqrt{2}y}{w(z)}$, then
\begin{equation}
\begin{split}
\frac{I}{P_{MEM}}=&\frac{I}{P_{MEM}}(u,v,z)=\frac{1}{\sum\limits_{m_1=0}^{N}\sum\limits_{n_1=0}^{N-m_1}|a_{m_1n_1}|^2}\frac{1}{w(z)^2}\exp{\left(-u^2-v^2\right)}\\
&\cdot\left(\sum\limits_{m_1=0}^{N}\sum\limits_{n_1=0}^{N-m_1}|a_{m_1n_1}|^2 c_{m_1n_1}^2H_{m_1}^2\left(u\right)H_{n_1}^2\left(v\right)\right.\\
& +2\sum\limits_{m_1=0}^{N}\sum\limits_{n_1=0}^{N-m_1} \sum\limits_{m_2=0}^{N}\sum\limits_{n_2=0}^{N-m_2}|a_{m_1n_1}||a_{m_2n_2}|c_{m_1n_1}c_{m_2n_2}\\
&\cdot H_{m_1}\left(u\right)H_{n_1}\left(v\right)H_{m_2}\left(u\right)H_{n_2}\left(v\right)\\
&\left. \cdot \cos{\left(\beta_{m_1n_1}-\beta_{m_2n_2}+\left(m_1+n_1-m_2-n_2\right)\zeta(z)\right)} \right) .\label{inten_uv}
\end{split}
\end{equation}
\begin{subequations}
\begin{align}
\int_{s_x(z)-w_x(z)}^{s_x(z)+w_x(z)}dx=\int_{\sqrt{2}(f_1(z)-f_3(z))}^{\sqrt{2}(f_1(z)+f_3(z))}\frac{w(z)}{\sqrt{2}}du,\\
\int_{s_y(z)-w_y(z)}^{s_y(z)+w_y(z)}dy=\int_{\sqrt{2}(f_2(z)-f_4(z))}^{\sqrt{2}(f_2(z)+f_4(z))}\frac{w(z)}{\sqrt{2}}dv
\end{align}\label{pr_limit}
\end{subequations}
Using \cref{inten_uv} and \cref{pr_limit} in the definition of the power ratio within the MSD spot range \cref{eq_pr}, we find $\epsilon_{P_{spot}}$ is obviously dependent on $z$
\begin{equation}
\epsilon_{P_{spot}}=\int_{\sqrt{2}(f_1(z)-f_3(z))}^{\sqrt{2}(f_1(z)+f_3(z))}\int_{\sqrt{2}(f_2(z)-f_4(z))}^{\sqrt{2}(f_2(z)+f_4(z))}\frac{w(z)^2}{2}\frac{I}{P_{MEM}}(u,v,z)dudv.
\end{equation}
The integrand $\frac{w(z)^2}{2}\frac{I}{P_{MEM}}(u,v,z)$ is still dependent on $z$ because of the Gouy phase part. According to the property of the $Arctan$ function, the Gouy phase always can be regarded as a constant when $z$ is large enough. If $z$ is large enough, $f_1(z)$, $f_2(z)$, $f_3(z)$ and $f_4(z)$, the integrand $\frac{w(z)^2}{2}\frac{I}{P_{MEM}}(u,v,z)$ and $\epsilon_{P_{spot}}$ will be independent from $z$.

\section{shifted HG beam}\label{sec:shift_HG}
Sometimes, it's not easy to find the center of the beam in the experiment. The paper by X. Xue et al. \cite{Xue2000} shows that people can use the intensity profiles to find the superposition of a finite number of HG modes which can represent the experimental beam. The centers of these HG modes are usually not the same as the center of the input experimental beam. If the center of these decomposed HG modes is set to the same as the origin of the experiment of the beam, we would do a more accurate component analysis. Here, we would like to test \cref{xbahg,ybahg,wxhg,wyhg}, especially the estimated value of beam center in the shifted HG Gaussian beam situation. The shifted beam means that the center of the beam is not the same as the centers of decomposed HG modes. For instance, we set the order of this shifted HG beam as $(1,0)$, and its origin is not located at the brightest point.  Assume the waist of this shifted HG beam is $w_0=\SI{1}{mm}$, the propagation direction is $(0,0,1)$ and the center is $(1,0)$. The electric field of this input shifted HG beam is
\begin{equation}
\begin{split}
E_{in}=u_{in}\exp{\left(i\omega t\right)}=&\sqrt{\frac{P}{\pi}}\frac{1}{w(z)}H_1\left(\frac{\sqrt{2}(x-1)}{w(z)}\right)H_0\left(\frac{\sqrt{2}y}{w(z)}\right)\exp{\left(-\frac{(x-1)^2+y^2}{w^2(z)}\right)}\\
&\times\exp{\left(-ikz-ik\frac{(x-1)^2+y^2}{2R(z)}+i2\zeta(z)+i\omega t\right)}.
\end{split}
\end{equation}
These symbols are defined in \cref{sec:method}. Here we use two different waists $\SI{0.4}{mm}$ and $\SI{0.5}{mm}$ for the finite number of HG modes. The direction of these HG modes is the same as the input beam. The centers of these HG modes are $(0,0)$. The decomposition happens in the waist plane of the shifted HG beam.
\begin{figure}[H]
	\centering
	\subfigure[]{
	\begin{minipage}[t]{0.45\linewidth}
		\centering
		\includegraphics[width=0.95\textwidth]{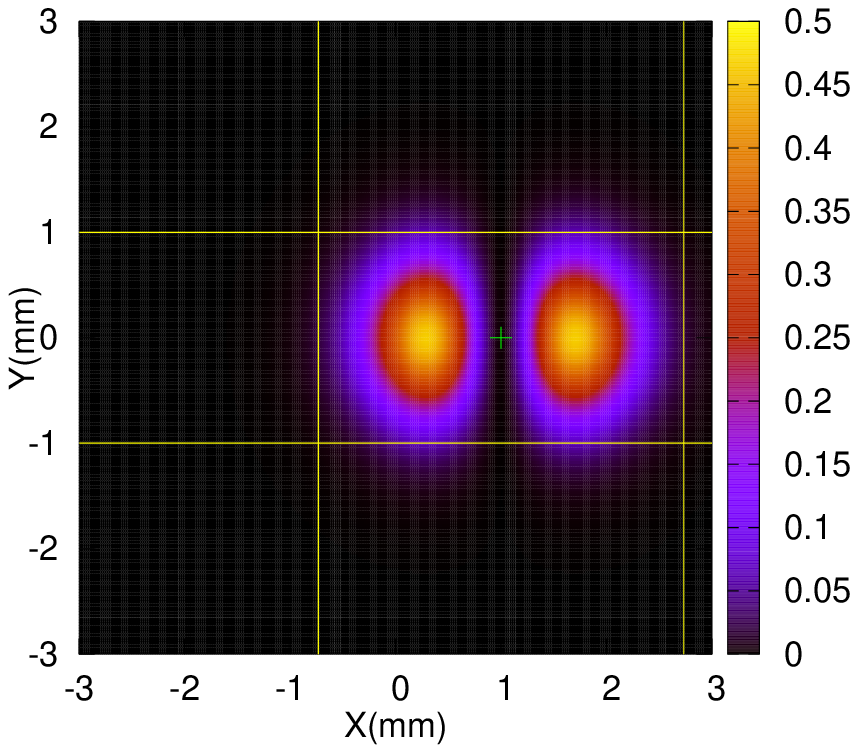}\label{sa10_100}
		
	\end{minipage}
	}
	\subfigure[]{
	\begin{minipage}[t]{0.45\linewidth}
		\centering
		\includegraphics[width=0.95\textwidth]{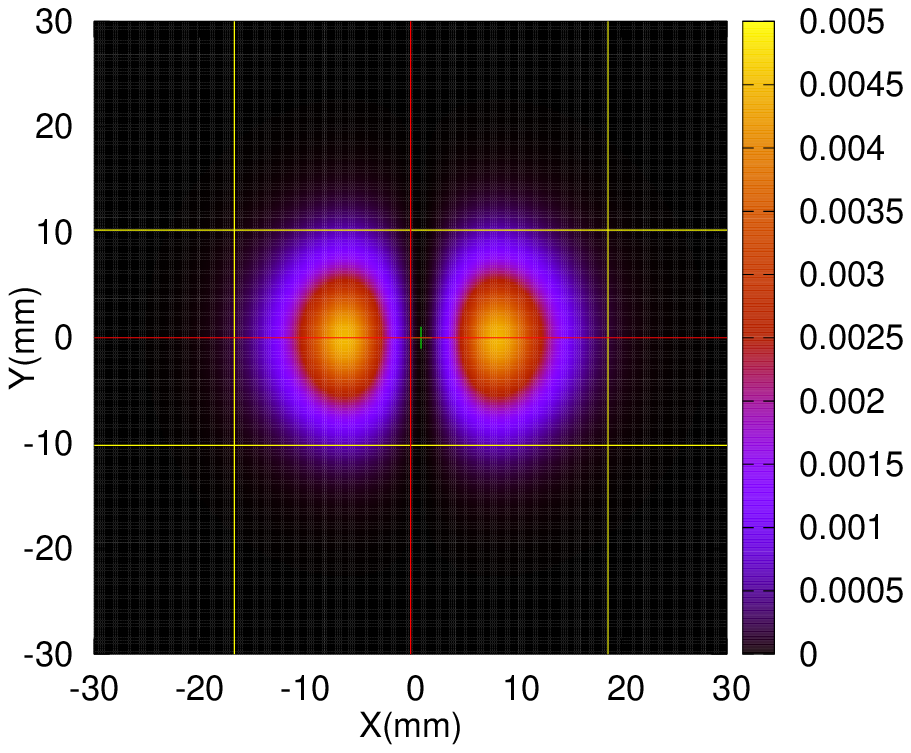}\label{sa10_30000}
		
	\end{minipage}
	}
	\centering
	\caption{(a). The intensity profile for $z= \SI{100}{mm}$. (b) .The intensity profile for $z= \SI{30000}{mm}$.}\label{sa10_inten}
\end{figure}
\begin{figure}[H]
	\centering
	\subfigure[]{
	\begin{minipage}[t]{0.35\linewidth}
		\centering
		\includegraphics[width=0.99\textwidth]{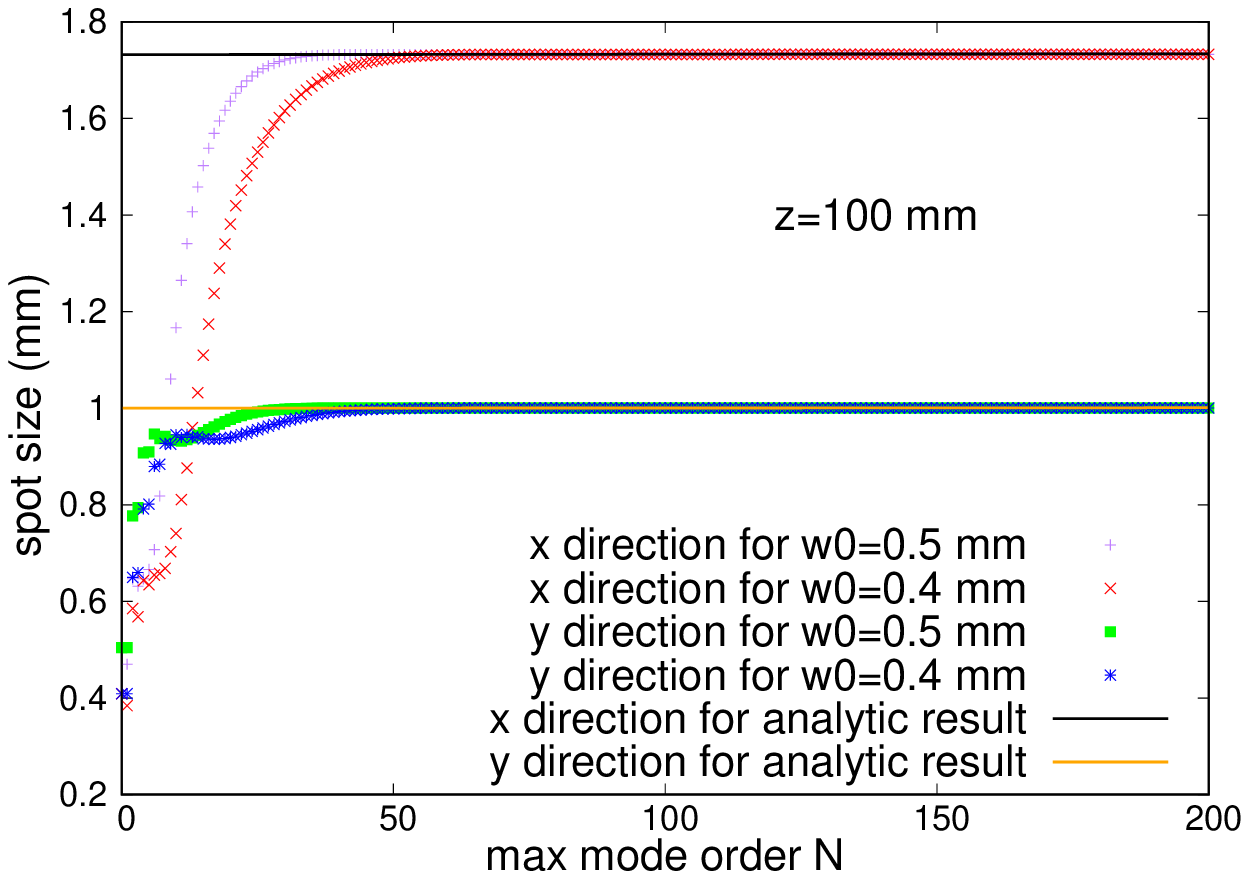}\label{spot_sahgb10_100_ord}
		
	\end{minipage}
	}
	\subfigure[]{
	\begin{minipage}[t]{0.35\linewidth}
		\centering
		\includegraphics[width=0.99\textwidth]{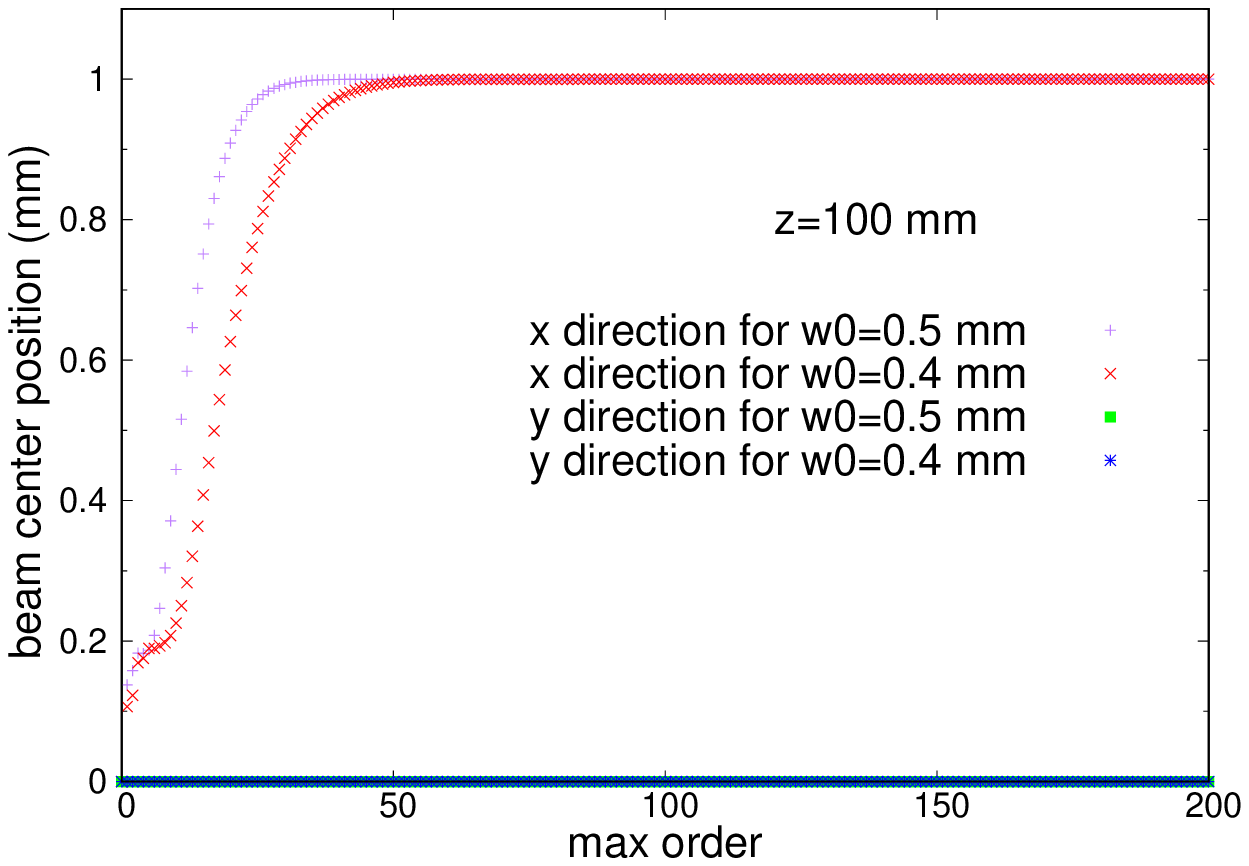}\label{cen_sahgb10_100_ord}
		
	\end{minipage}
	}
	\subfigure[]{
	\begin{minipage}[t]{0.35\linewidth}
		\centering
		\includegraphics[width=0.99\textwidth]{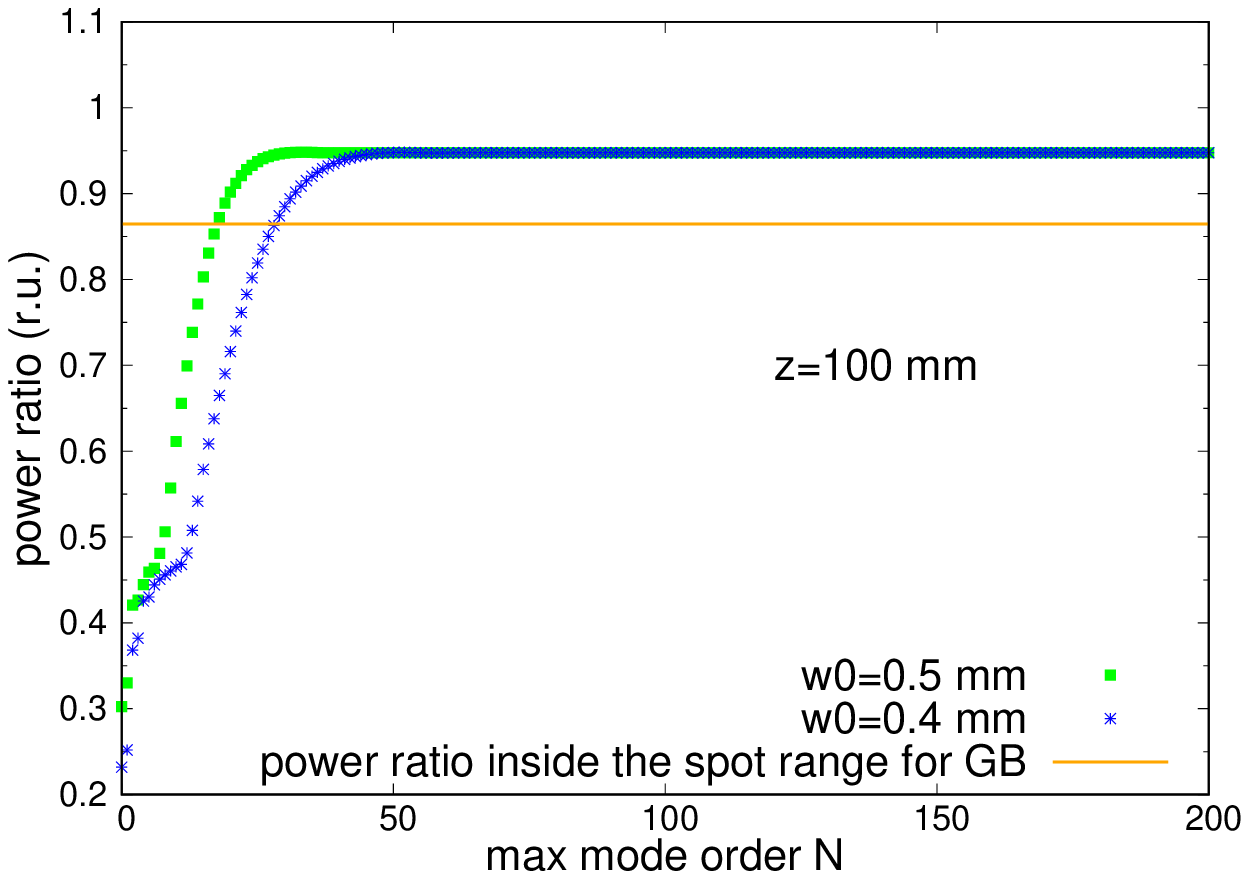}\label{pr_sahgb10_100_ord}
		
	\end{minipage}
	}
	\subfigure[]{
	\begin{minipage}[t]{0.35\linewidth}
		\centering
		\includegraphics[width=0.99\textwidth]{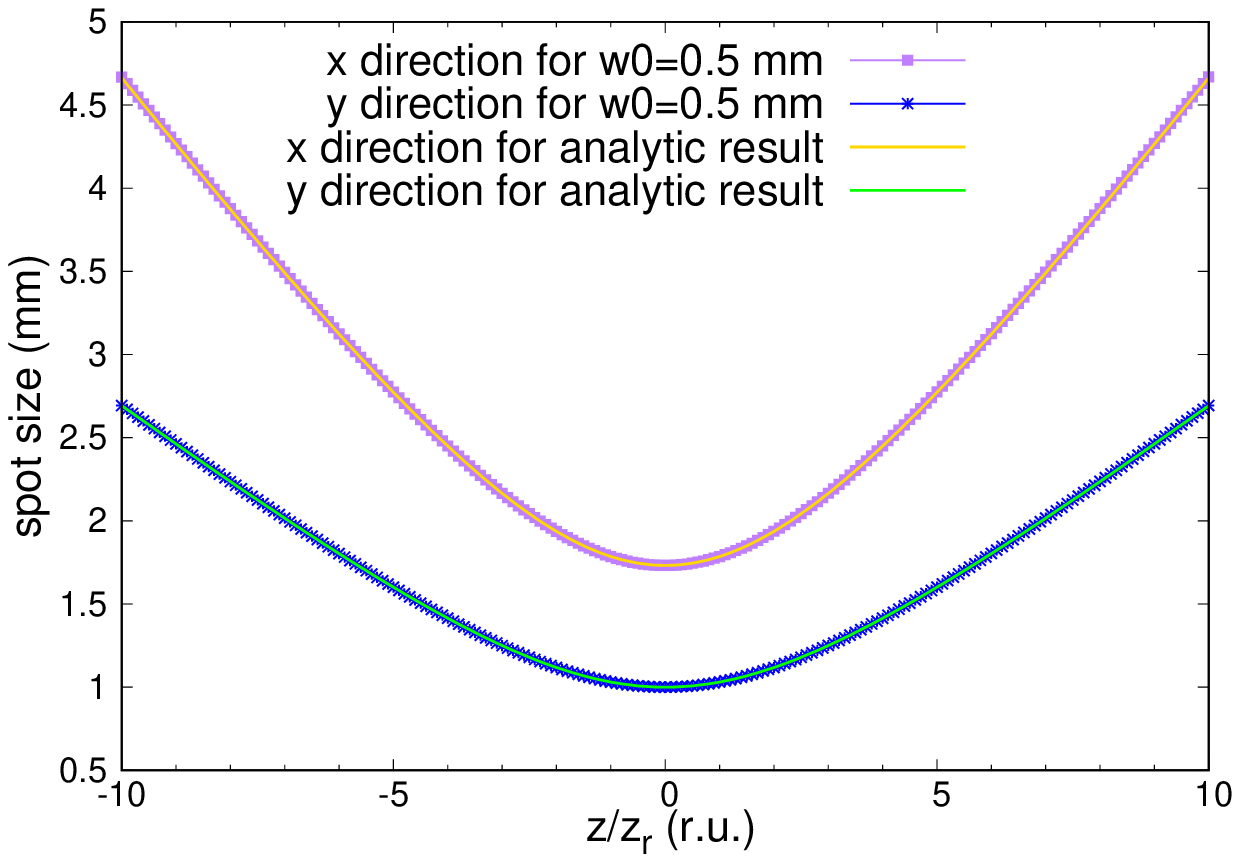}\label{spot_sahgb10_100_scan_z}
		
	\end{minipage}
	}
	\centering
	\caption{(a) The estimated values of spot size for different max decomposed mode order superpositions of $z= \SI{100}{mm}$. (b) The estimated values of beam center for different max decomposed mode order superpositions of $z= \SI{100}{mm}$.(c). The power ratio within the MSD spot size for different max mode order $N$ with $z= \SI{100}{mm}$. (d). The estimated values of spot size of MEM beam compared with the analytic spot size for different propagation distances.}\label{sa10}
\end{figure}
\Cref{sa10_100,sa10_30000} show the intensity profile for $z= \SI{100}{mm}$ and $z= \SI{30000}{mm}$ of the MEM beam which represent for the shifted $(1,0)$ HG beam. In \cref{sa10_inten}, the yellow line is the boundary of the MSD spot for the MEM beam, the green point is the estimated beam center $(1,0)$ and the red line is the coordinate axis. The waist of the basic HG modes for the MEM beam is $w_0=\SI{0.5}{mm}$ and the max mode order is $N=200$. The variations for the MSD spot size, energy center and the power ratio within the MSD spot range with different max mode order $N$ are shown in \cref{spot_sahgb10_100_ord,cen_sahgb10_100_ord,pr_sahgb10_100_ord}. The waist of the basic HG modes for the MEM beam is $w_0=\SI{0.4}{mm}$ or $w_0=\SI{0.5}{mm}$ and $z= \SI{100}{mm}$. As for \cref{spot_sahgb10_100_scan_z}, it clearly shows that the MSD spot size is coincide with the analytic result for a HG beam with different $z$. 

The MSD spot size also is the hyperbolic function of $z$. The fractional beam energy concentrated inside the MSD spot range of this situation is $94.77\%$. If the max mode order $N$ is large enough, the waist $w_0$ of the basic HG modes won't influence the estimated value of MSD spot size and the power ratio within the MSD spot range and the estimated value of MSD spot size and the power ratio within the MSD spot range coincident with the analytic result.

\section{ diffracted aberration Gaussian beam}\label{diffraction_aberr}
 Aberration and diffraction are common phenomena that occur in imaging systems. It is important to investigate the effects of the diffraction of optical aberrations of Gaussian beams. In this section, we calculate the MSD spot size for diffracted beams which are generated by passing aberrated Gaussian beams through an aperture. We then assume a non-perfect Gaussian beam with aberration $\phi_R$, which we model by using Zernike polynomials
\begin{equation}
\begin{aligned}
	E_{aberr}(x,y,z)&=u_{00}(x,y,z)\exp{\left(iwt\right)}\exp{\left(ik\phi_R\right)}\\
	 \phi_R &=\sum_{m}\sum_{n}Z_n^m
\end{aligned}
\end{equation}
Where  $Z_n^m$ is the Zernike polynomial. A circular aperture is put in the waist plane of the aberrated Gaussian beam and the beam passes through it vertically. The aperture center is located on the Gaussian beam axis of this aberrated beam and the coordinate value for aperture center is $(0,0,0)$. The diffracted beam on the aperture plane is
\begin{equation}
E_{diff}(x,y,0)=\left\{ 
\begin{aligned}
&E_{aberr}(x,y,0),\ x^2+y^2\leq r^2\\
&0,\ x^2+y^2>r^2
\end{aligned}
\right..
\end{equation}

For simplicity, here we set the waist for $u_{00}(x,y,z)$ is $w_0=\SI{1}{mm}$, $\lambda=\SI{1064}{nm}$, the aperture radius $r=w_0=\SI{1}{mm}$ and the aberration phase is $\phi_R=Z_{3}^{-3}=\sqrt{8}\rho^3\sin3\phi$, where $\rho=\sqrt{x^2+y^2}$ and $\phi$ is the corresponding azimuthal angle. We decomposed the aberrated beam by \cref{superhgdecom_finite} in the aperture plane and the propagation of the diffracted beam can be represented by the propagation of the decomposed MEM beam. We use five different waist settings for MEM to produce five MEM decomposed beams. The settings are $w_0=\SI{0.1}{mm}$, $w_0=\SI{0.2}{mm}$, $w_0=\SI{0.3}{mm}$, $w_0=\SI{0.4}{mm}$ and $w_0=\SI{0.5}{mm}$. All of these MEM beams have the same max mode order $N=200$. The beam centers and the propagation directions of the basic HG modes for MEM beams are all the same as the beam center and the propagation direction of the diffracted aberration beam. 

\begin{figure}[H]
	\centering
	\subfigure[]{
	\begin{minipage}[t]{0.33\linewidth}
		\centering
		\includegraphics[width=0.95\textwidth]{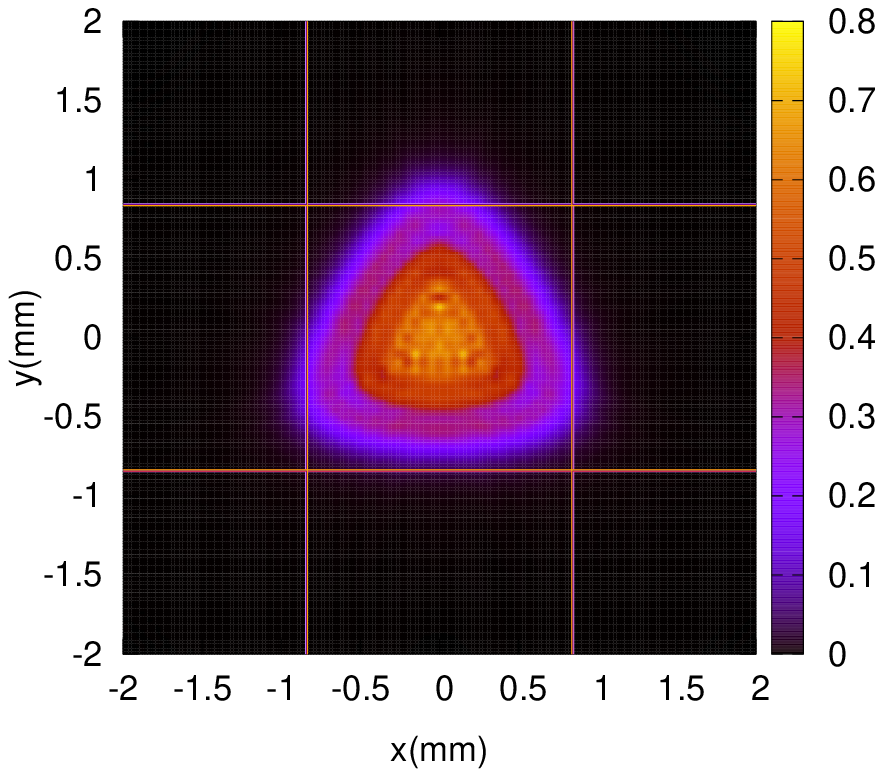}\label{aberr_inten_100}
		
	\end{minipage}
	}
	\subfigure[]{
	\begin{minipage}[t]{0.27\linewidth}
		\centering
		\includegraphics[width=0.95\textwidth]{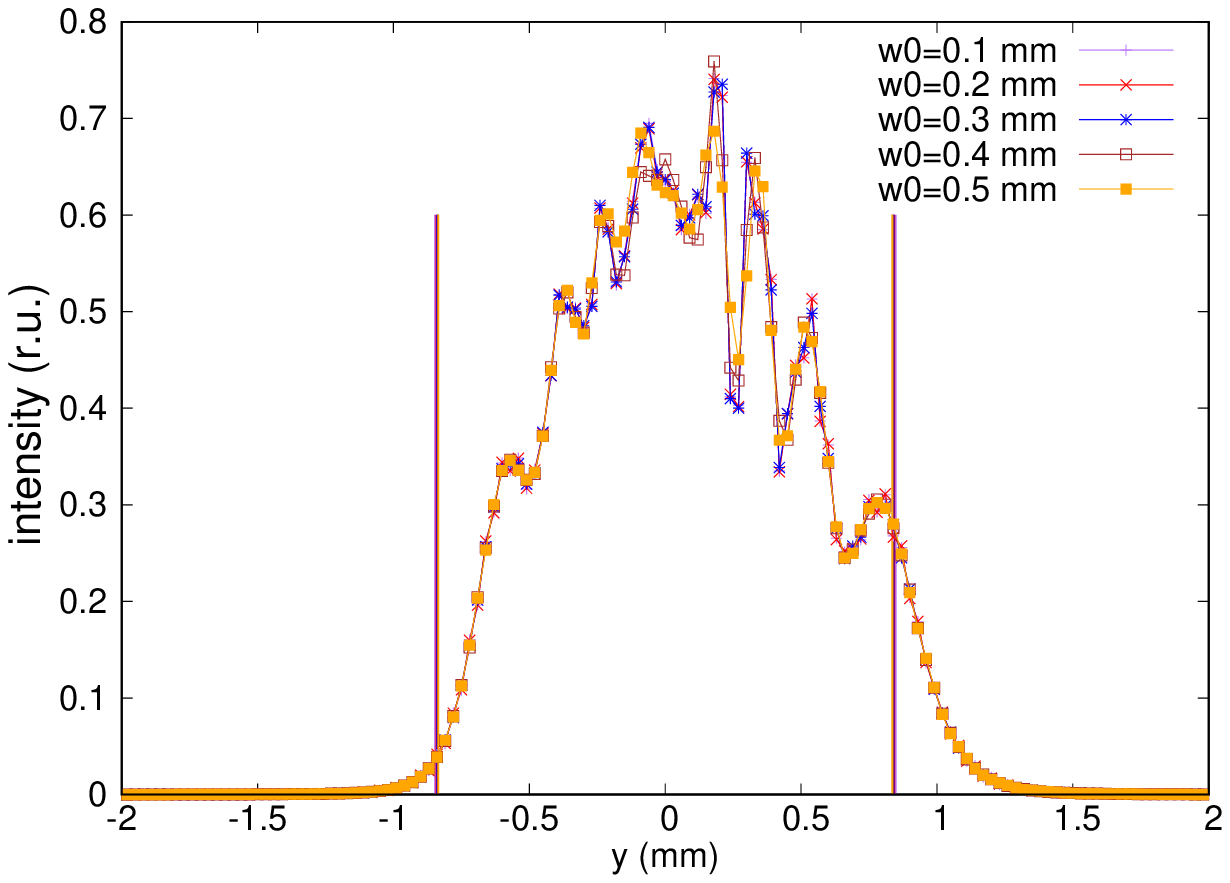}\label{aberr_inten_2d_100}
		
	\end{minipage}
	}\subfigure[]{
	\begin{minipage}[t]{0.27\linewidth}
		\centering
		\includegraphics[width=0.95\textwidth]{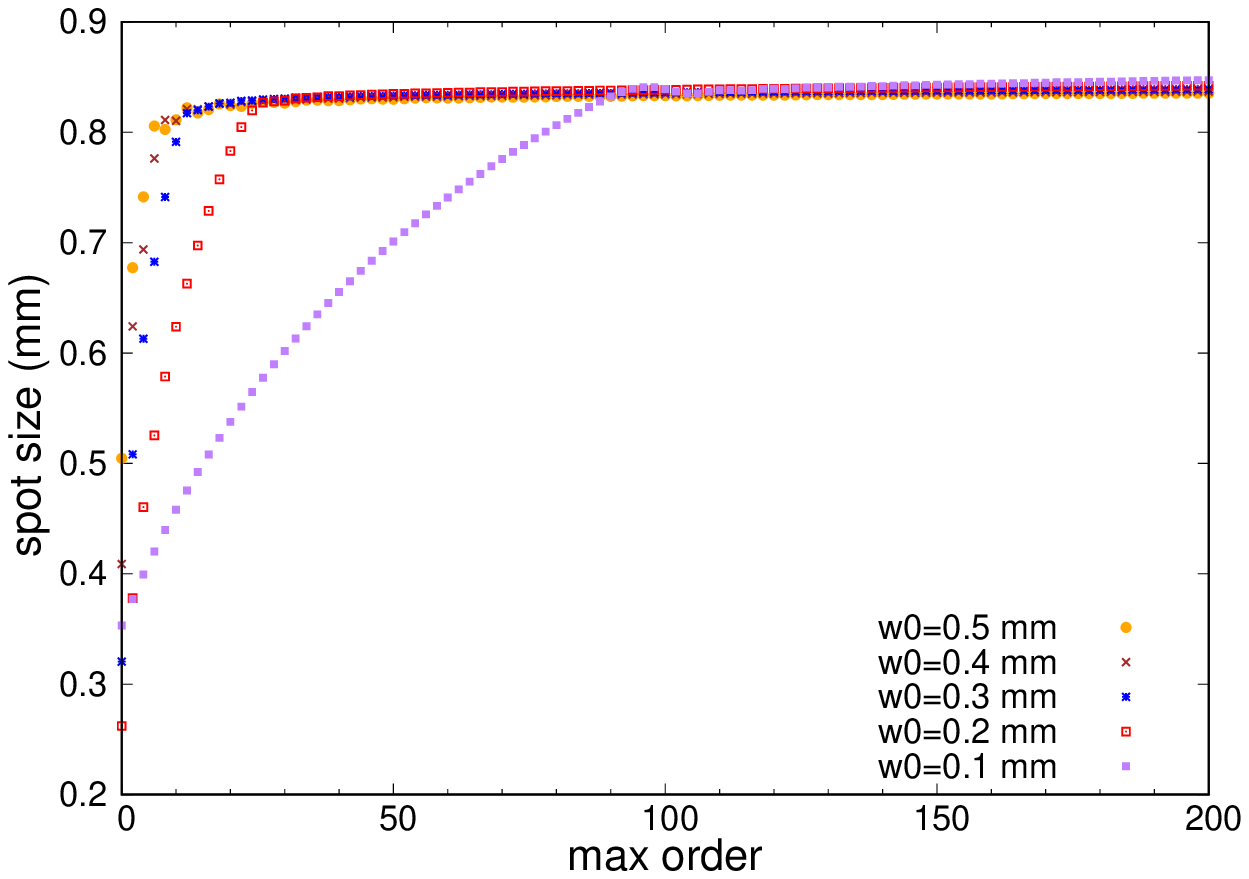}\label{aberr_spot_100}
		
	\end{minipage}
	}
	\subfigure[]{
	\begin{minipage}[t]{0.33\linewidth}
		\centering
		\includegraphics[width=0.95\textwidth]{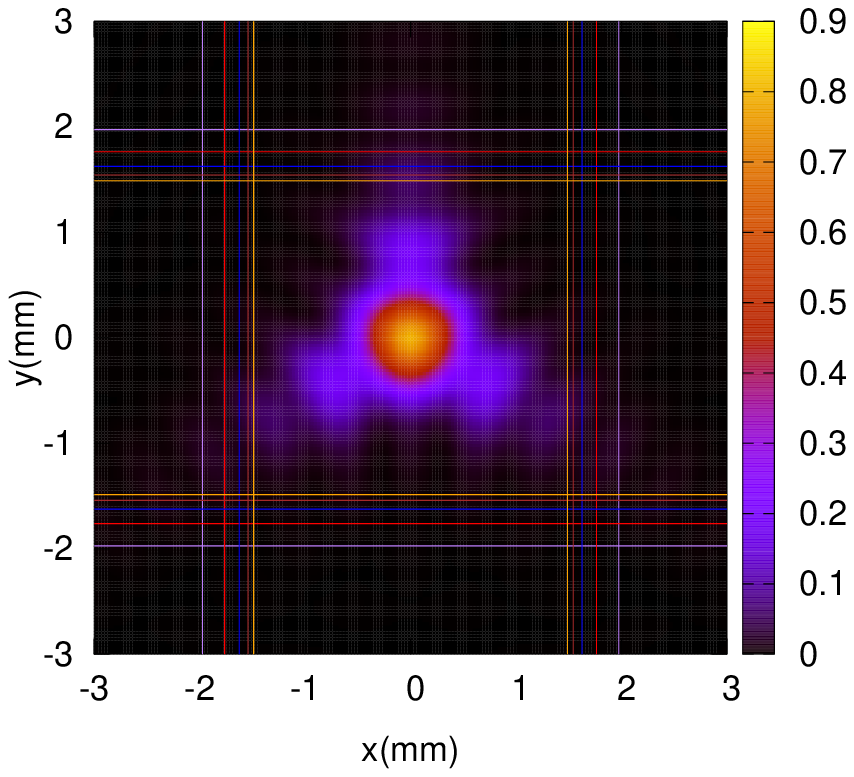}\label{aberr_inten_1000}
		
	\end{minipage}
	}
	\subfigure[]{
	\begin{minipage}[t]{0.27\linewidth}
		\centering
		\includegraphics[width=0.95\textwidth]{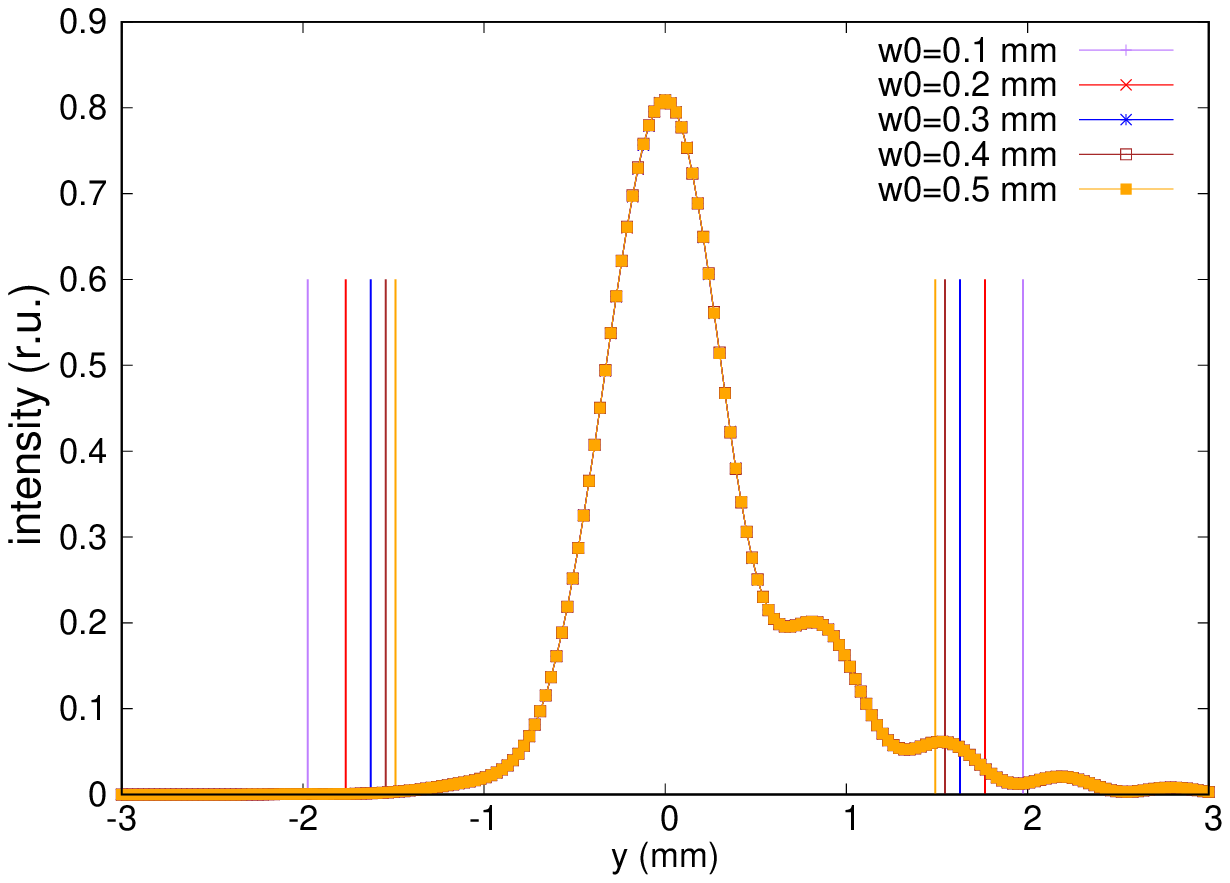}\label{aberr_inten_2d_1000}
		
	\end{minipage}
	}\subfigure[]{
	\begin{minipage}[t]{0.27\linewidth}
		\centering
		\includegraphics[width=0.95\textwidth]{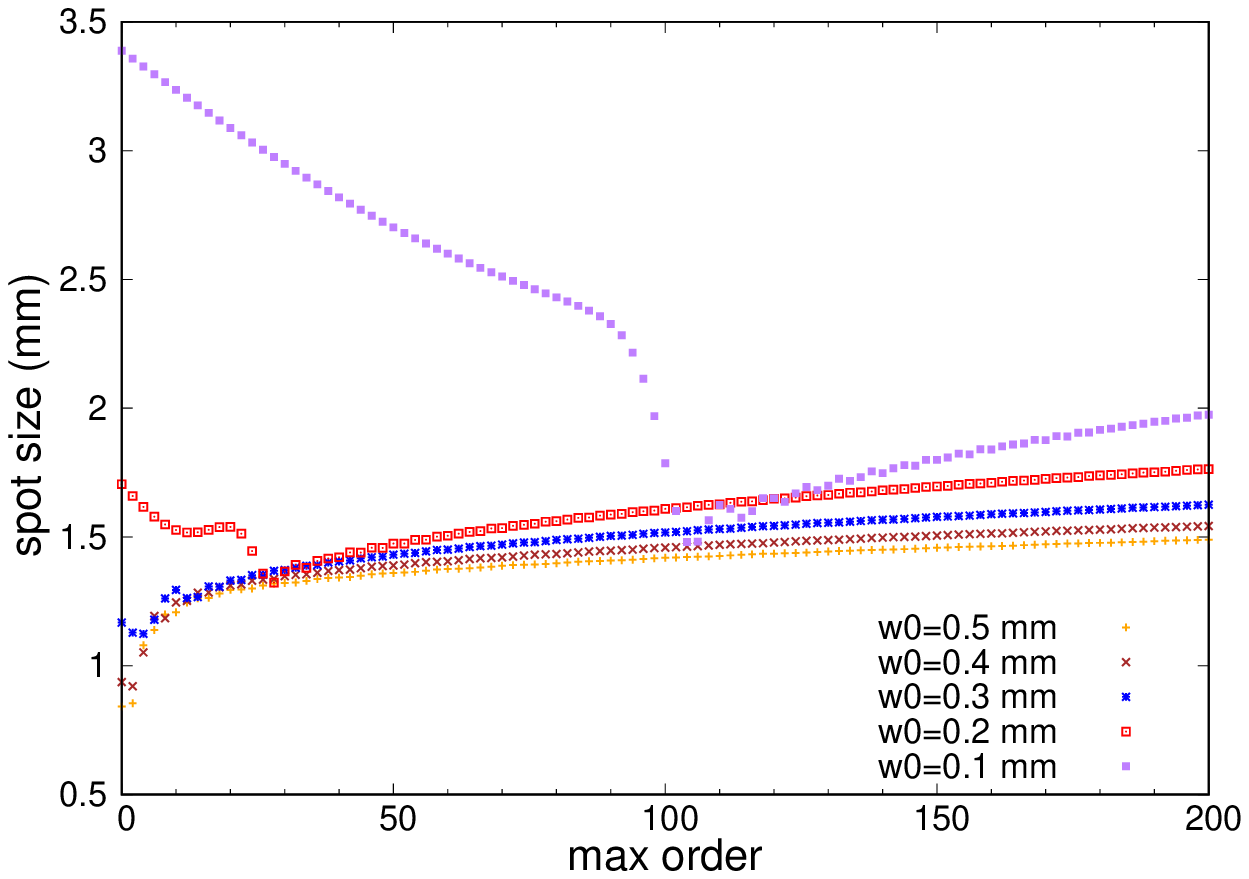}\label{aberr_spot_1000}
		
	\end{minipage}
	}
	\caption{The intensity profiles of the simulated gaussian beam with aberration $\phi_R=Z_{3}^{-3}$ for the propagation distance $z=\SI{100}{mm}$ (a) and $z=\SI{1000}{mm}$ (d). The intensity distributions of the aberration beam along y-axis for the propagation distance $z=\SI{100}{mm}$ (b) and $z=\SI{1000}{mm}$ (e). The MSD spot size of the simulated aberration beam for different max mode order $N$ with the propagation distance $z=\SI{100}{mm}$ (c) and $z=\SI{1000}{mm}$ (f).}\label{aberr}
\end{figure}

\Cref{aberr_inten_100,aberr_inten_1000} show the intensity profiles of the MEM beams which represent the aberration beam for $z=\SI{100}{mm}$ and $z=\SI{1000}{mm}$. The waist of the basic HG modes of the MEM beams is $w_0=\SI{0.3}{mm}$ and the max mode order $N$ of the MEM beams is $200$. \Cref{aberr_inten_2d_100,aberr_inten_2d_1000} show the intensity distributions of the MEM beams along the y-axis of \cref{aberr_inten_100,aberr_inten_1000}. The purple vertical lines in these subfigures represent the MSD spot size for the MEM beam with $w_0=\SI{0.1}{mm}$, the red lines represent $w_0=\SI{0.2}{mm}$, the blue lines represent $w_0=\SI{0.3}{mm}$, the brown lines represent $w_0=\SI{0.4}{mm}$ and the orange lines represent $w_0=\SI{0.5}{mm}$. The max mode orders $N$ in the above cases are fixed to 200. The variations of the MSD spot size with the max mode order $N$ for  $z=\SI{100}{mm}$ and $z=\SI{1000}{mm}$ are shown in \cref{aberr_spot_100,aberr_spot_1000} separately. 

The maximum mode order $N$ of the MEM beam and the waist size $w_0$ of the MEM basic HG modes affect the calculation of the spot size of the diffracted aberration beam.  As \cref{aberr} shows, all these spot ranges are big enough to contain the main part of the intensity patterns. The estimated spot ranges for these MEM beams are almost the same for near propagation ranges such as $z=\SI{100}{mm}$. As for $z=\SI{1000}{mm}$, the estimated spot ranges for different MEM beams are obviously different.  The larger the waist for the basic HG modes is, the smaller the MSD spot size and power ratio within the MSD spot range are. As mentioned before, the MSD spot size is divergent in hard-edge diffraction situations such as the diffracted aberration beam here. Using the MEM method, we can get a finite MSD spot size for diffracted aberration beam. However, this finite MSD spot size is dependent on the waist $w_0$ of the basic HG modes and the max mode order $N$ of the MEM beam. 
\newpage

\section{The parameters of the MEM beam for the halo beam}\label{sec:table_halo}
The waist of the basic HG modes of the MEM beam for the halo beam is $w_0=\SI{644.314}{\mu m}$. The corresponding complex coefficients for different HG modes are shown in \cref{table_halo}. The HG modes in \cref{table_halo} are ordered in the absolute amplitude of complex coefficient $a_{mn}$ decreasingly.

\begin{table}[H]
    \tbl{complex coefficients for different HG modes of halo beam}
    {
    \footnotesize
    \setlength{\tabcolsep}{5pt}
    \renewcommand{\arraystretch}{1.0}
    \centering
    \begin{tabular}{lccc}
        \hline
        mode  HG(m,n) & complex coefficient $a_{mn}$&mode  HG(m,n) & complex coefficient $a_{mn}$\\
        \hline
        HG(0,0)&1.000000+0.000000i &HG(1,2)&-0.043850+0.030069i\\
        \hline
HG(0,2)&-0.043689+0.208222i &HG(5,4)&-0.013644+0.044648i\\
        \hline
HG(0,8)&0.197328+0.079289i &HG(7,1)&0.041697+0.015981i\\
        \hline
HG(2,0)&0.059580+0.202141i &HG(2,3)&0.034655-0.027614i\\
        \hline
HG(8,0)&0.161853+0.090453i &HG(4,5)&0.023136-0.037789i\\
        \hline
HG(6,0)&0.017052+0.166904i &HG(3,6)&-0.015738+0.037032i\\
        \hline
HG(7,0)&-0.155352+0.053572i &HG(1,8)&-0.003275+0.039806i\\
        \hline
HG(2,6)&0.138949+0.064269i &HG(5,3)&0.033787+0.016435i\\
        \hline
HG(4,4)&0.131712+0.067135i &HG(3,2)&-0.035438-0.009784i\\
        \hline
HG(0,6)&-0.009895+0.144958i &HG(4,1)&0.035344-0.005482i\\
        \hline
HG(0,4)&0.143830-0.009943i &HG(3,5)&0.025634+0.022770i\\
        \hline
HG(6,2)&0.128264+0.065381i &HG(6,3)&0.021603-0.024983i\\
        \hline
HG(0,7)&0.121691-0.062886i &HG(0,11)&-0.030008-0.013461i\\
        \hline
HG(4,2)&-0.000903+0.128729i &HG(1,5)&-0.007792+0.030758i\\
        \hline
HG(2,4)&-0.007101+0.123730i &HG(1,7)&0.021235+0.020630i\\
        \hline
HG(0,10)&0.092926-0.069031i &HG(9,1)&0.024615-0.015661i\\
        \hline
HG(4,0)&0.105540-0.014269i &HG(5,1)&0.005574+0.027560i\\
        \hline
HG(0,9)&0.016402-0.104016i &HG(3,3)&0.006984+0.025667i\\
        \hline
HG(0,3)&0.092075-0.035122i &HG(8,1)&0.022251-0.014360i\\
        \hline
HG(5,2)&-0.092037+0.034858i &HG(1,4)&-0.018101-0.014466i\\
        \hline
HG(2,2)&0.089174-0.014776i &HG(11,0)&0.020614-0.010033i\\
        \hline
HG(2,5)&0.079592-0.040046i &HG(7,3)&0.017058-0.014068i\\
        \hline
HG(9,0)&-0.015792+0.087087i &HG(1,10)&0.015733-0.000945i\\
        \hline
HG(3,0)&-0.066284+0.043780i &HG(1,9)&0.009205-0.011677i\\
        \hline
HG(3,4)&-0.074522+0.020353i &HG(5,5)&0.009195-0.005310i\\
        \hline
HG(10,0)&0.072191-0.022591i &HG(3,7)&0.008557-0.001641i\\
        \hline
HG(2,8)&0.062475-0.038479i &HG(3,1)&0.007785-0.001349i\\
        \hline
HG(4,6)&0.062011-0.034334i &HG(2,9)&-0.002583-0.007210i\\
        \hline
HG(5,0)&-0.049725-0.043344i &HG(6,5)&0.004232-0.005767i\\
        \hline
HG(6,4)&0.059852-0.026250i &HG(1,1)&0.006285-0.001269i\\
        \hline
HG(4,3)&0.060351-0.023438i &HG(3,8)&0.005092+0.001007i\\
        \hline
HG(1,6)&-0.062262+0.016535i &HG(5,6)&0.003077+0.002502i\\
        \hline
HG(2,7)&0.020276-0.058881i &HG(10,1)&0.000208+0.003889i\\
        \hline
HG(6,1)&0.058091-0.010760i &HG(1,3)&-0.003512+0.000830i\\
        \hline
HG(8,2)&0.052511-0.025760i &HG(4,7)&-0.000359-0.003209i\\
        \hline
HG(2,1)&0.049257-0.030469i &HG(9,2)&-0.000481-0.003019i\\
        \hline
HG(0,5)&0.045250-0.033775i &HG(8,3)&0.001417+0.001375i\\
        \hline
HG(7,2)&-0.020618+0.052083i &HG(7,4)&-0.000059-0.000506i\\
        \hline

 \end{tabular}
    \label{table_halo}
    }
\end{table}
\newpage
\section{The parameters of the MEM beam for the Nearly-Gaussian beam \uppercase\expandafter{\romannumeral2}}\label{sec:table_near_gb2}
The waist of the basic HG modes of the MEM beam for the Nearly-Gaussian beam \uppercase\expandafter{\romannumeral2} in \cref{near_gb2} is $w_0=\SI{399.465}{\mu m}$. The corresponding complex coefficients for different HG modes are shown in \cref{table_near_gb2}. The HG modes in \cref{table_near_gb2} are ordered in the absolute amplitude of complex coefficient $a_{mn}$ decreasingly.

\begin{table}[H]
    \tbl{complex coefficients for different HG modes of Nearly-Gaussian beam \uppercase\expandafter{\romannumeral2}}
   { \footnotesize
    \setlength{\tabcolsep}{5pt}
    \renewcommand{\arraystretch}{1.0}
    \centering
    \begin{tabular}{lccc}
        \hline
        mode  HG(m,n) & complex coefficient $a_{mn}$&mode  HG(m,n) & complex coefficient $a_{mn}$\\
        \hline
HG(0,0)&1.000000+0.000000i &HG(8,0)&-0.019123+0.021579i\\
        \hline
HG(2,0)&0.278731+0.219901i &HG(1,6)&-0.003373+0.028556i\\
        \hline
HG(0,2)&-0.278731-0.219901i &HG(1,8)&0.025500-0.011701i\\
        \hline
HG(7,0)&0.091524+0.172152i &HG(1,5)&0.011388-0.025343i\\
        \hline
HG(5,0)&0.074602-0.159708i &HG(8,1)&0.026540-0.005864i\\
        \hline
HG(0,4)&0.045154+0.159202i &HG(6,4)&0.009602-0.024680i\\
        \hline
HG(2,2)&-0.077211-0.129553i &HG(4,6)&0.023531+0.011102i\\
        \hline
HG(3,0)&-0.147112-0.019083i &HG(6,1)&-0.024231-0.004293i\\
        \hline
HG(4,0)&-0.102501+0.088332i &HG(3,5)&0.023611+0.005056i\\
        \hline
HG(9,0)&-0.109524-0.077033i &HG(0,5)&-0.010220-0.020312i\\
        \hline
HG(6,0)&0.009151-0.097677i &HG(5,6)&-0.007497-0.021052i\\
        \hline
HG(7,2)&0.011396-0.087076i &HG(2,8)&-0.021737+0.002029i\\
        \hline
HG(1,1)&-0.046646-0.072063i &HG(3,6)&-0.010431+0.017490i\\
        \hline
HG(4,2)&0.068784-0.051119i &HG(0,11)&-0.007996-0.016383i\\
        \hline
HG(5,2)&-0.062910+0.053035i &HG(5,5)&-0.016094-0.005983i\\
        \hline
HG(2,4)&0.011725+0.073327i &HG(9,1)&0.001334+0.016860i\\
        \hline
HG(0,6)&0.035398-0.063483i &HG(6,3)&0.012951+0.010398i\\
        \hline
HG(11,0)&0.036618+0.061069i &HG(7,3)&-0.002000+0.015624i\\
        \hline
HG(3,2)&0.045392+0.043331i &HG(4,1)&0.012962+0.007516i\\
        \hline
HG(1,3)&0.013945+0.055936i &HG(1,10)&-0.010509-0.010347i\\
        \hline
HG(4,4)&-0.049109+0.017232i &HG(2,7)&-0.013172-0.002769i\\
        \hline
HG(1,2)&0.036656+0.036444i &HG(2,5)&0.001213-0.013254i\\
        \hline
HG(5,4)&0.049164+0.001326i &HG(4,3)&-0.008941-0.009230i\\
        \hline
HG(6,2)&-0.001349+0.044950i &HG(10,1)&-0.012455+0.002736i\\
        \hline
HG(2,6)&0.021935-0.038604i &HG(2,1)&0.012290+0.002338i\\
        \hline
HG(3,4)&-0.002739-0.040826i &HG(2,3)&0.011532+0.001631i\\
        \hline
HG(9,2)&0.001930+0.040420i &HG(1,7)&-0.008886+0.005939i\\
        \hline
HG(3,1)&0.021898-0.029556i &HG(8,3)&-0.007374-0.006575i\\
        \hline
HG(0,7)&-0.016107+0.032728i &HG(2,9)&0.000048+0.009559i\\
        \hline
HG(0,9)&0.036413-0.001986i &HG(3,7)&0.001158-0.009293i\\
        \hline
HG(1,4)&-0.019041-0.029166i &HG(0,10)&0.003370+0.008609i\\
        \hline
HG(5,1)&0.002507+0.033980i &HG(4,5)&0.006450+0.005334i\\
        \hline
HG(3,3)&-0.028347+0.016449i &HG(8,2)&-0.003821-0.004496i\\
        \hline
HG(7,4)&-0.030367+0.009496i &HG(3,8)&0.004060+0.001224i\\
        \hline
HG(10,0)&0.030037+0.008183i &HG(6,5)&0.000037-0.003933i\\
        \hline
HG(5,3)&0.018109-0.024213i &HG(0,3)&0.003511+0.001090i\\
        \hline
HG(7,1)&-0.019281-0.022492i &HG(1,9)&0.003354-0.000715i\\
        \hline
HG(0,8)&-0.028999-0.001569i &HG(4,7)&-0.000144-0.003408i\\
        \hline
\end{tabular}
    \label{table_near_gb2}}
\end{table}
\bibliography{library}
\bibliographystyle{ws-ijmpd}
\end{document}